\documentclass[aps,pre,twocolumn,nofootinbib]{revtex4}
\usepackage{amssymb,macros2erev4,graphicx,rotate,color}
\usepackage{times}
\newcommand{\be}{\begin{equation}}
\newcommand{\ee}{\end{equation}}
\newcommand{\bea}{\begin{eqnarray}}
\newcommand{\eea}{\end{eqnarray}}

\def\clip{T}  % this should be T=true or F=false
\newcommand{\fig}[2]{\includegraphics[width=#1\columnwidth]{./figures/#2}}
\newcommand{\Fig}[1]{\includegraphics[width=\columnwidth]{./figures/#1}}
\if\clip T

\newcommand{\clipfig}[2]{\includegraphics[width=#1\columnwidth,clip]{./figures/#2}}
\else

\newcommand{\clipfig}[2]{\includegraphics[width=#1\columnwidth]{./figures/#2}}
\fi

\newlength{\bilderlength}

\newcommand{\rme}{{\mathrm{e}}}
\newcommand{\rmd}{{\mathrm{d}}}
\newcommand{\nn}{\nonumber}

\newcommand{\E}{\epsilon}

\newcommand{\cgamma}{{\gamma}}
\newcommand{\ceta}{{\eta}}

\newcommand{\beq}{\begin{equation}}
\newcommand{\eeq}{\end{equation}}
\newcommand{\beqa}{\begin{eqnarray}}
\newcommand{\eeqa}{\end{eqnarray}}
\newcommand{\rp}{{\bf r}_\perp}
\newcommand{\rl}{{\bf r}_\parallel}
\newcommand{\un}{u_{n}}
\newcommand{\undot}{\dot{u}_{n}}

\begin{document}
\def\theequation{\arabic{section}.\arabic{equation}}
\bibliographystyle{../../macros/KAY.bst}
%\bibliographystyle{unsrt}
%%%%%%%%%%%%%%%%%%%%%%%%%%%%%%%%%%%%%%%%%%%%%%%%%%%%%%%%%%%%%%%%%%%%%%%%%%%%
%\begin{frontmatter}
%
\title{\sffamily\bfseries\large Depinning in a two-layer model of plastic flow}
\author{\sffamily\bfseries\normalsize  Pierre Le Doussal{$^1$},
M. Cristina Marchetti{$^2$}, Kay J\"org Wiese{$^1$}
 \vspace*{3mm}}
\affiliation{{$^1$} CNRS-Laboratoire de Physique Th{\'e}orique de
l'Ecole Normale Sup{\'e}rieure,
24 rue Lhomond, 75005 Paris, France\\
{$^2$} Physics Department, Syracuse University, Syracuse NY 13244, USA. \medskip }
\date{\today}

\begin{abstract}
We study a model of two layers, each consisting of a $d$-dimensional elastic object driven over a
random substrate, and mutually interacting through a viscous
coupling. For this model, the mean-field theory (i.e.\ a fully connected model) predicts a transition from
elastic depinning to hysteretic plastic depinning as disorder or viscous coupling is increased.
A functional RG analysis shows that any small inter-layer viscous coupling destablizes the standard (decoupled)
elastic depinning FRG fixed point for $d \leq 4$, while for $d>4$ most aspects of the mean-field theory
are recovered. A one-loop study at non-zero velocity indicates, for $d<4$, coexistence of a moving state
and a pinned state below the elastic depinning threshold, with hysteretic plastic depinning
for periodic and non-periodic driven layers.
A  2-loop analysis of quasi-statics unveils the possibility of more subtle effects, including
a new universality class for non-periodic objects.
We also study the model in $d=0$, i.e.\ two coupled particles, and show that hysteresis
does not always exist as the periodic steady state with coupled layers
can be dynamically unstable. It is also
proved  that stable pinned configurations
remain dynamically stable in presence of a viscous coupling in any dimension $d$. Moreover,
the layer model for periodic objects is stable to an infinitesimal commensurate
density coupling. Our work shows that a careful study of attractors in phase space
and their basin of attraction is necessary to obtain a firm conclusion for dimensions $d=1,2,3$.
\end{abstract}

\maketitle
%%%%%%%%%%%%%%%%%%%%%%%%%%%%%%%%%%%%%%%%%%%%%%%%%%%%%%%%%%%%%%%%%%%%%%%%%%%%
%%%%%%%%%%%%%%%%%%%%%%%%%%%%%%%%%%%%%%%%%%%%%%%%%%%%%%%%%%%%%%%%%%
%%%%%%%%%%%%%%%%%%%%%%%%%%%%%%%%%%%%%%%%%%%%%%%%%%%%%%%%%%%%%%%%%%
%%%%%%%%%%%%%%%%%%%%%%%%%%%%%%%%%%%%%%%%%%%%%%%%%%%%%%%%%%%%%%%%%%
%%%%%%%%%%%%%%%%%%%%%%%%%%%%%%%%%%%%%%%%%%%%%%%%%%%%%%%%%%%%%%%%%%
%%%%%%%%%%%%%%%%%%%%%%%%%%%%%%%%%%%%%%%%%%%%%%%%%%%%%%%%%%%%%%%%%%
%%%%%%%%%%%%%%%%%%%%%%%%%%%%%%%%%%%%%%%%%%%%%%%%%%%%%%%%%%%%%%%%%%
%%%%%%%%%%%%%%%%%%%%%%%%%%%%%%%%%%%%%%%%%%%%%%%%%%%%%%%%%%%%%%%%%%
%%%%%%%%%%%%%%%%%%%%%%%%%%%%%%%%%%%%%%%%%%%%%%%%%%%%%%%%%%%%%%%%%%
%%%%%%%%%%%%%%%%%%%%%%%%%%%%%%%%%%%%%%%%%%%%%%%%%%%%%%%%%%%%%%%%%%
%                                                                %
%                        Introduction                            %
%                                                                %
%%%%%%%%%%%%%%%%%%%%%%%%%%%%%%%%%%%%%%%%%%%%%%%%%%%%%%%%%%%%%%%%%%
%%%%%%%%%%%%%%%%%%%%%%%%%%%%%%%%%%%%%%%%%%%%%%%%%%%%%%%%%%%%%%%%%%
%%%%%%%%%%%%%%%%%%%%%%%%%%%%%%%%%%%%%%%%%%%%%%%%%%%%%%%%%%%%%%%%%%
%%%%%%%%%%%%%%%%%%%%%%%%%%%%%%%%%%%%%%%%%%%%%%%%%%%%%%%%%%%%%%%%%%
%%%%%%%%%%%%%%%%%%%%%%%%%%%%%%%%%%%%%%%%%%%%%%%%%%%%%%%%%%%%%%%%%%
%%%%%%%%%%%%%%%%%%%%%%%%%%%%%%%%%%%%%%%%%%%%%%%%%%%%%%%%%%%%%%%%%%
%%%%%%%%%%%%%%%%%%%%%%%%%%%%%%%%%%%%%%%%%%%%%%%%%%%%%%%%%%%%%%%%%%
%%%%%%%%%%%%%%%%%%%%%%%%%%%%%%%%%%%%%%%%%%%%%%%%%%%%%%%%%%%%%%%%%%
%%%%%%%%%%%%%%%%%%%%%%%%%%%%%%%%%%%%%%%%%%%%%%%%%%%%%%%%%%%%%%%%%%

\section{Introduction}

\subsection{Overview}

Nonequilibrium transitions from stuck to moving states underlie the physics of a wide range of phenomena
\cite{DSFisher1998}, from fracture and earthquake rupture \cite{LawnBook1993,RamanathanFisher1997,RamanathanFisher1998}
to flux flow in type-II superconductors~  \cite{BlatterFeigelmanGeshkenbeinLarkinVinokur1994,GiamarchiBhattacharya2002} and sliding of charge density
waves in metals \cite{FukuyamaLee1978,LeeRice1979,Gruner1988,Thorne1996}. The rich collective nonequilibrium dynamics of this
broad range of phenomena can be modeled as  an extended medium driven over quenched disorder.
One can distinguish two main classes depending on whether the description allows or not for plastic deformations of the medium. Within each class one may restrict to microscopic overdamped dynamics or allow for more complicated, e.g.\ inertial, effects.

The first class of models, overdamped {\it elastic} media pulled by an applied force $f$, has been studied extensively.
By definition, the driven medium can be deformed by disorder but is not allowed to tear, and topological defects are excluded, the only degrees of freedom being elastic deformations. The question of applicability of this model to realistic situations is still debated in the static case ($f=0$), and even more so in the driven dynamics. The general expectation is that such a model is relevant to describe real systems in some range of lengthscales, a range which becomes broad (and potentially infinite, depending on space dimension) in weak-disorder, strong-elasticity situations. Indeed one may conceive that, even if topological defects can be generated by the competition of elasticity, disorder and drive, they may remain bounded, and confined to shorter scales and thus unimportant for the effective large-scale description. This is known to happen in the statics, e.g.\ for interfaces in random ferromagnets. Even when topological defects are relevant at large scale, the elastic description may still apply at shorter scales. Hence the overdamped elastic model is a necessary first step to understand the collective dynamics in more complex situations. Many results were obtained for this model, although some questions remain open: At zero temperature
($T=0$) the elastic model exhibits a nonequilibrium
phase transition from a pinned to a  sliding state at a critical value $f_\mathrm{c}$ of the driving force, first studied in the context of charge density waves \cite{MiddletonFisher1991,MiddletonFisher1993}. Starting from mean-field theory \cite{DSFisher1985}, an analogy with standard critical phenomena was developed, with the medium's mean velocity $v$ acting as the order parameter,  and a diverging correlation length \cite{DSFisher1985,DSFisher1998}. A functional extension of usual RG was developed to treat quenched disorder and obtain the roughness and dynamical exponents at the threshold $v=0^{+}$ to 1-loop accuracy \cite{NarayanDSFisher1992b,NattermanStepanowTangLeschhorn1992}. Extensions at non-zero $v$ emphasized the differences with standard critical phenomena \cite{ChauveGiamarchiLeDoussal2000}. It was shown that a two-loop Functional RG (FRG) approach is necessary to fully describe the difference between statics and $v=0^{+}$ quasi-static depinning \cite{ChauveLeDoussalWiese2000a,LeDoussalWieseChauve2002,LeDoussalWieseChauve2003} and to reach satisfactory agreement with numerical simulations \cite{Middleton1995,MyersSethna1993,RossoHartmannKrauth2002}. Universality classes were identified, which are
distinguished, for example, by the range of interactions or by the periodicity (or nonperiodicity) of the
pinning forces. A key feature of the overdamped elastic model is that for one component displacements $N=1$, i.e.\ interfaces, the sliding state is unique, the $v(f)$ curve is single-valued, and no hysteresis can occur in the moving state at $v>0$. This property, based on Middleton's theorem \cite{Middleton1992}, which also leads to simplifications \cite{LeDoussalWieseChauve2002} in the FRG description for $N=1$ is not expected to hold for $N>1$. As a result, the understanding of the $N>1$ depinning transition for e.g.\ lines or vortex lattices, is still not satisfactory despite some attempts\cite{ErtasKardar1994,ErtasKardar1996,Kardar1997}. Furthermore, there is a second type of universality classes for depinning (e.g.\ anisotropic depinning) which does not obey the so-called statistical tilt symmetry (or rotation symmetry) and where non-linear terms become relevant (e.g.\ Kardar-Parisi-Zhang (KPZ) like terms) \cite{TangKardarDhar1995}. Despite efforts \cite{BarabasiGrinsteinMunoz1996,LeDoussalWiese2002a}, a complete theory for this class is still lacking, and even the value of the upper critical dimension is a matter of debate. The question of non-linear terms may be of importance to experiments of contact line depinning and cracks \cite{LeDoussalWieseRaphaelGolestanian2004,KatzavAdda-BediaBenAmarBoudaoud2007,BouchbinderBregmanProcaccia2007}. Away from depinning, well into the uniformly sliding state at $v>0$, it was found that the dynamics can be surprisingly rich
\cite{Hwa1992}, especially for $N>1$ component periodic objects ~\cite{KoshelevVinokur1994,GiamarchiLeDoussalBookYoung}. New terms can be generated in the equation of motion, a linear convective term, a static random-force term, and a host of possible non-linear, KPZ-type terms \cite{Krug1995,BalentsFisher1995,GiamarchiLeDoussal1996,MoonScalettarZimanyi1996,ChenBalentsFisherMarchetti1996,BalentsMarchettiRadzihovsky1997,BalentsMarchettiRadzihovsky1998,LeDoussalGiamarchi1997,PardoDeLaCruzGammelBucherBishop1998}.
For $N>1$, a distinct, ``moving glass'' fixed point was found in the FRG, with persistence of transverse order and transverse pinning, leading to the prediction of a moving Bragg-glass and a moving smectic state~\cite{GiamarchiLeDoussal1996,BalentsMarchettiRadzihovsky1998,LeDoussalGiamarchi1997}. In both states the flow is organized in static-like channels, in a layered fashion. Extensions to correlated disorder was studied, and a moving Bose-glass state predicted  \cite{ChauveLeDoussalGiamarchi2000,OliveSoretDoussalGiamarchi2003}.
Although clear evidence of these effects were found in numerics and experiments ~\cite{PardoDeLaCruzGammelBucherBishop1998}, no systematic study of finite-size corrections was made. Since the simultaneous analytical treatment of all the terms allowed by symmetry within a FRG approach is a problem of formidable complexity~\cite{LeDoussalGiamarchi1997}, even a fully consistent theory of the elastic flow at large velocity is still lacking. Hence the question of which moving state is stable in the thermodynamic limit is still open. Finally, once the elastic system is understood, one may hope to construct arguments for or against stability of the elastic flow to defects. These however are even more delicate than in the statics, where the stability of the Bragg glass was debated, and the validity of the driven elastic model has only been assessed qualitatively ~\cite{BalentsMarchettiRadzihovsky1998,LeDoussalGiamarchi1997,AransonScheidlVinokur1998,ScheidlVinokur1998}.
Hence, as one can see, despite being well studied, the overdamped elastic model is still far from being  understood.  Extensions to include inertial effects and stress overshoots\footnote{Note that in the mean-field limit the stress-overshoot model and the model studied here (see below) are identical, aside from the fact that in the crack model nonperiodic disorder has been considered.} have also been considered
\cite{SchwarzFisher2002}, but much work remains to be done.

There are many experimental situations where the elastic medium model seems insufficient and
one needs to take into account plastic deformations, as e.g.\ topological defects in the medium.
In a wide class of experiments strong disorder yields large deformations of the
driven medium that make a strictly elastic model of the extended structure
inapplicable~\cite{BhattacharyaHiggins1993,HigginsBhattacharya1996,Nori1996,Tonomura1999}. In contrast, the medium tears as topological
defects are constantly generated and healed by the interplay of drive, disorder and
interactions \cite{JensenBrassBrechetBerlinsky1988,HellerqvistEphronWhiteBeasleyKapitulnik1996,HellerqvistEphronWhiteBeasleyKapitulnik1996b,HendersonAndreiHigginsBhattacharya1996,FaleskiMarchettiMiddleton1996,HendersonAndreiHiggins1998,Tonomura1999,PekkerBarankovGoldbart2007}. At slow
average flow rates the dynamics near depinning is spatially and temporally inhomogeneous, with coexistence of
pinned and sliding degrees of freedom. The depinning
transition may become discontinuous (first order), possibly with a macroscopic hysteresis and switching between
pinned and sliding states~\cite{MaedaNotomiUchinokura1990,ThorneCicakONeillLemay2002}. Experiments on charge density waves show that varying
the temperature leads to a transition from continuous depinning to hysteretic depinning with
sharp switching between pinned and sliding states~\cite{MaedaNotomiUchinokura1990}. Whether such phase slip effects
occur in the bulk or only at the contacts~\cite{LemayVanWijngaardenAdelmanThorne1998}, remains to be clarified. Related
slip effects or plastic behavior have
been proposed to explain the complex dynamics of many other dissipative systems, including vortex arrays in
type-II superconductors. Lorentz
microscopy images of driven vortex arrays in irradiated thin films of Niobium show
vortex rivers flowing past each other at the boundaries of pinned regions of
the lattice~\cite{Tonomura1999}. Scanning tunneling microscopy, which can resolve individual vortices at high
density, also reveals the evolution of the vortex dynamics with disorder strength~\cite{TroyanovskiAartsKes1999}.
There too, there are edge contamination effects, and they may be responsible for the coexistence of a metastable disordered phase and a stable ordered phase~\cite{PaltielZeldovMyasoedovShtrikmanBhattacharyaHigginsXiaoAndreiGammelBishop2000,MarchevskyHigginsBattacharya2002}. It is clear that more work is needed to understand the rich dynamics of
driven systems in experiments.

It was ubiquitously found in numerical studies of interacting particles driven on a random substrate at $T=0$ (away from the
weak-disorder limit) that near the onset of mean sliding the motion occurs along filamentary channels or rivers that are determined by the spatial disorder of the random medium. Such channels are preferentially aligned along the direction of mean motion, but can exhibit large transverse excursions. At higher mean-flow rates the rivers coalesce into a more coherent structure that eventually results in a uniform flow. Hence the plastic flow takes, at a qualitative level, a variety of forms with increasing correlations: (i) filamentary flow with a single well-defined channel or several uncoupled
channels~\cite{Nori1996,BasslerPaczuskiAltshuler2001,BasslerPaczuskiReiter1999,BasslerPaczuski1998,OliveSoret2006}  to coupled or synchronized channels, to a layered
smectic type structure to a moving
lattice which may or not still contain frozen or moving dislocations ~\cite{MoonScalettarZimanyi1996,GiamarchiLeDoussal1996,BalentsMarchettiRadzihovsky1997,BalentsMarchettiRadzihovsky1998,LeDoussalGiamarchi1997,PardoDeLaCruzGammelBucherBishop1998,GiamarchiBhattacharya2002}.
While one may hope that at large velocity, where
the effective disorder is smaller, the flow is closer to the one described by an elastic model, it is clear that one needs to take into account plastic deformations  to describe these various regimes.

The theoretical understanding of the dynamics of such plastic systems is much less developed than that of driven elastic
media. It is not even clear how to characterize the various moving states which are observed by some order parameter, and to properly define steady states and their large-size limit. One can measure the distribution of time-averaged velocities $P(v)$ of the individual particles. A non-trivial $P(v)$ exists for instance in the filamentary regime where some particle seem permamently pinned while others are moving along channels. In small systems with periodic boundary conditions a periodic steady state is observed near the threshold with a non-trivial $P(v)$. Whether this feature persists in the infinite system limit, and how it depends e.g.\ on the geometry and aspect ratio of the sample, is not known. As was recently pointed out \cite{OliveSoret2006}, it is fruitful to apply tools and ideas from the theory of dynamical systems and chaos. It was found that upon increasing $f$,  the system undergoes a transition from periodic to a fully chaotic flow with positive Lyapunov exponents and a non-trivial attractor. The dimension of this attractor, which is low, may also provide a tool to characterize the phases of plastic flow. These ideas remain to be explored, in particular whether the elastic flow could exhibit some chaotic regime. At larger drive, $P(v)$ becomes more peaked around a single velocity and some degree of spatial coherence in the phase accross the layers may arises. Whether $P(v)$ eventually becomes a delta function in the large-size limit, and whether the phase-coherence lengths diverge or not, has not been systemetically studied numerically. There has been some efforts to use numerical simulations to
correlate the spatial and temporal structure of the dynamics with the shape of the macroscopic
response~\cite{FaleskiMarchettiMiddleton1996}, for instance the IV characteristics due to flux flow in a type-II superconductors.
A number of mean-field models of driven extended systems with locally underdamped relaxation or phase
slips have been proposed in the literature~\cite{StrogatzMarcusWesterveltMirollo1988,StrogatzWestervelt1989,MontakhabCarlsonLevy1994,VinokurNattermann1997,MarchettiMiddletonPrellberg2000,SchwarzFisher2001,MarchettiDahmen2002,MarchettiMiddletonSaundersSchwarz2003,SchwarzFisher2003,SaundersSchwarzMarchettiMiddleton2004}. Whether or not these dynamical models
exhibit truly collective behavior and universality in finite dimensions remains an open question
which motivated this work, as discussed below. A model which attempts to describe filamentary flow away from mean field
was proposed in ~\cite{WatsonFisher1996}.

\subsection{Layered Model}

Given the difficulty in describing topological defects, a simpler approach consists in considering layers such that
deformations within a layer are only elastic. Since the relative displacements between layers can be arbitrarily large, inter-layer plastic deformations are allowed. Whether they occur or not depend on the interaction between the layers.
This approach was successful to treat disorder in the statics, where it lead to solvable limits for e.g.\ the Bragg glass phase \cite{KierfeldNattermannHwa1997,CarpentierLedoussalGiamarchi1996}, the decoupling transition for magnetically coupled superconductors
\cite{HorovitzLeDoussal2005}. It is also studied to describe interacting quantum systems such as the sliding Luttinger liquid
\cite{VishwanathCarpentier2001}.
Recently a similar strategy was applied to describe plastic flow and depinning (see Refs.\ \cite{MCMPramanaStatPhys05,MCMSitges06} for a review), and coupling phenomena in the driven dynamics \cite{VinokurNattermann1997}. There it is even more natural since the flow naturally tends to be along layers (which can be channels) in the direction of the applied force. In one version of the model, introduced by one of us and collaborators,
the layers are only {\it viscously} coupled  in at least one of the directions transverse to the
mean motion. Although this is a much simplified version of plastic flow, for instance there are no convective terms in the equation of motion, it is motivated by the moving smectic phase in driven vortex lattices
mentioned above. It incorporates elastic responses to compressional deformations and allows for local slips
of neighboring degrees of freedom due to shear deformations. It is also relevant to
experiments on driven superconducting vortices in narrow channels and other controlled geometries
\cite{PruymboomKesVanderdriftRadelaar1988,TheunissenVanderdriftKes1996,BesselingNiggebruggeKes1999,AndersSmithBesselingKesJaeger2000,KokuboBesselingVinokurKes2002}. One possible realization is a
a layered (e.g.\ high $T_c$) superconductor when the vortices are aligned with the magnetic field within the $\mathrm{Cu}_2\mathrm{O}$, $ab$-plane layers and move along these layers under a $c$-axis current. In the limit where the intrinsic pinning potential from the $\mathrm{Cu}_2\mathrm{O}$ planes is strong compared to the weak isotropic disorder from point impurities, the vortex dynamics may be modeled in terms of $2d$ elastic layers or ``channels'' coupled  viscously along the $c$ axis.
The fact that only the viscous coupling between layers is retained makes it more tractable. It is expected to be valid in situations where the commensurate density-density interlayer interaction, studied in Ref.\ \cite{VinokurNattermann1997} , which couples the displacements in each layer, can be neglected. The general case can be defined as follows: Consider a $d=d_\parallel+d_\perp$-dimensional medium composed of
elastic $d_\parallel$-dimensional channels  coupled via viscous interactions in the remaining $d_\perp$
directions. The medium is driven by a uniform force $\bf f$ applied along one of the directions in the
$d_\parallel$-dimensional channels. Here we only consider the dynamics of a scalar displacement field
$u(\rl,\rp,t)$ describing deformations in the direction of the driving force at position ${\bf r}=(\rl,\rp)$,
with ${\bf r}$, $\rl$ and $\rp$ vectors in $d$, $d_\parallel$ and $d_\perp$ dimensions, respectively. To index
the channels one discretizes spatial coordinates in the direction normal to the layers ($\rp\rightarrow {\bf
r}_{ n}$, where ${\bf r}_{ n}$ denotes the $n$-th layer) and let $\rl\equiv x$. The dynamics of the
displacement $\un(x,t)$  of each degree of freedom is governed by the equation,
\begin{widetext}
\begin{equation}
\label{model_discrete} \gamma\partial_t \un(x) =\int_{x'} K(x-x') (u_n(x) - u_n(x'))
   +\sum_{n=1}^{M}
   \eta_{{ n},{ m}}[\dot{u}_{m}(x) -\undot(x)]
+ f + F_n(\un(x),x) \label{Mlayermodel}
\end{equation}
\end{widetext}
This is the $M$-layer model. Among the various universality classes of disorder, the one of most interest here is the random periodic class where the pinning force has the form:
\begin{equation}
F_n(u_{n},x) = h_{n}^iY(\un(x) -\beta_{n}^i)\;,
\end{equation}
with $Y(u)$ a periodic function. The pinning strengths $h_{n}^i$ are independent random variables
distributed with probability $\rho(h_{n}^i)$ and $\beta_{n}^i$ are random phases uniformly and
independently distributed in $[0,1)$. This  models the
dynamics of driven periodic media, such as vortex lattices, charge density waves, or Wigner crystals.
In these systems substrate impurities couple to the density of the lattice which, in the absence of in-layer topological defects, has the periodicity of the ordered lattice. As a result, the pinning force contains periodic components at all
reciprocal lattice vectors \cite{BlatterFeigelmanGeshkenbeinLarkinVinokur1994,Nattermann1990}.
In the bare model one can retain the lowest Fourier components only,  since as is well
known from FRG studies of statics and depinning, all Fourier components are generated by coarse-graining
and should be included to describe the properly
renormalized disorder correlator. Such a correlator develops cusp-like singularities at large scales that
control the dynamics. The other type of disorder, the non-periodic or random-manifold class, which at the elastic depinning was shown to give rise to a single universality class encompassing both random bond (i.e.\ short-ranged) and random field (i.e.\ long-ranged) disorder, will also be studied. This is done by choosing a non-periodic correlator for the random forces $F_n(u,x)$. Physical realizations are less obvious since the above velocity coupling is local in $x$ space only while a realistic coupling e.g.\ between two directed polymers would also depend on the field $u$. Two possible realizations are: (i) manifolds with internal disorder, as studied in Ref.~\cite{CuleHwa1996,CuleHwa1998}. (ii) periodic systems for which the correlation length of the disorder $r_f$ is small compared to the lattice spacing $a$. Then the two scales for pinning, the Larkin length $R_c$, and  $R_a$ for the decay of translational order, can be very different, and it is known that for scales $R_c \ll L \ll R_a$ all harmonics of the disorder correlator are important and the system behaves effectively as a random manifold within
this range of scales
\cite{BlatterFeigelmanGeshkenbeinLarkinVinokur1994,GiamarchiLeDoussal1995,GiamarchiLeDoussalBookYoung}. Hence, below we also consider the nonperiodic or random-manifold model and discuss the different behaviors in the two cases.

\subsection{Aim and outline of the paper: Two-layer model} \label{intro}

The layered model (\ref{Mlayermodel}) with viscous couplings  was proposed as a generic coarse-grained model representative of a class of dissipative driven systems with strong disorder that encompasses many of the models considered in the literature. It was solved within mean-field theory and shown to predict a qualitative change from continuous to discontinuous and hysteretic dynamics as a function of disorder strength, consistent with experimental observations in a variety of systems \cite{MarchettiMiddletonSaundersSchwarz2003}. It has also been studied numerically in finite dimensions. The numerical studies show evidence of hysteresis in 2+1 dimensions above a critical value of the interlayer coupling. Hysteresis was not clearly evident, however, in 1+1 dimensions nor for the two-layer model studied below, although it could also not be conclusively ruled out on the basis of finite-size scaling \cite{hhh1}.

The aim of this paper is to go beyong the mean-field treatment of model (\ref{Mlayermodel}) and explore using functional RG whether hysteretic dynamics also occurs  and whether universal features emerge in low dimensions where one usually does not expect the mean-field approximation to be accurate. Since generalization to $M$ layers is straightforward and not expected to bring important qualitative changes, we study in detail the technically simpler case of two viscously coupled layers $M=2$. We start by recalling in Section \ref{sec:mft} the main features of the mean-field solution so as to provide a basis for comparison. In Section \ref{sec:oneloop} we study the $M=2$ model first by direct perturbation theory and next using 1-loop FRG. We prove that the elastic single-layer (i.e.\ decoupled layer) quasi-static $v=0^+$ depinning fixed point is {\it always unstable} to a small viscous inter-layer coupling. A partial one-loop analysis at $v>0$ shows the generic co-existence of a pinned and a moving state below the single-layer depinning threshold. The resulting $v(f)$ curves show similarities with the mean-field ones, and in some regimes the agreement can even be made quantitative. We estimate the velocity $v_c$ at which the $v(f)$ curve becomes vertical (and a jump may occur in the fixed force ensemble).
A key feature of the one-loop study is that the inter-layer viscous coupling $\gamma_{12}$ is {\it not corrected} by disorder. To determine whether this is maintained to higher order, we carry in Section \ref{sec:twoloop} the FRG to two loops, near the uncoupled elastic quasi-static depinning. The analysis gives a strong correction as compared to one loop in the non-periodic class, i.e.\ random manifolds, with a new universality class for plastic depinning whose $z$ exponent is computed in Section \ref{subsec:nonper}. The analysis in the periodic case is quite subtle and presented in Section \ref{subsec:per}. Finally, to get a better understanding of possible behaviors and of the connection between dynamical hysteresis and attractors in phase space in simpler cases, we study in Section \ref{sec:toy} two $d=0$ toy models of two viscously coupled particles on, respectively, a smooth (Section \ref{subsec:smooth}) and a discontinuous (Section \ref{subsec:scalloped}) force landscape. Finally, in Section \ref{sec:discussion} the main results are summarized, and extensions and future directions discussed. In particular it is proved that a small interlayer commensurate coupling is irrelevant at depinning. Since such a term is always present in real systems, this shows that the viscous model is consistent. The Appendices contain the details of the two-loop calculation and a proof that stable static configurations where decoupled layers are independently pinned remain dynamically stable in presence of the inter-layer viscous coupling.

Let us now define the two-layer model studied here, and fix notations.
We consider the overdamped dynamics of two layers coupled by a viscous coupling in a random potential. Each
layer is an elastic system parameterized by a one-component ($N=1$) displacement field $u^i(x)$, also denoted $u^i_x$, for $i=1,2$, or
$u^i_{x,t}$ to indicate explicitly the dependence on time. The equation of motion of one layer  is
\begin{equation}\label{1.1}
\cgamma_{0} \dot  u^{1}_{x,t} = \ceta_{0} \left(\dot  u^{2}_{x,t}-\dot u^{1}_{x,t} \right)  + c \nabla^2
u^{1}_{x,t} + F^1 ( u^{1}_{x,t})+f\ ,
\end{equation}
where $\cgamma_0$ is the in-layer friction coefficient. Hence, in addition to  elastic intra-layer restoring
forces (elastic coefficient $c$) and the quenched random pinning force, one layer is also pulled by the other
layer through a velocity (or viscous) coupling $\ceta_0$. Here we focus on the case of uncorrelated disorder in
each layer, and denote the second cumulant of the pinning forces by
\begin{equation}\label{y1}
\overline{F^{i} (x,u) F^{j} (x',u')} = \delta^{ij} \delta^{d}(x-x') \Delta_0 (u-u')\ .
\end{equation}
The equation of motion for the system of two layers driven by an external force $f$ can then be written as:
\begin{equation}\label{y2}
\!\left(\begin{array}{cc}
\cgamma_{11}&\cgamma_{12} \\
\cgamma_{12}&\cgamma_{22}
\end{array} \right) \frac{\rmd}{\rmd t}\left(\begin{array}{c}
 u^{1}_{x,t}\\
 u^{2}_{x,t}
\end{array} \right) = c \nabla^2 \left(\begin{array}{c}
 u^{1}_{x,t}\\
 u^{2}_{x,t}
\end{array} \right) + \left(\begin{array}{c}
F^{1} (x,u^{1}_{xt}) + f \\
F^{2} (x,u^{2}_{xt}) + f
\end{array} \right).
\end{equation}
The bare values for the friction matrix are
\begin{eqnarray}\label{y5}
\cgamma_{11}=\gamma_{22}&=&  \cgamma_{0}+\ceta_{0}\\
\cgamma_{12}&=&  - \ceta_{0}\ .
\end{eqnarray}

\section{Mean-field theory}

\label{sec:mft}

To set up the mean-field theory for the multi-layered model, it is convenient to discretize space in both the
transverse and longitudinal directions, using integer vectors $\ell,m$ for the $d_\parallel$-dimensional
intra-layer index.  The local displacement along the direction of motion at time $t$ is $u^i_\ell(t)$, with
$i=1,\ldots,M$ the layer index and $\ell=1,\ldots,N$ labeling the degrees of freedom within each layer.  Its dynamics
is governed by the equation (in this section we drop the subscript '$0$' on the bare frictions),
\begin{eqnarray}
\label{MF.1} \gamma \dot u^i_\ell(t)&=&\sum_{m}K_{\ell m}[u^i_{m}(t)-u^i_\ell(t)]
   +\sum_{j}
   \eta_{ij}[\dot{u}^j_{\ell}(t)-\dot{u}^i_{\ell}(t)]\nonumber\\
&&+f+h^i_\ell Y(u^i_\ell(t)-\beta^i_\ell)\;,
\end{eqnarray}
where $Y(u)=Y(u+n)$ is a periodic function and $K_{\ell m}$ and $\eta_{ij}$ have constant values $c$ and $\eta$,
respectively, for nearest-neighbor pairs and vanish otherwise.
 The random pinning strengths $h^i_\ell$
are chosen independently with probability distribution $\rho(h^i_\ell)$ and the random phases $\beta^i_\ell$ are
distributed uniformly and independently in $[0,1)$.

\subsection{Fully connected mean-field theory}
\label{mft}

 One mean-field approximation is obtained by assuming that all sites are coupled with uniform
strength, both within each layer and across the layers, i.e., $K_{\ell m}=c/N$ for all  $\ell$ and $m$ and
$\eta_{ij}=\eta/M$ for all  $i$ and $j$. The mean displacement and velocity are given by
\begin{eqnarray}\label{MF.2}
\overline{u}(t)&=&\frac{1}{NM}\sum_\ell \sum_iu^i_\ell(t)\\
\label{MF.3}v&=&\frac{1}{NM}\sum_\ell \sum_i \dot u^i_\ell(t)
\end{eqnarray}
and we look for solutions moving with a uniform velocity so that (up to a choice of the origin of time)
\begin{equation}\label{MF.3b}
\overline{u}(t)=vt\;.
\end{equation}
Since the displacements  are coupled only through the mean fields, they can be indexed by their disorder
parameters $\beta$ and $h$, rather than by the spatial indices $\ell$, $i$, i.e., $u^i_\ell(t)\rightarrow
u(t;\beta,h)$. The mean-field dynamics is governed by the equation
\begin{equation}
\label{MF.4} (\gamma+\eta)\dot{u}(t;\beta,h)= c\big(v t-u\big)
    +f
   +\eta v+hY(u-\beta)
\end{equation}
that must be solved with the self-consistency condition that determines the mean field,
\begin{equation}\label{MF.5}
\langle u(t;\beta,h)-v t\rangle_{\beta,h}=0\;,
\end{equation}
where $\langle ...\rangle_{\beta,h}=\int_0^1 d\beta\int dh...\rho(h) $ denotes the average over disorder.

The long-time steady-state solution to Eq.~(\ref{MF.4}) can be written as
\begin{equation}\label{MF.6}
u(t;\beta,h)=vt+\hat{u}\;,
\end{equation}
with
\begin{equation}
\label{MF.7} (\gamma+\eta)\dot{\hat{u}}= -c\hat{u}
    +f-\gamma v
  +hY(\hat{u}+vt-\beta)\;,
\end{equation}
to be solved with the condition $\langle\hat{u}\rangle_{\beta,h}=0$. It is apparent from Eq.~(\ref{MF.7}) that
$\hat{u}$ is a periodic function of time with period $1/v$ and depends on time and phase $\beta$ only through
the combination $\hat{u}=\hat{u}(t-v/\beta;h)$. This will allow us to carry out the average over $\beta$ by
averaging over time.

We display below the solution for a parabolic scalloped pinning potential, corresponding to a piecewise linear
pinning force with jumps of size $h$ at the boundaries of each period,
\begin{equation}\label{MF.8}
Y(u)=n+\frac12-u,\hspace{0.2in} n\leq u\leq n+1\;,
\end{equation}
with $n$ an integer. The mean-field equation (\ref{MF.4}) is formally identical to the mean-field equation for a
purely elastic medium, with friction coefficient $\Gamma=\gamma+\eta$ and an effective drive $F=f+\eta v$. The
solution of the mean-field equation for a scalloped pinning potential and $\eta=0$ was obtained by Narayan and
Fisher \cite{NarayanDSFisher1992b} and is easily adapted to our case. The solution for finite $\eta$ is described in
Ref.\ \cite{MarchettiMiddletonSaundersSchwarz2003,MCMSitges06}  and will be summarized here for completeness.

The pinning force has a jump discontinuity $h$ at the end of each period. The displacement $\hat{u}$ is
continuous across neighboring periods, but the local velocity $\dot{\hat{u}}$ has jumps of size $h/\Gamma$ at
$t_J+n/v$. The solution of Eq.~(\ref{MF.7}) for $t_J(\beta)+n/v\leq t\leq t_J(\beta)+(n+1)/v$ is
\begin{equation}
\label{MF.9} \hat{u}=
  Ae^{-\lambda t}+\frac{h(\beta-vt)+F+h(n+1/2)}{\Gamma\lambda}
  -\frac{cv}{\Gamma\lambda^2}\;,
\end{equation}
where $\lambda=(c+h)/\Gamma$ and $t_J(\beta)$ is the ``jump time''. The constant $A(\beta)$ and the jump time
$t_J(\beta)$ are determined by  requiring
\begin{eqnarray}
&&\label{bc1}\hat{u}(t=t_J+n/v)=n+\beta\;,\\
&&\label{bc2}\hat{u}(t=t_J+(n+1)/v)=n+1+\beta\;.\qquad
\end{eqnarray}
It is important to appreciate a crucial difference between the mean-field theory of the purely viscous model
($c=0$, or $d_\parallel =0$) discussed in Refs.~\cite{MarchettiMiddletonPrellberg2000,MarchettiDahmen2002} and the mean-field theory
of the model considered here that includes additional elastic couplings within the channels. In the purely
viscous case, each degree of freedom is coupled only to the local velocities (which exert an additional
effective driving force) and can slide with its own period. In contrast, when $c\not=0$ each degree of freedom
couples to the average displacements via a spring-type interaction that  forces all periods to be the same,
independent of $h$.

After inserting $A$ and $t_J$ obtained from the solution of  Eqs.~(\ref{bc1}) and (\ref{bc2}) in
Eq.~(\ref{MF.9}), it is straightforward to impose the self-consistency condition as
\begin{equation}
\Big\langle \hat{u}(t-\beta/v;h)\Big\rangle_{\beta,h}=v\Big\langle\int_{t_J+\frac{n}{v}}^{t_J+\frac{n+1}{v}} dt\,
\hat{u}(t-\beta/v;h)\Big\rangle_h=0\;.
\end{equation}
This yields an implicit solution  for the mean velocity as
\begin{eqnarray}
\label{MFvFApp}
F(v)=f(v)+\eta v&=&f_c+\Gamma v\big[1-M(c)\big]\nonumber\\
&&
  +\Big\langle\frac{h^2}{(c+h)}\frac{1}{e^{\lambda/v}-1}
  \Big\rangle_h\qquad
\end{eqnarray}
with $f_c$ the threshold force for the onset of uniform sliding,
\begin{equation}
f_c=\Big\langle\frac{h^2}{2(c+h)}\Big\rangle_h\;,
\end{equation}
and
\begin{equation}
M(c)=\Big\langle\frac{h^2}{(c+h)^2}\Big\rangle_h \ .
\end{equation}
The threshold force depends only on the elastic coupling $c$, but not on the viscous coupling $\eta$. This
follows because in mean field the viscous coupling becomes effective only when the system is moving as a whole,
while away from mean field one expects additional fluctuation effects.

\begin{figure}[t]
\centerline{\fig{0.8}{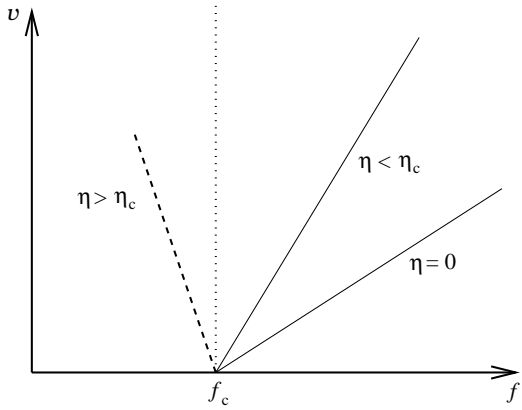}} \caption{The mean-field  velocity $v(f)$ as a function of $f$ near threshold, as
given by Eq.~(\ref{vflinear}). The slope of the linear function diverges at $\eta=\eta_c(c)$ and is negative for
$\eta>\eta_c(c)$. } \label{f:MFstraight}
\end{figure}

Near threshold the  term on the second line of Eq.~(\ref{MFvFApp}) gives contributions of order $\sim e^{-1/v}$
and can be neglected. It is then easy to invert Eq.~(\ref{MFvFApp}) to obtain $v(f)$, with the result
\begin{equation}\label{vflinear}
v(f)\sim\frac{(f-f_c)^{\beta_{\mathrm{MF}}}}{\gamma-(\gamma+\eta)M(c)}\ .
\end{equation}
The mean velocity vanishes linearly for $f\rightarrow f_c^+$ , with a MF exponent $\beta_{\mathrm{MF}}=1$ which is
generic for discontinuous pinning forces~\cite{NarayanDSFisher1992b}. When $\eta=0$ the slope of the linear curve is
always positive as $M(c)<1$. The slope  diverges, however,  at a critical value of $\eta$,
\begin{equation}\label{etac}
\eta_c(c)=\gamma\Big[\frac{1}{M(c)}-1\Big]\ ,
\end{equation}
and becomes negative for $\eta>\eta_c$, as shown in Fig.\ \ref{f:MFstraight}. For $\eta>\eta_c(c)$ the velocity
curve is multivalued, yielding hysteretic behavior.

\begin{figure}[t]
\centerline{\fig{1}{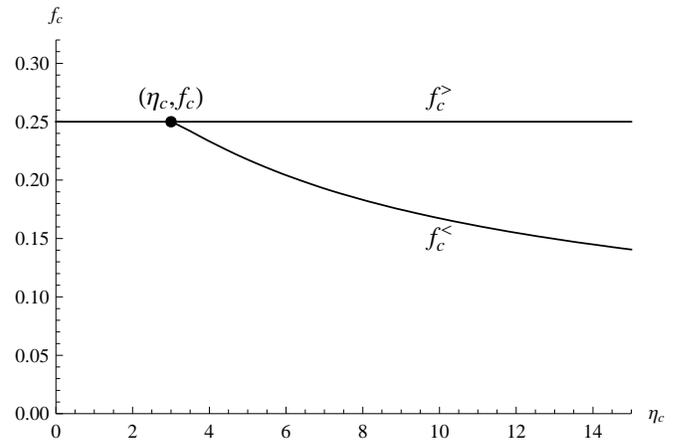}} \caption{ Phase diagram obtained from the fully-connected mean-field
solution for $\rho(h)=\delta(h-1)$, for $\gamma=1$ and $c=1$, corresponding to $\eta_c=3$. There is a critical
point at $(\eta_c,f_c)$ separating continuous from discontinuous depinning. } \label{f:MFPhaseDiagram}
\end{figure}

\begin{figure}[t]
\centerline{\fig{1}{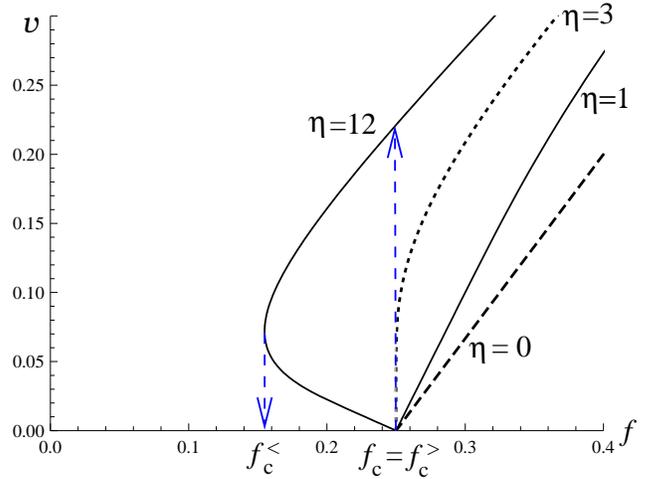}} \caption{Typical velocity-force curves obtained from the fully-connected
mean-field solution for $\rho(h)=\delta(h-1)$, $c=1$ and $\eta_c=3$. For $\eta=0$ (dashed curve) and $\eta=1$
(solid line)  the system depins continuously at $f_c$. At  $\eta=\eta_c=3$ (dotted line) the slope diverges at
threshold. For $\eta=12$ (solid line) the velocity-force curve is multivalued. This corresponds to a hysteretic
depinning transition as
 the system depins at $f_c^>$ when the force is ramped up from zero and repins at the lower value $f_c^<$ when the force is ramped down from the sliding state. } \label{f:MFvfcurves}
\end{figure}

The  phase diagram and typical velocity-force curves are shown in Figs.~\ref{f:MFPhaseDiagram} and
\ref{f:MFvfcurves} for $\rho(h)=\delta(h-h_0)$. The finite long-time elasticity ($c\not=0$) guarantees that the
behavior is \emph{independent of the shape of the pinning-force distribution }$\rho(h)$. The phase diagram for
$\rho(h)=e^{-h}$ was shown in \cite{MarchettiMiddletonSaundersSchwarz2003} and has the same form as the one shown here. The point
 $(\eta_c,f_c)$ is a {\it tricritical point} separating single-valued from multi-valued velocity curves.
For $\eta<\eta_c$, a continuous depinning transition at $f_c$ separates a pinned state from a sliding state with
{\em unique} velocity. A question addressed below is whether $\eta_c$ remains non-zero in finite dimension and
if so, whether the depinning transition for $\eta < \eta_c$ is in the same universality class as the depinning
of an elastic medium ($\eta=0$)~\cite{DSFisher1985}.

In our mean-field example, the linear response diverges at $\eta_c$ as $v(\eta=\eta_c)\sim 1/\ln(f-f_c)$. For
$\eta>\eta_c$ the solution is multivalued.  In this case when the force is ramped up from zero the system depins
at $f_c^>=f_c$. When the force is decreased from a value above $f_c$ the system gets stuck at the lower value
$f_c^<$, yielding hysteretic $v(f)$ curves.

For $\eta>\eta_c$ the mean velocity has a jump discontinuity. The value $v_c$ of this jump is given by the
solution of
\begin{equation}\label{J1}
\Big(\frac{\partial f(v)}{\partial v}\Big)_{v=v_c}=0\ ,
\end{equation}
where $f(v)$ is given by Eq.~(\ref{MFvFApp}). An explicit solution for the jump $v_c$ can be obtained for the
case of a sharp disorder distribution $\rho(h)=\delta(h-h_0)$. In this case the condition (\ref{J1}) for the
jump becomes
\begin{equation}\label{J2}
\frac{\eta-\eta_c}{\gamma+\eta}=\frac{(\lambda/v)^2}{4\sinh^2(\lambda/(2v))}\;.
\end{equation}
For $\eta\gg\gamma,\eta_c$ this gives
\begin{equation}\label{J4}
v_c\approx \frac{c+h_0}{2\sqrt{3}\eta}\sqrt{\frac{\eta}{\gamma+\eta_c}}\;.
\end{equation}
Finally, we note that the mean-field theory for a smooth periodic pinning potential gives qualitatively the same
phase diagram, although with mean-field exponent $\beta_{\mathrm{MF}}=1/2$.

The fully connected mean-field theory discussed here for the layered visco-elastic model is formally identical
to the  mean-field  limit of a model of crack propagation with stress overshoot studied by Schwarz and
Fisher \cite{SchwarzFisher2001,SchwarzFisher2003,MaimonSchwarz2004}, although the crack model contains random force disorder
instead of the periodic disorder considered here~\cite{MCMPramanaStatPhys05,MCMSitges06}.

\subsection{Self-consistent single-layer approximation}
\begin{figure}[t]
\centerline{\fig{0.9}{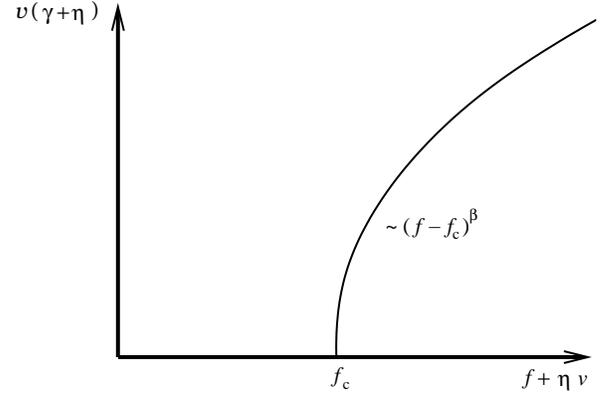}} \caption{The mean velocity $(\gamma+\eta)v$ plotted as a function of $F=f+\eta
v$. When $f+\eta v$ is used as the independent variable, the velocity-force characteristic  has the same
functional form as that of a single elastic layer that depins at a threshold $f_c$ with $v\sim(f-f_c)^\beta$ and
$\beta<1$ as $f\rightarrow f_c^+$. } \label{f:elasticvf}
\end{figure}

\begin{figure}[t]
\centerline{\fig{0.9}{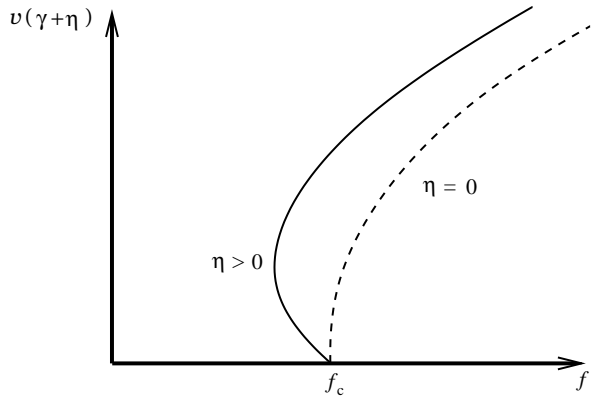}} \caption{The velocity-force curve for finite $\eta$ can be obtained fro the
single-layer curve of Fig.~\ref{f:elasticvf} corresponding to $\eta=0$ (dashed line) by a change of the
independent variable. Since $\beta<1$, the resulting $v(f)$ will be multivalued for any finite $\eta$. }
\label{f:plasticvf}
\end{figure}

An alternative, ``partial'' mean-field approximation treats only one direction of space using mean field, and
reduces the problem to an effective single-layer model. It is obtained by assuming uniform, i.e.\ infinite-range
couplings  of strength $\eta_{ij}=\eta/M$ across the layers for each in-layer site $\ell$. The corresponding
mean field is given by
\begin{eqnarray}\label{AMF.3.1}
v_\ell=\frac{1}{M} \sum_i \dot u^i_\ell(t)\;.
\end{eqnarray}
In the thermodynamic limit of an infinite number of layers, assuming the system is self-averaging, the mean field
$v_\ell$ will not depend on $\ell$ and this label can be dropped. The mean-field dynamics is then described by
the equation
\begin{eqnarray}
\label{AMF.4}
(\gamma+\eta) \dot u^i_\ell(t)&=&\sum_mK_{\ell m}[u^i_m(t)-u^i_\ell(t)]\nonumber\\
 &&  +f
 +  \eta v+h^i_\ell Y(u^i_\ell(t)-\beta^i_\ell)\;,\qquad
\end{eqnarray}
which must be solved with the condition $\langle \dot{u}^i_\ell\rangle_{\beta,h}=v$. It is illuminating to
rewrite Eq.~(\ref{AMF.4}) by replacing the discrete in-layer index $\ell$ by the original continuum variable
$x$,
\begin{equation}
\label{AMF.4.1} (\gamma+\eta) \dot u^i_{x,t}=c\nabla^2u^i_{x,t}
  +f + \eta v+h^i_x Y(u^i_{x,t}-\beta^i_x)\;,
\end{equation}
to be solved with the self-consistency condition $\frac{1}{M}\sum_i \dot{u}^i_{x,t}=v$. It is  apparent that
Eq.~(\ref{AMF.4.1}) describes the dynamics of $M$  identical elastic layers coupled only through the mean field
$v$. Each layer is a dissipative elastic medium of   friction $\Gamma=\gamma+\eta$,  driven by a force $F=f+\eta
v$.  For $\eta=0$ the layers are decoupled, with $\Gamma=\gamma$ and $F=f$. The velocity-force curve $v_{\mathrm{sl}}(f)$
of one decoupled layer has been studied in details~\cite{DSFisher1985,NarayanDSFisher1992a,NarayanDSFisher1992b,NattermannScheidl2000,ChauveGiamarchiLeDoussal2000,ChauveLeDoussalGiamarchi2000}.  Each layer is pinned with $v=0$
for  $f<f_c$. It depins at $f=f_c$ and slides for $f>f_c$ with mean velocity $v_{\mathrm{sl}}(f)={\cal G}(f)/\gamma$,
and ${\cal G}(f)\sim (f-f_c)^\beta$ as $f\rightarrow f_c^+$ and $\beta<1$ a critical exponent that depends
only on the system's dimensionality. It is clear from the form of Eq.~(\ref{AMF.4}) that the velocity-force
characteristic of the coupled layers has the same functional form as that of an individual layer, with the
replacement $f\rightarrow F$, i.e, $v(F)={\cal G}(F)/\Gamma$. A sketch of this velocity-force characteristic is
shown in Fig.~\ref{f:elasticvf}. The velocity-force characteristic $v(f)$ of the coupled layers can then be
obtained simply by performing a shift in the independent variable in the known result for a single layer. The
result is shown in Fig.~\ref{f:plasticvf}. Near threshold $v(f)\sim(f+\eta v-f_c)^\beta$, with
$\beta=1-(4-d_\parallel)/6+{\cal O}[(4-d_\parallel)^2]<1$ and $v(f)$ will be multivalued for every finite value
of $\eta$, yielding a hysteretic depinning transition \cite{MarchettiMiddletonSaundersSchwarz2003}. The hysteresis for any $\eta>0$, for
$d<4$, obtained here in the approximation of a global transverse coupling, will be confirmed below within a
one-loop FRG analysis which incorporates inter-layer fluctuations neglected here. In both cases it is a
consequence of the non-trivial renormalization of $\gamma$ within a single layer, responsible for $\beta<1$ and
$z>2$ for elastic depinning.

As pointed out in Ref. \cite{MarchettiMiddletonSaundersSchwarz2003}, the self-consistent single-layer approximation, with uniform
couplings across the layers, is equivalent to a model of charge density waves (CDWs) that incorporates the
coupling of the CDW to normal carriers via the addition of a global velocity coupling to the equation of motion
for the phase \cite{Littlewood1988,LevySherwinAbrahamWiesenfeld1992,MontakhabCarlsonLevy1994}.

Finally, another ``partial'' mean-field theory is obtained by assuming uniform couplings of strength $K_{\ell
m}=K/N$ for every $\ell,m$ within each layer. This model will be discussed elsewhere. The two-particle toy model
described in Section V corresponds to the $K=0$ limit of this mean-field theory.

%, the mean field is defined as
%
%\begin{eqnarray}\label{AMF.1.1}
%\overline{u}^i(t)=\frac{1}{N} \sum_\ell u^i_\ell(t)=vt
%\end{eqnarray}
%
%In the thermodynamic limit, assuming each layer is self-averaging,  the mean field defined in Eqs.~(\ref{AMF.1.1})  will not depend on $i$ and this label has been dropped. The dependence on the in-layer index $\ell$ enters only through the disorder $h^i_\ell,\beta^i_\ell$. Assuming these do not depedn on $\ell$, the dynamics reduces to that of $M$ layers described by a displacement field $u^i(t)$, each coupled elastically to its own center of mass velocity, and coupled viscously to its nearest neighbors:
%
%\begin{eqnarray}
%\label{AMF.2}
%\gamma_0 \dot u^i(t)&=&K(vt-u^i(t))
%   +\sum_{j}\eta_{ij}(\dot{u}^j(t)-\dot{u}^i(t))\nonumber\\
%&&+f+h^i Y(u^i(t)-\beta^i)\;.
%\end{eqnarray}
%
%The case $M=2$ is solved explicitly in Section.

\section{Functional RG to one loop}

\label{sec:oneloop}

To go beyond mean field we now develop a Functional RG approach.

\subsection{Perturbation theory and length scales}
Consider model (\ref{y2}) driven by a force $f$ and assume that it reaches a time-translational invariant
steady state (e.g.\ with periodic boundary conditions for each layer). There are  two modes:
\begin{equation}
u^+ =(u^1+u^2)/2 \quad , \quad u^- = u^1-u^2 \ . \label{twomodes}
\end{equation}
For a system of finite size $L$,  because of fluctuations in the pinning force, the velocity in each layer
will be different. However this effect should disappear in the infinite-$L$ limit, and can be supressed
using appropriate boundary conditions. Hence we define $v$ to be the velocity of the center of mass
$v=\overline{\dot u^+}$,  perform the shift to the comoving frame,
\begin{equation}
u^i = v t + \hat u^i\ ,
\end{equation}
and immediately drop the hat. We can now write the dynamical action associated to the resulting equation of
motion (i.e.\ (\ref{y2}) shifted):
\begin{widetext}
\begin{eqnarray}\label{action}
{\cal S}[u,\tilde u] &=& \int_{x,t}\left(\begin{array}{c}
 \tilde u^{1}_{x,t}\\
 \tilde u^{2}_{x,t}
\end{array} \right)\left[ \left(\begin{array}{cc}
\cgamma^0_{11}&\cgamma^0_{12} \\
\cgamma^0_{12}&\cgamma^0_{11}
\end{array} \right) \frac{\rmd}{\rmd t} \left(\begin{array}{c}
 u^{1}_{x,t}\\
 u^{2}_{x,t}
\end{array} \right)  - c \nabla^2 \left(\begin{array}{c}
 u^{1}_{x,t} \nonumber \\
 u^{2}_{x,t}
\end{array} \right)\right]
- \int_{x,t} (f - \cgamma^0_+ v) \left(\begin{array}{c}
 \tilde u^{1}_{x,t}\\
 \tilde u^{2}_{x,t}
\end{array} \right) \\
&&
- \frac{1}{2} \int_{x,t,t'} \left(\begin{array}{c}
 \tilde u^{1}_{x,t}\\
 \tilde u^{2}_{x,t}
\end{array} \right) \left(\begin{array}{cc}
\Delta_0 (u^{1}_{x,t}-u^{1}_{x,t'} + v(t-t')) &0 \\
0& \Delta_0 (u^{2}_{x,t}-u^{2}_{x,t'} + v(t-t'))
\end{array} \right) \left(\begin{array}{c}
 \tilde u^{1}_{x,t}\\
 \tilde u^{2}_{x,t}
\end{array} \right)
\end{eqnarray}
\end{widetext}
The subscript $0$ indicates that these are quantities for the bare
model (and it is often dropped in the following). The matrix of
friction coefficients is diagonal in the basis (\ref{twomodes}) and
we denote the frictions associated to the center of
mass and relative motion as
\begin{eqnarray} \label{y5p}
\cgamma_{+} = \cgamma_{11}+\cgamma_{12} \quad , \quad
\cgamma_{-} &=& \cgamma_{11}-\cgamma_{12}\ .
\end{eqnarray}
The bare values are $\cgamma^0_{+} = \cgamma_{0}$ and $\cgamma^0_{-} =\cgamma_{0}+2\ceta_{0}$.
The bare response functions, i.e.\ those in the absence of disorder, read:
\begin{equation}\label{y3}
R^{ij}_{k,t} := \left< u_{k t}^{i}  \tilde u^{j}_{-k,0} \right>_0
\end{equation}
\begin{eqnarray}\label{y4}
R^{11}_{k,t} \equiv R^{22}_{k,t} &=& \Theta (t)\left[ \frac{\rme^{-
t k^{2}/\cgamma_{+} }}{2\cgamma_{+}}
+\frac{\rme^{- t k^{2}/\cgamma_{-} }}{2\cgamma_{-}}\right]\\
R^{12}_{k,t} \equiv R^{21}_{k,t} &=&\Theta (t)\left[ \frac{\rme^{- t
k^{2}/\cgamma_{+} }}{2\cgamma_{+}} -\frac{\rme^{- t k^{2}/\cgamma_{-}
}}{2\cgamma_{-}}\right]\ .\qquad
\end{eqnarray}
The case of a single layer is reproduced upon setting
$\ceta_{0}=0$, or equivalently $\cgamma_{-}= \cgamma_{+}$ (then
$R^{11}_{kt} \to R_{kt}$, the standard single-layer response
function, and $R^{12}_{k,t} \to 0$).

Writing $S=S_0+S_{\mathrm{int}}$ where $S_{\mathrm{int}}$ contains only the disorder,
i.e.\ the second line in (\ref{action}), the effective action
$\Gamma[u]$ of the system can be computed perturbatively in the disorder:
\begin{equation}
\Gamma[u,\tilde u] = S_0[u,\tilde u] + \langle S_{\mathrm{int}}[u,\tilde u]
\rangle_{S_0} + \frac{1}{2} \langle S_{\mathrm{int}}[u,\tilde u]^2
\rangle^c_{S_0} + \cdots \label{gamma}
\end{equation}
In the average over $S_0$ only 1-particle irreducible (1PI)
diagrams (i.e.\ containing loops) are kept. The quadratic part
of the effective action yields the exact disorder-averaged response
and correlation functions:
\begin{eqnarray}
&& {\cal R}^{ij}_{q,t-t'} = \frac{\delta^2 \Gamma[u,\tilde
u]}{\delta \tilde u^i_{-q t'} \delta
u^j_{q t}}\bigg|_{u=\tilde u=0} \\
&& \overline{u^i_{qt} u^j_{-qt'}} := {\cal C}^{ij}_{q,t-t'} = {\cal
R}^{ik}_{q,t-t_1} {\cal R}^{jl}_{q,t-t_2} \frac{\delta^2
\Gamma[u,\tilde u]}{\delta \tilde u^k_{-q t_1} \delta \tilde u^l_{q
t_2}}\bigg|_{u=\tilde  u= 0}\ . \nonumber
\end{eqnarray}
Both functions are  symmetric in $i,j$ and in $q$. The effective action has a complicated form but contains
terms similar to those in the above action (\ref{action}) with renormalized (i.e.\ ``corrected'') values for the
friction matrix $\cgamma_{ij}$, and the second cumulant of disorder $\Delta(u)$. The elastic term is
unrenormalized (i.e.\ the zero frequency part of the $\tilde u u$ term in $\Gamma$ is the same as in $S_0$)
thanks to the statistical tilt symmetry \footnote{The invariance of the non-linear (i.e. disorder) terms of ${\cal S}$ under $u^i_{xt} \to
u^i_{xt} + \phi^i(x)$ for two arbitrary static functions $\phi^i(x)$, $i=1,2$, should persist for $\Gamma$.} which holds independently in each layer. Other terms are generated in perturbation theory,
such as higher disorder cumulants, higher frequency corrections to the self energy or non-linear terms such as
$\int\tilde u (\partial_t u)^2$. In each case their relevance should be assessed carefully. These terms are
usually irrelevant near $d=4$. A simplifying feature is that the coupling between the layers is purely
dynamical. Therefore the static part of the theory (i.e.\ the 0-frequency part of the effective action) consists
of two decoupled static layers. This implies, among others, that no outer-diagonal elements of the disorder
correlator are generated in perturbation theory.

Let us now examine perturbation theory and power counting. The
effective action contains the term:
\begin{eqnarray}
&& - \tilde f  \int_{xt} \sum_i \tilde u^i_{xt} \quad , \quad \tilde
f=f - \cgamma_0 v + \delta f(v)
\end{eqnarray}
where $\delta f(v)$ contains all corrections due to disorder. On average these
are the same  for each layer, and depend on $v$. The
equation of motion is obtained from the condition $\tilde f=0$
equivalent to $\langle u_{xt} \rangle=0$ (in shifted variables).
From (\ref{gamma}) one finds that to lowest order in $\Delta$ (i.e.
to one loop) the corrections to friction and force are:
\begin{eqnarray}
 \delta \cgamma_{12} &=& 0  \\
 \delta \cgamma_{11} &=& - \int_q \int_0^\infty \rmd\tau\, \tau \Delta''(v
\tau) R^{11}_{q \tau}\qquad
\\
 \delta f &=&  \int_q \int_0^\infty \rmd\tau\, \Delta'(v \tau) R^{11}_{q
\tau}\ , \label{deltaf}
\end{eqnarray}
where the index $0$ is implicit if one studies perturbation theory on the bare action. The correction to the
disorder $\Delta$ is of order $\Delta^2$ and, at $v=0$, is identical to the one for a single-layer model, while
at finite $v$ it has a complicated expression (even in the single-layer case, as given in \cite{ChauveGiamarchiLeDoussal2000} not displayed here. As is well known, for $v=0^+$, $\Delta(u)$ acquires a cusp
for scales larger than the Larkin length $L_c$.

\begin{figure}[t]
\centerline{\fig{0.8}{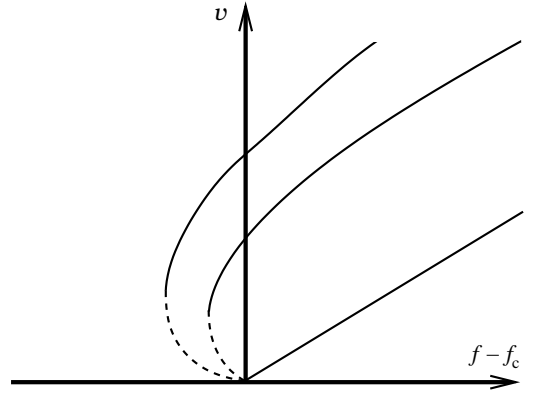}} \caption{schematic $v-f$ curve corresponding to Eq. (\ref{v2}) } \label{f:XXX1}
\end{figure}

\begin{figure}[t]
\centerline{\fig{0.6}{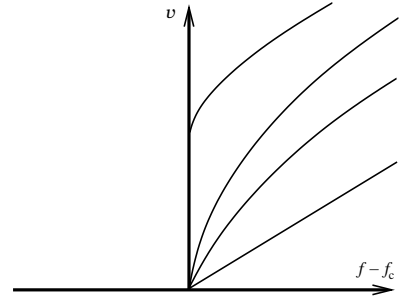}} \caption{schematic $v-f$ curve corresponding to a sharp transition}
\label{f:XXX2}
\end{figure}

Before obtaining the 1-loop FRG equations let us make some general
qualitative comments on the stability of the 1-layer elastic
quasi-static depinning to the viscous inter-layer coupling. The
absence of 1-loop corrections to $\cgamma_{12}$ implies that to this
order $\cgamma_{12}=\cgamma_{12}^0=-\ceta_0$. Consider
quasi-static depinning $v=0^+$. Then one finds
\begin{eqnarray}
\delta \cgamma_{11} = - \cgamma_{11} \Delta''(0^+) \int_q \frac{1}{q^4}\ ,
\end{eqnarray}
where a UV cutoff is implicit everywhere. This is the same correction as for the single-layer problem (i.e.\ for
$v=0$ it does not depend on $\cgamma_{12}$); hence under coarse graining $\cgamma_{11}$ is reduced compared to
its bare value (above the Larkin scale $L_c$, $\Delta''(0^+)$ is strictly positive). The intra-layer friction
$\cgamma_{11} < \cgamma_{11}^0=\cgamma_0+\ceta_0$ remains finite and non-zero for $d>4$ (where the above
integral converge at small $q$) while for $d \leq 4$ it becomes dependent on the system size $L$, $\cgamma_{11}
\approx (\gamma_0 + \eta_0) (L/L_c)^{z-2}$, $z<2$ being the single-layer dynamical exponent for elastic depinning in $d<4$. Since $\cgamma_{12}$ is uncorrected (it is negative) and $\cgamma_{11}$ is reduced,  it is clear that
the friction coefficient  of the center of mass of the system $\cgamma_+=\cgamma_{11}+\cgamma_{12}$ may become
negative at some scale, denoted $L_{\mathrm{pl}}$. When this occurs the fixed point of elastic quasi-static depinning
becomes unstable (and inconsistent). This {\it always} occurs for $d<4$, but only for $\ceta_0$ larger than a
critical value $\ceta_{c}$ for $d>4$. The qualitative picture is then as follows:

(i) $d > 4$: To lowest order the equation of motion reads:
\begin{eqnarray}
&& \left[ \kappa (\cgamma_{0}+ \ceta_0) - \eta_0 \right] v =  f - f_c + O(v_1^2)\qquad \\
&& \kappa = \cgamma_{11}/\cgamma^0_{11} = 1 + \Delta''(0^+) \int_q \frac{1}{q^4}\ , \label{kappa}
\end{eqnarray}
where we denote by $0 < \kappa < 1$ the usual reduction factor in friction in the single-layer problem. Elastic
quasi-static depinning exists, with velocity
\begin{eqnarray}
 v \approx  \frac{f - f_c}{\kappa (\cgamma_{0}+ \ceta_0 ) - \ceta_0 }\ ,
\end{eqnarray}
until the critical value of the interlayer coupling is reached,
\begin{eqnarray}
 \ceta_0 =  \ceta_c = \frac{\kappa}{1-\kappa} \cgamma_0\ .
\end{eqnarray}
Here the reentrant (or hysteretic) branch appears. This is qualitatively similar to the mean-field picture. One
can relate formally $(1-\kappa) \rightarrow M(c) \sim \overline{h^2}/c^2$ which for small disorder has the same
form as (\ref{kappa}), if the elastic coefficient (set to one is this Section) is restored and one identifies
$\Delta''(0) \rightarrow \overline{h^2}$.  An interesting question is the nature of the elastic to hysteretic
transition at $\eta_c$. Expanding (\ref{deltaf}) in powers of $v$ yields the equation of motion near the
critical point:
\begin{equation}
 (\ceta_c - \ceta) v = f -f_c + 2 v^2 (\gamma_{11}^2+\gamma_{12}^2)  \Delta'''(0^+) \int_q \frac{1}{q^6} +
O(v^3) \nonumber \\ \label{v2}
\end{equation}
As one can see on Figure \ref{f:XXX1}, the transition is continuous if $\Delta'''(0^+) <0$ and $v \sim
\sqrt{f-f_c}$ at the transition \footnote{From the factor $\int_q q^{-6}$ one could identify $d=6$ as a critical
dimension for the tricritical point, and find that the terms $D \hat u
\partial_t^2 u$ and $B \hat u (\partial_t u)^2$ both become relevant there. However one should
remember that $\Delta'''(0)$ is irrelevant in $d=6$. Whether this modifies the exponents and leads to new
universality class is left for future study}. Such a scenario may hold for the non-periodic, random manifold
class \footnote{For $d>4$, if the bare disorder is strong enough, $\Delta(u)$ develops a cusp, see Appendix in  \cite{BalentsLeDoussal2004}.}. A series of higher multicritical
points should exist, associated to correlators with leading behaviour $\Delta^{(n+1)}(0^+) v^n$. For the
periodic scalloped potential $n=\infty$ and the transition exhibits a jump, or a quasi-jump (inverse logarithm)
as in mean-field, illustrated in Fig \ref{f:XXX2}.

(ii) $d < 4$: the friction coefficient of the center of mass decreases with scale as:
\begin{eqnarray}
 \gamma_{+}(L) \approx (\gamma_0 + \eta_0) (L_c/L)^{2-z} - \eta_0\ .
\end{eqnarray}
It reaches values near zero at a scale
\begin{eqnarray}
 L_{\mathrm{pl}} = L_c \left(\frac{\eta_0}{\gamma_0 + \eta_0}\right)^{\!\!- \frac{1}{2-z}}
\end{eqnarray}
which diverges as $\eta_0 \to 0$, and which we term the ``plastic length''. Thus the depinning
transition of a system of size $L < L_{\mathrm{pl}}$ remains similar to the standard (finite-size) elastic depinning of a
single layer, while systems with $L > L_{\mathrm{pl}}$ cannot be described by single-layer elastic depinning. It is then
likely that the system breaks into domains which can depin and move independently. The full collective dynamics at scales $L
> L_{\mathrm{pl}}$ however remains to be understood. This instability of the elastic depinning at finite scale
is an effect beyond mean field.

Another important length scale is associated to a non-zero velocity. For single-layer elastic depinning it is
\begin{eqnarray}
 L_v \equiv L_{v}(\gamma_0) := L_c \left(\frac{\Lambda^2 r_f}{\gamma_0 v}\right)^{\!\!\frac{1}{z}}\ . \label{defL0}
\end{eqnarray}
It is such that $v \tau=r_f$, where $\tau$ is the time scale diverging at depinning and $r_f$ the correlation
length of the disorder, equal to the period $a$ (here set to unity) for the simplest CDW class. Beyond that
scale the effect of quenched disorder is washed out into an effective thermal noise and the motion is
uncorrelated. Equating the two scales $L_v(\gamma_0)=L_{\mathrm{pl}}$ defines a characteristic velocity scale:
\begin{eqnarray}
&& \frac{\gamma_0 v_{\mathrm{pl}}}{\Lambda^2 r_f} = \left(\frac{\eta_0}{\gamma_0 + \eta_0}\right)^{\frac{z}{2-z}} \label{vpl}
\end{eqnarray}
below which plastic effects cannot be neglected. The behaviour of the system at and beyond that scale still
needs to be elucidated.

Since we found that the FRG fixed point of elastic depinning  is dynamically {\it unstable} (to one loop) to the viscous coupling we now investigate the phase diagram of the moving phase.

\subsection{Functional RG}
Let us now derive and analyze the FRG equations to 1-loop order at non-zero velocity. For pedagogical purposes,
we use a Wilson scheme i.e.\ we compute $\Gamma[u,\tilde u]$ to one loop using a cutoff $\Lambda_l=\Lambda
e^{-l}$ and write RG equations as the cutoff is varied (i.e.\ integrating over a shell using $\int_{\Lambda_l
e^{-dl} < q <\Lambda_l} f(q) = \tilde S_d \Lambda_l^d f(\Lambda_l) \rmd l$). A method which can handle higher
loops, based on a non-zero mass scheme, is presented in the next section. Here we restrict to the periodic
problem and choose units such that the period is one.

The standard single-layer result for the correction to disorder upon
integration over the shell can be expressed as
\begin{eqnarray}
 \partial_l \tilde \Delta(u) = \epsilon \tilde \Delta(u) -
\frac{1}{2}  \left[\left(\tilde \Delta(u)-\tilde \Delta(0)\right)^2\right]''\ ,
\end{eqnarray}
where one has defined $\tilde \Delta(u)= \tilde S_d
\Lambda_l^{-\epsilon} \Delta(u)$. This result holds in the limit of
zero velocity $v=0^+$. As is well known, it results in a non-analytic
correlator  beyond the Larkin length $L_c$. We note that a non-analytic $\Delta(u)$ decreases $\gamma$, while  an analytic correlator would increase it. We
denote
\begin{eqnarray}
\sigma = \Delta''(0^+) \quad , \quad \tilde \sigma = \tilde
\Delta''(0^+)
\end{eqnarray}
The family of quadratic correlators:
\begin{eqnarray}
 \Delta(u)= \frac{\sigma}{2} \left[\frac{1}{6} - u(1-u)\right]
\label{scalloped}
\end{eqnarray}
for $0<u<1$, periodically continued to all $u$, is preserved by the
FRG flow, with $\partial_l \tilde \sigma =\epsilon \tilde \sigma - 3
\tilde \sigma^2$. It is realized by a scalloped potential, or
more generally by uncorrelated periodic shocks, and contains the
universal fixed point of the random periodic (RP) class: For $\epsilon=4-d
>0$ it flows to $\tilde \sigma^*=2-z_{\mathrm{el}}=\epsilon/3$. For $d=4$
the fixed point is at zero but the slow asymptotic decay $\tilde
\sigma \sim 1/(3 l)$ is universal.

Inserting formula (\ref{y4}) for the response function and the Fourier
series $\Delta(u)=\sum_p e^{i 2 \pi p u} \Delta_p$ (over integer
$p$) one finds the correction:
\begin{eqnarray}\label{3.23}
 \delta \cgamma_{11} &=& \sigma \int_q \left[ \frac{1}{2 v^2 \cgamma_+}
\Phi(\frac{q^2}{\cgamma_+ v}) + \frac{1}{2 v^2 \cgamma_-}
\Phi(\frac{q^2}{\cgamma_- v})\right] \\
\Phi(x)&:=&-  \int_0^\infty du \frac{\Delta''(u)}{\Delta''(0^+)} u e^{- x u} = \sum_p \frac{(2 \pi p)^2}{(x- i 2
\pi p)^2} \frac{\Delta_p}{\sigma} \nonumber
\end{eqnarray}
For the scalloped potential family (\ref{scalloped}) $\Delta_p=(1-\delta_{p0})
\sigma/(2 \pi p)^2 $, and $\Phi(x)$ reads
\begin{eqnarray}
 \Phi(x)= - \frac{1}{x^2} + \frac{1}{\left[2 \sinh(x/2)\right]^2}\ . \label{Phi}
\end{eqnarray}
In the sequel, we use the scalloped family (\ref{scalloped}) and the form (\ref{Phi}).
From (\ref{3.23}) and (\ref{Phi}) one obtains the RG equations
\begin{equation}
\partial_l \cgamma_{11} = - \tilde \sigma \cgamma_{11}  + \frac{\tilde
\sigma}{8 v^2} \left[\frac{1}{\cgamma_+}
\frac{\Lambda_l^4}{\sinh^2(\frac{\Lambda_l^2}{2 \cgamma_+ v})} +
\frac{1}{\cgamma_-} \frac{\Lambda_l^4}{\sinh^2(\frac{\Lambda_l^2}{2
\cgamma_- v})}\right] \label{rgeta11}
\end{equation}
and of course $\partial_l \cgamma_{12}=0$. For $v=0^+$ it reproduces
the elastic depinning RG equation $\partial_l \cgamma_{11} :=
(z_{\mathrm{el}}-2) \cgamma_{11} = - \tilde \sigma \cgamma_{11}$ which at the fixed
point yields the dynamical exponent $z_{\mathrm{el}}=2-\epsilon/3$.

One can see from formula (\ref{rgeta11}) that a non-zero velocity $v>0$ tends to cut the flow of $\cgamma_{11}$.
This is a usual effect in the case of elastic depinning ($\eta_0=0$) associated, in that case, to the single
length scale $L_v(\gamma_0)$ defined in (\ref{defL0}). Here there are a priori two length scales, associated to
the two modes $u^+$ and $u^-$. The effect of disorder is washed out only for scales larger than both lengths,
i.e.\ if
\begin{eqnarray}
\gamma_{\pm} v e^{2 l}/\Lambda^2 \gg 1 \quad , \quad L > L_v^\pm\ .
\end{eqnarray}
Then the equation reduces to $\partial_l \cgamma_{11} = - \frac{\tilde \sigma}{12} \frac{\Lambda^4}{\cgamma_+
\cgamma_- v^2} e^{-4 l}$. The difficulty is that $L_v^\pm$ are not simply equal to $L_v(\gamma^0_{\pm})$
since the $\gamma_{\pm}(l)$ do not behave as the single-layer coupling (it does only for scales $L < L_{\mathrm{pl}}$).
In fact, $\gamma_{+}(l)$ may vanish at some scale, hence the condition $\gamma_{+}(l) v e^{2 l}/\Lambda^2 \gg 1$
may never be fulfilled, at {\it any} scale. A more careful analysis, performed below, is thus required.

One notes that (\ref{rgeta11}) is the derivative $\partial_l
\cgamma_{11} = \partial_l \cgamma_+ = - \partial_v \partial_l \tilde f$
with:
\begin{equation}
\partial_l \tilde f = \cgamma_{11} \tilde \sigma v - \frac{1}{4}
\tilde \sigma \Lambda_l^2 \left[\coth(\frac{\Lambda_l^2}{2 \cgamma_+ v})+
\coth(\frac{\Lambda_l^2}{2 \cgamma_- v})\right]
\end{equation}
from which the velocity-force characteristics is obtained as:
\begin{equation}
f(v) = \cgamma_0 v - \int_0^\infty \rmd l\, \partial_l \tilde f
\label{vfcurve}
\end{equation}
In the limit $v=0^+$ one recovers $\partial_l \tilde f = - \frac{1}{2} \tilde \sigma \Lambda_l^2$ which
integrates to $-f_c^{\mathrm{el},\mathrm{sl}}=-\frac{1}{4} \tilde \sigma \Lambda^2$, the critical force of a single elastic layer.
One notes the general relation,
\begin{equation}
\partial_v f(v) = \cgamma_+  \label{slope}\ ,
\end{equation}
valid for $l=\infty$, or for any intermediate scale, if one defines
a finite-scale curve for  $f(v)$ by setting the upper integration bound to
$l$ in (\ref{vfcurve}).

\begin{figure}[t]
\centerline{\fig{1}{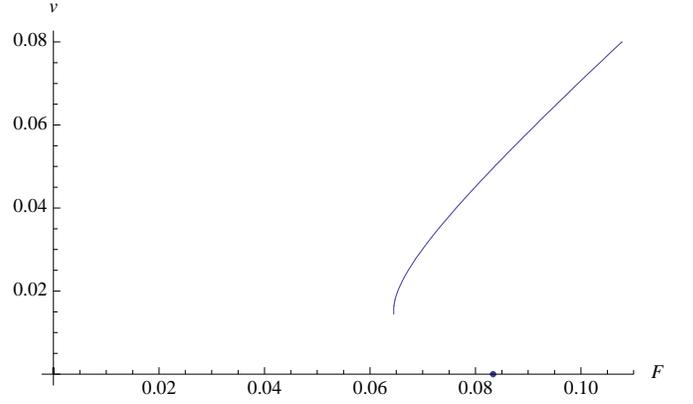}} \caption{$v(F)$ curve obtained by integration of the one-loop equations in the
text, for $\Lambda=1$, $\tilde \sigma=1/3$ ($\epsilon=1$) $\eta_0=10$, $\gamma_0=1$} \label{f:oneloop}
\end{figure}

We now study the flow of $\cgamma_{11}$ which depends on $\cgamma_+=\cgamma_{11}-\ceta_0$ and
$\cgamma_-=\cgamma_+ + 2 \ceta_0$. We recall that the starting value is $\cgamma_{11}^0=\cgamma_0+\ceta_0$. If
the velocity is large enough, although $\cgamma_+$ decreases upon renormalization, the corrections may be weak
enough so that it remains positive, even for $d<4$. In the latter case,  there should always be a critical
velocity $v_c$ such that $\cgamma_+(l=\infty)=0$. For $v>v_c$ the $v(f)$ curve is well-defined and continuous.
For $v<v_c$ there is no moving solution such that $v(f)$ has a positive slope. Hence in a fixed-applied-force
ensemble there is a jump to the pinned phase as the force is decreased. At $v=v_c^+$, from (\ref{slope}) the
slope of the $v(f)$ curve is infinite. This corresponds to the minimal force $f_c^{<}$ at which the jump must
occurs. By contrast, when the force is increased in the pinned phase the critical force is
$f_c^{>}=f_c^{\mathrm{el},\mathrm{sl}}$. It corresponds to the maximal force at which the jump upward in velocity to the moving
state must occur \footnote{In some cases it was observed that the jump can occur before these extremal values,
either due to finite-size effects or due to a dynamical instability, which is beyond the present description.}. At $v=v_c^+$ the
length $L_v^+$ is {\it infinite}. This suggests that motion should be correlated on all scales, and that this
point is very much like a critical point where scale invariance holds. An example of a $v(f)$ curve predicted by
the one-loop FRG is given in Fig.\ \ref{f:oneloop}.

To estimate the jump velocity $v_c$ it is simpler to first study a
model where the bare value of $\cgamma_+$, $\cgamma_0$, is already small
compared to $\cgamma_-$, i.e.\ $\cgamma_0 \ll \ceta_0$. Then equation (\ref{rgeta11})
can be approximated by:
\begin{equation}
\partial_l \cgamma_{+} = - \tilde \sigma \ceta_0  + \frac{\tilde
\sigma \Lambda^4 \rme^{- 4 l}}{16 \ceta_0 v^2 \sinh^2(\frac{\Lambda^2
e^{- 2 l}}{4 \ceta_0 v})}\ . \label{approx}
\end{equation}
This is integrated into:
\begin{eqnarray}
 \cgamma_{+}(l=\infty) &=& \cgamma_0 - \frac{1}{2} \tilde \sigma \ceta_0
H\!\left(\frac{\Lambda^2}{4 \ceta_0 v}\right) \\
 H(x)&=&\int_0^x \frac{\rmd y}{y}\left[1 - \frac{y^2}{\sinh^2(y)}\right] \nonumber\\
 &=& -1+x \coth(x) + \ln\left(\frac{x}{\sinh(x)}\right)\ , \qquad
\end{eqnarray}
with $H(x) \approx \ln(2x)-1$ at large $x$ and $H(x) \approx
\frac{x^2}{6}$ at small $x$.
The critical velocity $v_c$ is hence determined by
\begin{eqnarray}
 H\left(\frac{\Lambda^2}{4 \ceta_0 v_c}\right) = \frac{2}{\tilde \sigma}
\frac{\cgamma_0}{\ceta_0}\ ,
\end{eqnarray}
which gives the asymptotic behaviour:
\begin{eqnarray}
 \frac{\ceta_0 v_c}{\Lambda^2}  &\approx& \frac{1}{2 e} \exp\left(-
\frac{2 \cgamma_0}{ \tilde \sigma \ceta_0 }\right) \quad , \quad
\frac{\tilde \sigma \ceta_0}{\cgamma_0} \ll 1
\\
 \frac{\ceta_0 v_c}{\Lambda^2} &\approx& \frac{1}{8} \left(\frac{\tilde
\sigma \ceta_0}{3 \cgamma_0}\right)^{\!\!1/2} \quad , \quad  \frac{\tilde \sigma \ceta_0}{\cgamma_0} \gg 1
\label{largeeta}
\end{eqnarray}
Here we have assumed $\tilde \sigma$ to be scale independent, hence a reasonable value for it \footnote{Given
the assumption $\cgamma_0 \ll \ceta_0$ the first regime exists only for small $\tilde \sigma$ which is either
$\epsilon \to 0$, or if bare disorder is very small until the scale which controls the jump.} is $\tilde
\sigma=\sigma^*=2-z_{\mathrm{el}}=\epsilon/3 + O(\epsilon^2)$. At the fixed point in $d=3,2,1$ the second regime is the
relevant one and gives the value of the critical velocity for large $\ceta_0/\cgamma_0$.

To estimate the critical velocity when $\ceta_0/\cgamma_0$ is small, one must first integrate the flow up to
scale $l_1$ at which $\cgamma_+(l_1)= k \ceta_0 = k/(k+2) \cgamma_-$ and $k$ a number smaller than unity. Within
this scale we can approximate $\cgamma_{11}(l)= (\ceta_0+\cgamma_0) e^{- \tilde \sigma l}$, which yields
$(\ceta_0+\cgamma_0) e^{- \tilde \sigma l_1} = (k+1) \ceta_0$. The length scale $L_c e^{l_1}$ is of the order of the
plastic length $L_{\mathrm{pl}}$ introduced above. Beyond that scale one can apply the previous analysis
\begin{eqnarray}
 \cgamma_+(l=\infty)=\cgamma_+(l_1) - \frac{1}{2} \tilde \sigma \ceta_0
H\left(\frac{\Lambda^2}{4 \ceta_0 v e^{2 l_1}}\right)\ ,\qquad
\end{eqnarray}
which yields the estimate
\begin{eqnarray}
\frac{\ceta_0 v_c}{\Lambda^2} \sim  \left(\frac{(k+1)
\ceta_0}{\cgamma_0 + \ceta_0}\right)^{2/\tilde \sigma} \frac{1}{4 H^{-1}(2
k /\tilde \sigma)}\ .
\end{eqnarray}
Hence we find that the critical velocity vanishes as $\ceta_0 v_c \sim (\ceta_0/\cgamma_0)^{2/(2-z)}$ in the
limit of small viscous coupling, consistent with the estimate (\ref{vpl}) for the scale at which plastic effects
become important. The present 1-loop analysis indicates however that the jump is always non-zero \footnote{One
notes that the flow of the disorder correlator, which is too complicated to analyze here, is also cut by
velocity at the scale $\max(L_v^+,L_v^-)$. Hence above that scale the parameter $\tilde \sigma$ cannot be assumed
to take its fixed-point value and instead will decrease to zero. Since the effects computed here occur below
these scales, one expects at most a change in the prefactors from taking these effects into account.}.

It is instructive to compare with the predictions from mean-field theory (MFT) recalled in Section \ref{mft}. In
the regime of large viscous coupling $\ceta_0 \gg \cgamma_0$, one sees that formula (\ref{largeeta}) is very
similar to the mean-field prediction
\begin{eqnarray}
 \ceta_0 v_c = \frac{(c+h_0)}{2} \left(\frac{\ceta_0}{3 \cgamma_0}\right)^{\!\!1/2} \ ,
\end{eqnarray}
if one identifies $c+h_0 \to \sqrt{\tilde \sigma}/4$. Hence the 1-loop FRG result, taken in the limit of large
$\ceta_0$, is very similar to mean-field theory (MFT) even for $d<4$, with the difference that the disorder
parameter flows to a universal fixed value $\tilde \sigma^*$. In the other limit of small ratio
$\ceta_0/\cgamma_0$, the result is very different from MFT because of the strong renormalisation of the in-layer
friction coefficient, and the threshold $\ceta_c$ which exists in mean field is zero for $d<4$.

It is also instructive to study the FRG flow for $d=4$ and $d>4$. For $d=4+\epsilon$ and a scalloped potential
one has $\tilde \sigma = \tilde \sigma_0 e^{- \epsilon l}$, hence one finds at zero velocity
$\cgamma_{11}(l)=(\cgamma_0+\ceta_0) \exp\left(- \frac{\tilde \sigma_0}{\epsilon} (1-e^{- \epsilon l})\right)$
and $\cgamma_+(l)=\cgamma_{11}(l)-\ceta_0$. There is thus a threshold for the jump in the $v(f)$ curve; it
occurs only for $\ceta_0 > \ceta_c$ with
\begin{eqnarray}
\ceta_c = \frac{\cgamma_0}{e^{\tilde \sigma_0/\epsilon}-1}\ .
\end{eqnarray}
$\eta_c$ becomes very small as $d \to 4^+$. For $d=4$ one has $\tilde \sigma = 1/(3 l)$, hence
$\cgamma_{11}=(\cgamma_0+\ceta_0)/l^{1/3}$ and there is no threshold, $\ceta_c=0$. The plastic length scale
however diverges extremely fast as $L_{\mathrm{pl}} = L_c \exp((\cgamma_0/\ceta_0)^3)$ for small $\ceta_0$.

The analysis of this section used that $\cgamma_{12}$ is not corrected.
We now turn to a two-loop analysis to check whether this holds to
higher orders.

\section{Analysis including 2-loop corrections}

\label{sec:twoloop}

In this Section we compute the corrections which arise at two loop around the quasi-static elastic depinning
transition of the single layer at $f=f_c^{\mathrm{el},\mathrm{sl}}$. The calculation is performed in the limit $v \to 0^+$. The
FRG flow is discussed separately for the non-periodic and for the periodic cases. Possible consequences at non-zero $v$ are discussed in each section.

The natural setting for higher-loop calculations is to use a mass as an infrared cutoff. It amounts to adding the
force vector $m^2 (w(t) - u^i_{x,t}) $ to the r.h.s of the equation of motion (\ref{y2}). It describes two layers
both pulled by a spring attached to a point at position $w(t)$ which performs quasi-static forward motion. In
that setting, it was shown \cite{LeDoussalWiese2006a} that the force correlator $\Delta(u)$ computed in the FRG is an observable
related to the mean-square center-of-mass fluctuation around $w(t)$ in each layer. One introduces the rescaled
correlator
\begin{eqnarray}
\tilde \Delta(u) = C_d m^{- \epsilon+ 2 \zeta} \Delta(u m^{- \zeta})\ ,
\end{eqnarray}
where $C_d = \epsilon \tilde I_2=\epsilon \int_k (k^2+1)^2$ for $\epsilon=4-d >0$. One finds that $\tilde
\Delta(u)$ converges to a fixed point, and to 1-loop order it reproduces the Wilson approach.

\subsection{2-loop FRG equations}

The 2-loop FRG flow-equation for the disorder is taken to be the same as the one derived in
\cite{LeDoussalWieseChauve2002} at the quasi-static depinning transition:
\begin{eqnarray}\label{k20}
\partial_{\ell} \Delta (u) &=& (\epsilon -2 \zeta)\Delta (u) + \zeta  u \Delta' (u)\nonumber \\
&& -\frac{1}{2}\left[\left(\tilde \Delta (u)-\tilde \Delta (0) \right)^{2} \right]''\nonumber \\
&&+\frac{1}{2} \left[\left(\tilde \Delta (u)-\tilde \Delta (0) \right)\tilde \Delta' (u)^{2} \right]''\nonumber \\
&&+\frac{1}{2}\tilde \Delta' (0^{+})^{2}\tilde \Delta'' (u)\ .
\end{eqnarray}
where $\partial_{\ell}:=- m \partial_m$. As explained there, the derivation of this FRG equation at two loop (especially the last term) relies on the Middleton
theorem \cite{Middleton1992} which states that if all local velocities are positive at some time, they remains so at all times. In
the two-layer viscous problem this property does not hold stritly, as backward motion of one layer is sometimes
observed. The present calculation hence assumes that these effects can be neglected at large scale near the
quasi-static depinning, and to this order.

The corrections to the friction coefficients are computed in Appendix \ref{a:2loop}. They read:
\begin{eqnarray}\label{k2}
\partial_{\ell }\, \cgamma_{12} &=& \frac{ \cgamma _{12} \tilde \Delta'(0^+)
\tilde \Delta'''(0^+)}{2 }\log \left|\frac{\cgamma _{11}+\cgamma _{12}}{\cgamma _{11}-\cgamma _{12}}\right|
\\
\label{k3}
\partial_{\ell }\, \cgamma_{11} &=&  \cgamma_{11} \left[- \tilde \Delta'' (0) +
\tilde \Delta'' (0)^{2} + \tilde \Delta''' (0)\tilde \Delta' (0)\left({\textstyle \frac{3}{2}}-\ln 2 \right)
\right]
\nonumber \\
&&+
 \frac{3 \cgamma _{12} \tilde \Delta'(0^+)
\tilde \Delta'''(0^+)}{2 }\log \left|\frac{\cgamma _{11}+\cgamma _{12}}{\cgamma
_{11}-\cgamma _{12}}\right| \nonumber\\
&&+\cgamma _{11} \tilde \Delta'(0^+) \tilde \Delta'''(0^+) \log \left|1-\frac{\cgamma _{12}^2}{\cgamma
_{11}^2}\right|
\end{eqnarray}
The calculation was performed in the physical domain $\cgamma_{+},\cgamma_{-}>0$. For mainly illustrative
purpose, an analytical continuation was performed to the domain with negative friction coefficients, which yield
the absolute values above. We find however that whenever the coefficient of the $\log$ terms are non-zero, the
solution of the flow, obtained below, remains in the physical region.

It turns out that the two combinations $\tilde \Delta''(0^+)$ and $\tilde \Delta'(0^+) \tilde \Delta'''(0^+)$
which appear in these equations are universal numbers which can be related to the roughness exponent, $\zeta$
(using derivatives of (\ref{k20}) at $u=0$), independently of the precise form of the fixed point:
\begin{eqnarray}
&& \tilde \Delta''(0) = \frac{1-\zeta_1}{3} \epsilon+\frac{\zeta_1^2-3 \zeta_1-3
   \zeta_2+2}{9} \epsilon^2+O(\epsilon^3) \nonumber \\
&& \tilde \Delta'(0) \tilde \Delta^{(3)}(0) = \frac{(1- \zeta_1) \zeta_1}{12} \epsilon^2 + O(\epsilon^3)\ .
\end{eqnarray}
Here we have defined
\begin{eqnarray}
 \zeta = \zeta_1 \epsilon + \zeta_2 \epsilon^2 + O(\epsilon^3)\ .
\end{eqnarray}

\subsection{Non-periodic problem}

\label{subsec:nonper}

\begin{figure}[t]
\centerline{\Fig{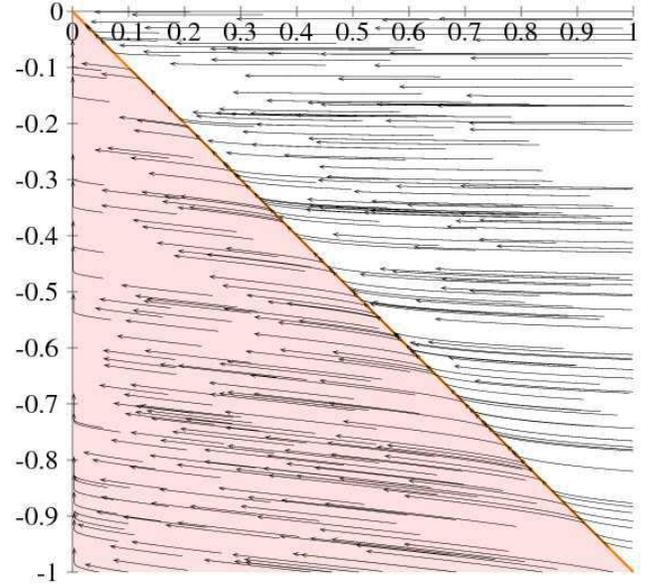}} \caption{Flow of $\cgamma_{11}$ and $\cgamma_{12}$, for $\epsilon =1$, as a
function of $\cgamma_{11}$ (x-axis) and $\cgamma_{12}$ (y-axis). The separatrix is the diagonal line (orange)
$\cgamma_{+}=\cgamma_{11}+\cgamma_{12}=0$. All physical initial conditions , corresponding to  $\cgamma_{+}>0$, remain
physical. The shaded pink
region corresponds to unphysical initial conditions $\cgamma_{+}<0$.} \label{f:floweta1}
\end{figure}

\begin{figure}[t]
\centerline{\Fig{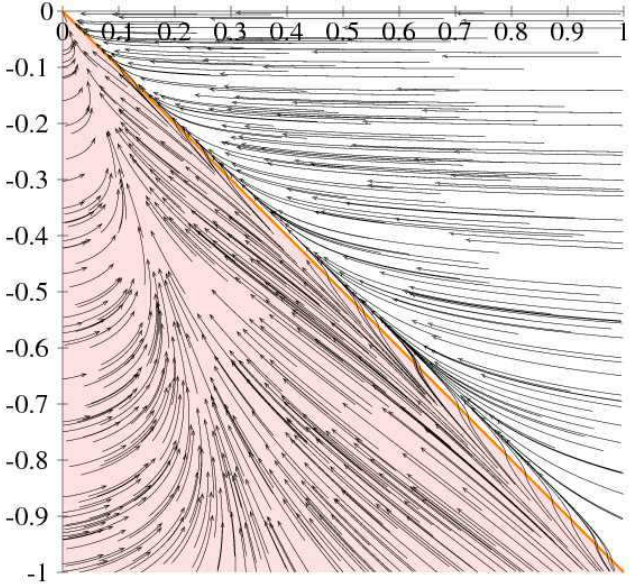}} \caption{Flow of
$\cgamma_{11}$ and $\cgamma_{12}$, for $\epsilon =4$, as a function of
$\cgamma_{11}$ (x-axis) and $\cgamma_{12}$ (y-axis). The separatrix (orange)
is the line $\cgamma_{+}=\cgamma_{11}+\cgamma_{12}=0$. The pink region
corresponds to unphysical initial conditions $\cgamma_{+}<0$.} \label{f:floweta4}
\end{figure}

As was shown in \cite{LeDoussalWieseChauve2002} for a wide range of microscopic disorders, there is a unique
elastic-depinning fixed-point, calculated there, and identical at one loop to the random-field (RF) disorder
class. At this fixed point one finds $\zeta_1=1/3$ and $\zeta_2=1/(27 \sqrt{2} \gamma)$ with
$\gamma=0.5482228..$. This yields:
\begin{eqnarray}\label{k25}
\partial_{\ell} \cgamma_{11} &=& \left(-0.0432087 \epsilon ^2-\frac{2 \epsilon
}{9}\right) \cgamma _{11}\nonumber \\
&& +\frac{1}{54} \log \left(1-\frac{\cgamma
_{12}^2}{\cgamma _{11}^2}\right) \cgamma _{11} \epsilon ^2\nonumber \\
&& +\frac{1}{36} \log
\left(\frac{\cgamma _{11}+\cgamma _{12}}{\cgamma _{11}-\cgamma _{12}}\right) \cgamma
_{12} \epsilon ^2\\
\partial_{\ell}\cgamma_{12}&=& \frac{1}{108} \epsilon ^2 \log
\left(\frac{\cgamma _{11}+\cgamma _{12}}{\cgamma _{11}-\cgamma _{12}}\right) \cgamma
_{12}
\end{eqnarray}
We integrated the flow-equations numerically. The result is given in Fig.\ \ref{f:floweta1} for $\epsilon =1$ and
in Fig.\ \ref{f:floweta4} for $\epsilon =4$, to illustrate how the flow changes with $\epsilon$. Looking
carefully, one sees that starting in the physical region $\cgamma_{+}>0$, the unphysical region $\cgamma_{+}< 0$
is avoided. One also sees that $\cgamma_{+}$ approaches zero quickly, at least for small $\epsilon$.
%\begin{figure}[b]
%\centerline{\Fig{etafloweps=2-imp}} \caption{Flow of
%$\cgamma_{11}$ and $\cgamma_{12}$, for $\epsilon =2$, as a function of
%$\cgamma_{11}$ (x-axis) and $\cgamma_{12}$ (y-axis). The separatrix (orange)
%is the line $\cgamma_{+}=\cgamma_{11}+\cgamma_{12}=0$. The pink region
%corresponds to unphysical initial conditions $\cgamma_{+}<0$.} \label{f:floweta2}
%\end{figure}%
%
%\begin{figure}[b]
%\centerline{\Fig{etafloweps=3-imp}} \caption{Flow of
%$\cgamma_{11}$ and $\cgamma_{12}$, for $\epsilon =3$, as a function of
%$\cgamma_{11}$ (x-axis) and $\cgamma_{12}$ (y-axis). The separatrix (orange)
%is the line $\cgamma_{+}=\cgamma_{11}+\cgamma_{12}=0$. The pink region
%corresponds to unphysical initial conditions $\cgamma_{+}<0$.} \label{f:floweta3}
%\end{figure}%
We now confirm these two findings analytically. To do so, we change
variables to $\cgamma_{+}=\cgamma_{11}+\cgamma_{12}$ and
$\cgamma_{-}=\cgamma_{11}-\cgamma_{12}$, see Eq.~(\ref{y5}). We are interested
in  $\cgamma_{+}\approx 0$.  There the flow-equations become
\begin{eqnarray}\label{k5}
\partial_{\ell} \cgamma_{+}&=& \epsilon \left(- 0.009259 \log
\left(\frac{\cgamma _+}{\cgamma _-}\right) \epsilon - 0.008768 \epsilon -\frac{1}{9}
\right) \cgamma_{-} \nonumber\\
\partial_{\ell} \cgamma_{-}&=& -\left( \frac{1}{9}+ 0.008768 \epsilon\right)
   \epsilon  \cgamma_{-}\ .
\end{eqnarray}
The second equation has the solution
\begin{equation}\label{3.2}
\cgamma_{-} (\ell) = \rme^{-( 1/9+ 0.008768 \epsilon)
   \epsilon \ell} \cgamma_{-} (0)\ .
\end{equation}
The solution for $\cgamma_{+}$ is easiest expressed as a function of $\cgamma_{-}$, instead of $\ell$:
\begin{equation}\label{3.3}
\frac{\rmd \cgamma_{+} (\cgamma_{-})}{\rmd \cgamma_{-}} = 1+ \frac{\epsilon \log \left(\frac{\cgamma
_+\left(\cgamma _-\right)}{\cgamma _-}\right)}{12+0.94697 \epsilon }\ .
\end{equation}
The ratio
\begin{equation}\label{k6}
r (\cgamma_{-}):= \frac{\cgamma _+\left(\cgamma _-\right)}{\cgamma_{-}}
\end{equation}
satisfies a closed flow equation as a function of $\cgamma_{-}$:
\begin{equation}\label{k7}
- \frac{\rmd r (\cgamma_{-})}{\rmd \ln \cgamma_{-}} = r (\cgamma_{-}) -1-
  \frac{\epsilon \log
r (\cgamma_{-})}{12+0.94697
\epsilon }\ .
\end{equation}
For all relevant values of $\epsilon$ ($0\le \epsilon \le 4$), there are two solutions: $r=1$ (unstable) and a
non-trivial ($r^*\ll 1$) solution of
\begin{equation}\label{k8}
r^{*}-1 =\frac{\epsilon \log r^{*}}{12+0.94697 \epsilon }
\end{equation}
which yields:
\begin{equation}\label{k9}
r^{*}\approx  \exp \left(-\frac{12+0.94697\epsilon}{\epsilon} \right)\ .
\end{equation}
The eigenvalue of the flow close to $r^{*}$ is at leading order
\begin{equation}\label{k11}
y \approx - \frac{\epsilon^{2} \rme^{12/\epsilon}}{108}\ .
\end{equation}
Thus for $\epsilon$ small, this fixed point is very attractive. This is the fixed point obtained numerically
above. It has the property that $\cgamma_{+}$ remains strictly positive.

From (\ref{3.2}) we extract the dynamical exponent associated with $\cgamma_{-}$:
\begin{equation}\label{k12}
z_{\mathrm{plastic}}=2 - \frac{\epsilon}9 - 0.008768 \epsilon^{2}
\end{equation}
Since $\cgamma_{+}\approx r^{*}\cgamma^{-}$, it has the same dynamical scaling, and the above
$z_{\mathrm{plastic}}$ is indeed the critical exponent for the dynamics of both modes.

Hence within the two-loop analysis, and the stated assumptions, one finds a fixed point for the case of
non-periodic disorder. The dynamical exponent at this new fixed point deviates even at leading order in
$\epsilon$ from the standard elastic depinning value:
\begin{equation}\label{k13}
z_{\mathrm{elastic}}=2 - \frac{2\epsilon}{9}- 0.0432087 \epsilon^{2}\ .
\end{equation}
Compared to one loop, the two-loop corrections appear singular, as seen from the $\ln(\gamma_-/\gamma_+)$
factors in the corrections to friction. As a result their magnitude is drastically enhanced above the plastic
length $L_{\mathrm{pl}}$ from an expected $O(\epsilon^2)$ to an actual $O(\epsilon)$. The term $\tilde \Delta'(0^+)
\tilde \Delta'''(0^+) \ln(\gamma_-/\gamma_+)$ in the correction to $\gamma_{12}$ in Eq. (\ref{k2}) is in effect
replaced, upon integration of the flow, by $\tilde \Delta''(0)$. This results in a value for $2-z$ twice
smaller, to leading order, than the usual elastic fixed point.

\medskip
To summarize, the 1-loop analysis showed that $\gamma_+$ becomes very small near the plastic length, and
provided a scenario for scales larger than $L_{\mathrm{pl}}$ which could sustain only a moving state at $v>v_c$. Although
we did not perform the analysis for the non-periodic case in detail we do not expect a difference at one loop.
The present analysis - in the non-periodic case - shows that additional physics occurs at two loop. It suggests
that a $v=0^+$ state may still be possible. From the above analysis, one could surmise that it results in a very
abrupt, almost vertical $v(f)$ curve (since $\gamma_+$ is found to converge very rapidly to a very small value)
which is not strictly a jump, although it may look like one in a numerical calculation or an experiment. This
``quasi-jump'' would occur near the critical force of the elastic system, at variance with one loop. To
confirm or infirm this scenario one would need to include the effects of a non-zero $v$, and a possible violation
of the Middleton theorem within the two-loop theory, a challenge left for future work.

It also remains to be investigated to which extent the present analysis can be trusted in the region where
$\gamma_+$ becomes very small, i.e.\ the region where $\ln(\gamma_-/\gamma_+)$ becomes of order $1/\epsilon$. We
expect that in that region terms such as $\dot u^2$ in the equation of motion may become important. Such effects are presumably
correctly resummed in the two-loop corrections and may explain why $\gamma_+$ remains positive. However since
the counting of order in $\epsilon$ becomes unconventional if one follows the flow further in that region, there
is no guarantee that higher loops may not lead to even more singular terms. In the best-case scenario only the
$O(\epsilon^2)$ term in (\ref{k13}) would be changed by higher-loop corrections. Although the present results
hint at a new depinning universality class with a dynamical exponent $z=z_{\mathrm{pl}}$, a deeper understanding of the
behaviour of the system in the plastic region seems necessary before a firm conclusion can be drawn.

\subsection{Periodic problem}

\label{subsec:per}

The case of periodic disorder is also challenging. The quasi-static depinning fixed point has the form $\tilde
\Delta(u) \sim u (1-u)$, as in (\ref{scalloped}) with
\begin{equation}\label{k21}
\tilde \sigma = \frac{\epsilon}{3} +\frac{\epsilon^{2}}{18}+\dots\ ,
\end{equation}
and is expected to maintain that form to any order in $\epsilon$. If the system is {\it exactly} at its fixed
point, then, since $\tilde \Delta''' (0^{+})=0$ at this fixed point, the flow-equations for the $\cgamma$'s read
\begin{eqnarray}\label{k22}
\partial_{\ell }\, \cgamma_{12} &=& 0\\
\partial_{\ell }\, \cgamma_{11}  &=&  \cgamma_{11} \left ( - \tilde \sigma  +
\tilde \sigma^{2} \right) = - \cgamma_{11} (2-z_{\mathrm{el}}) \\
z_{\mathrm{el}} & = & 2 - \left(\frac{\epsilon}{3}+\frac{\epsilon^{2}}{9} \right)\ .
\end{eqnarray}
Hence there are no drastic effects of the two-loop corrections, apart from changing the value of $z$:
$\cgamma_{11}$ always decreases as in 1-loop, $\gamma_+$ vanishes at some scale, and the 1-loop analysis
remains at least qualitatively correct. Hence this confirms the 1-loop approach.

It is less obvious to understand the situation where the system is not exactly at its fixed point, but converges
to it, i.e.\ $\tilde \Delta'''(0^+) \sim e^{- \epsilon l}$. Inserting this behaviour in the above two-loop
equations still results in drastic effects, i.e.\ $\gamma_+$ never crossing zero, again due to the logarithmic
divergence of the corrections in that region, as for a non-periodic problem. The discontinuous behaviour between
a zero and a small non-zero $\tilde \Delta'''(0^+)$ remains to be understood. One scenario which would save the
agreement with the 1-loop approach is that other irrelevant operators than $\Delta'''(0^+)$, neglected in the
two-loop treatment of the periodic class, are equally important and modify the result back to (\ref{k22}). More
work is clearly needed to settle these issues.

\section{Toy Models with 2 particles}

\label{sec:toy}

To gain insight on some of the issues arising in the dynamics of coupled elastic layers it is instructive to
study the model in $d=0$ i.e.\ a toy model with two particles. This approach has proved useful for the elastic-depinning problem \cite{LeDoussalWiese2006a}, in particular in clarifying the information contained in the FRG functions. As we show
below, a variety of behaviors arises already for two viscously coupled particle. Here we focus on the simplest
situation of two particles in a periodic one-dimensional landscape driven by a force, and leave for future work
the interesting non-periodic case, as well as driving by a spring (which is more suitable for comparison with
FRG). The model is thus the $d=0$ version of (\ref{MF.1}), with a pinning force $h^i Y(u^i-\beta^i)$;  we
choose $h^1=h^2$ for simplicity. The random phase can be eliminated by a shift of the $u^i$, hence it is
sufficient to study the case of two particles in the same landscape (up to a change in initial conditions). We
first study smooth disorder, and then a scalloped landscape\footnote{If an additional self-consistency condition
is imposed, these models can also be used to implement a third mean-field approach, discussed at the end of
Section \ref{mft}.}.

\begin{figure*}
\centerline{\clipfig{0.7}{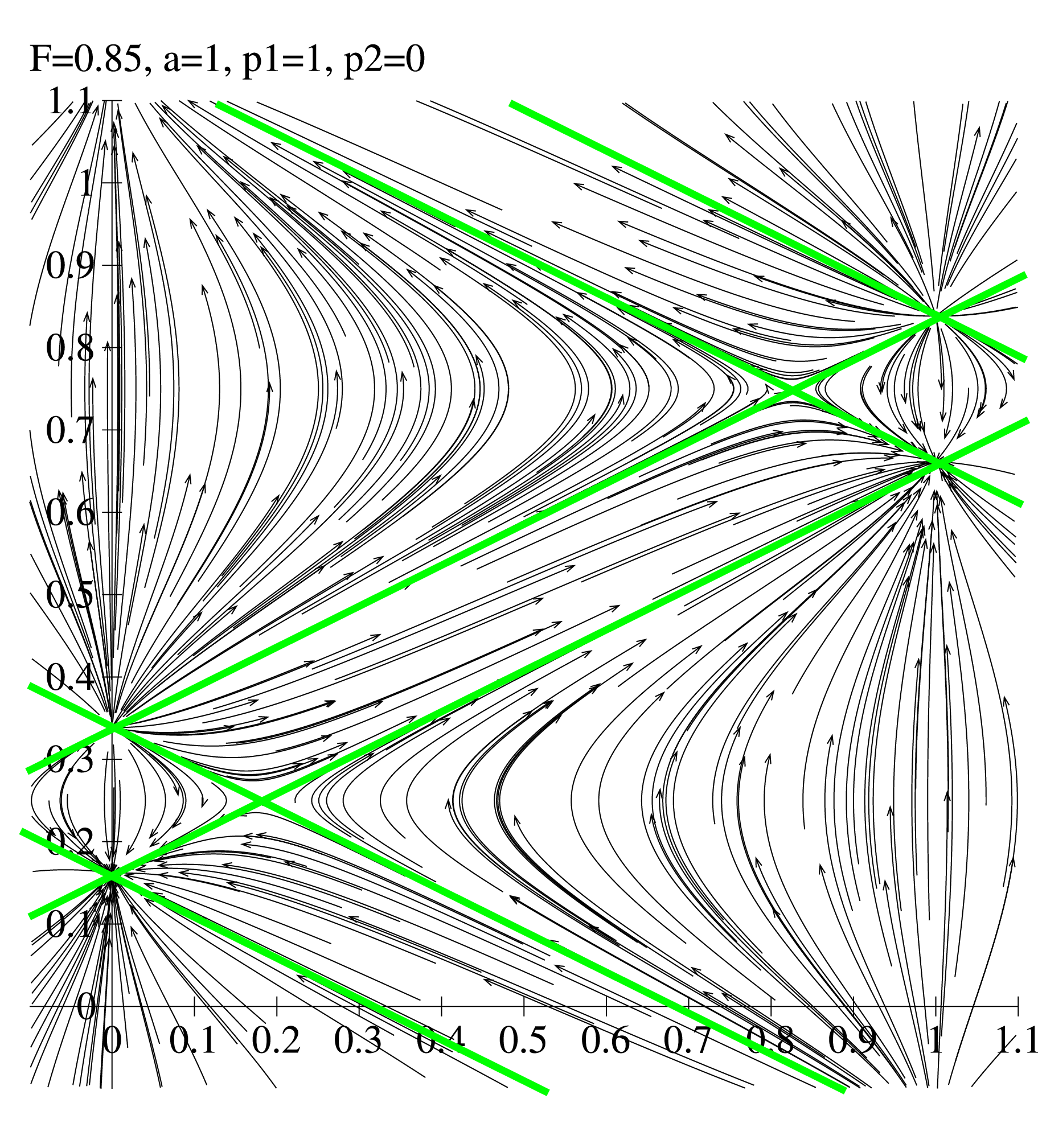}\clipfig{0.7}{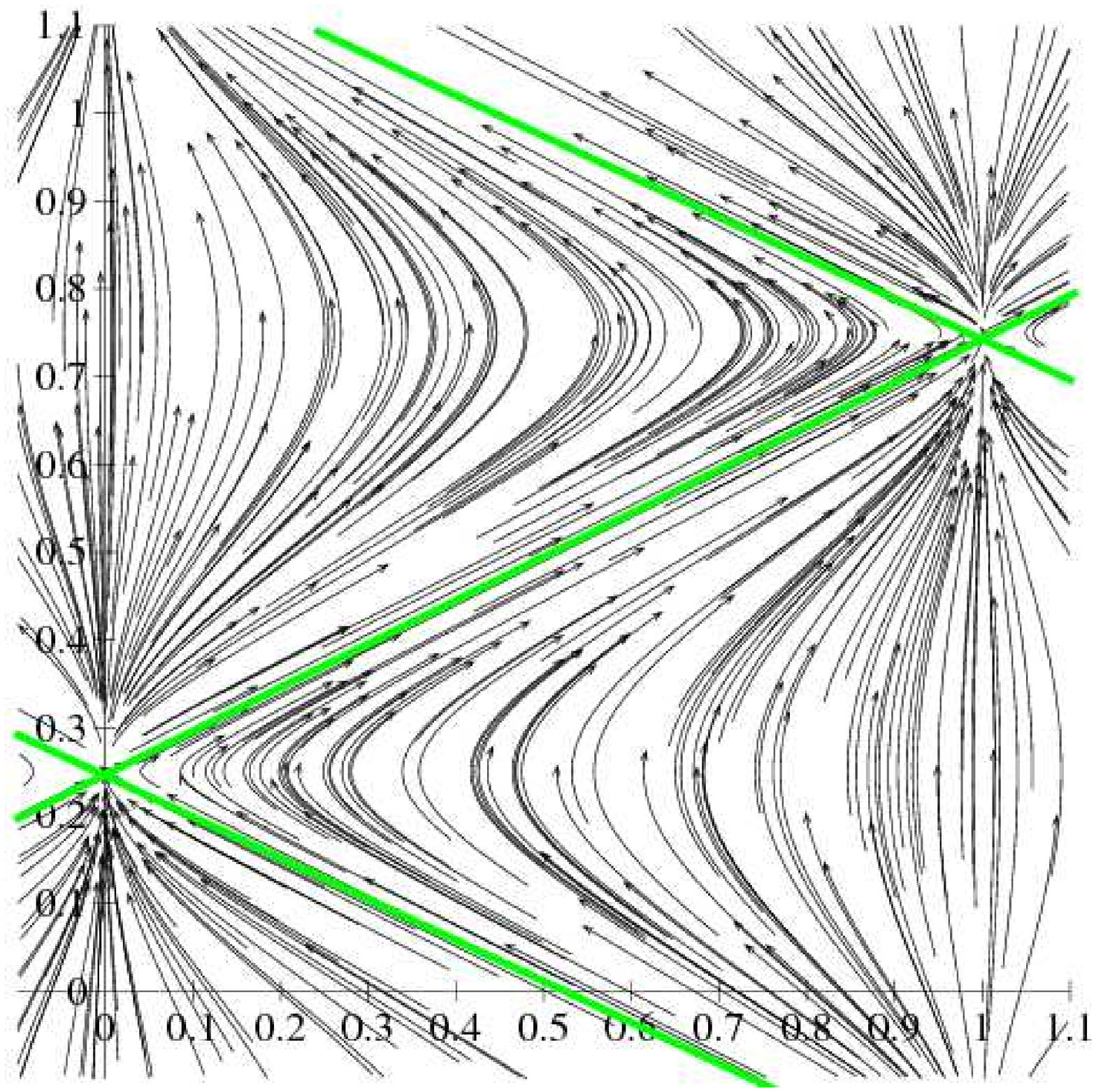}\clipfig{0.7}{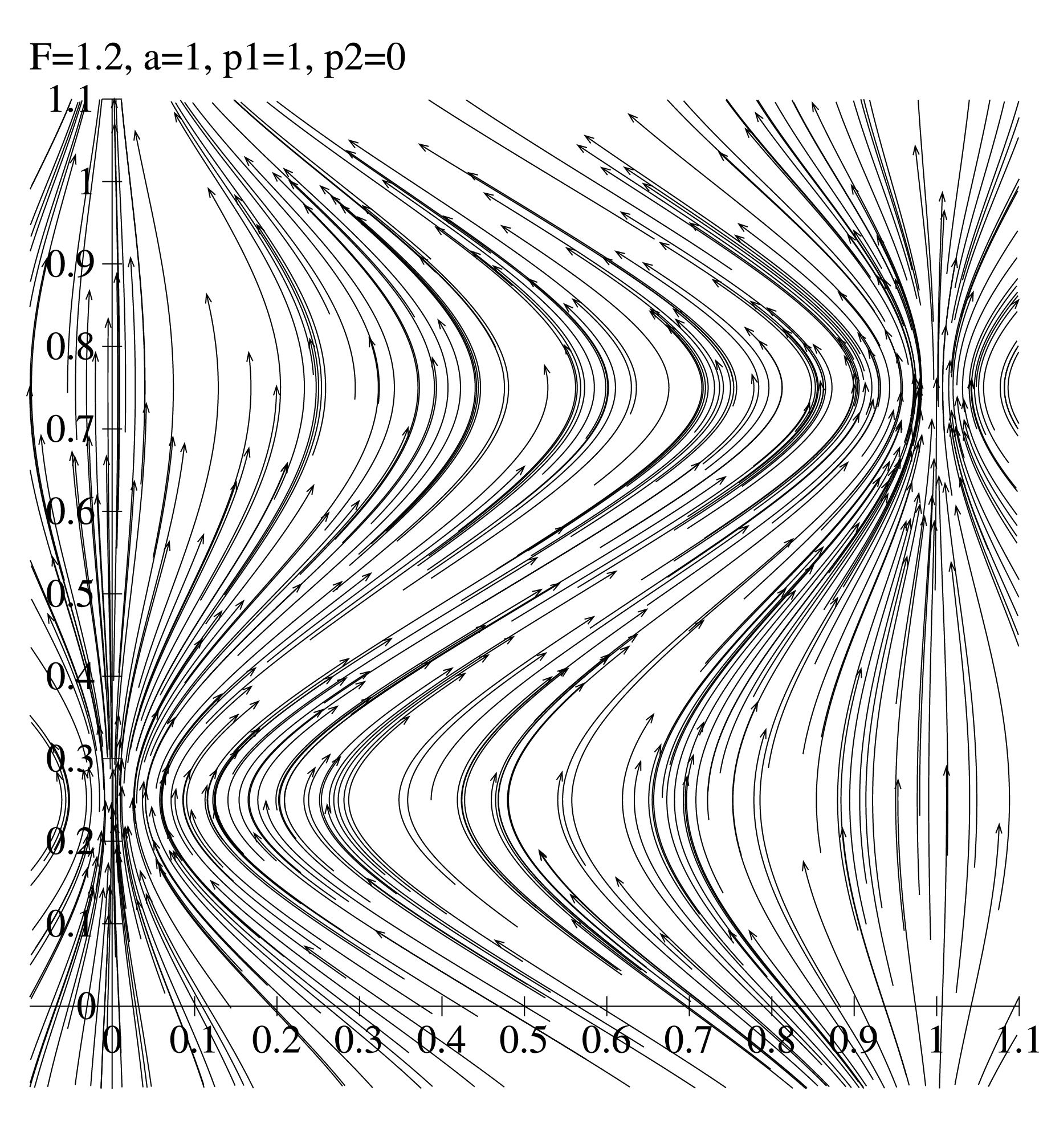}}
\caption{Uncoupled particles below depinning ($F=0.85$, left), at depinning ($F=1$, middle)
and above depinning ($F=1.2$, right); $a=1$, $p_{1}=1$,
$p_{2}=0$. We always plot $x$ to the right and $y$ to the
top. Separatrices for the different attractive regions below threshold
are drawn in green.}
\label{f:1}
\end{figure*}
\begin{figure*}
\centerline{\clipfig{0.7}{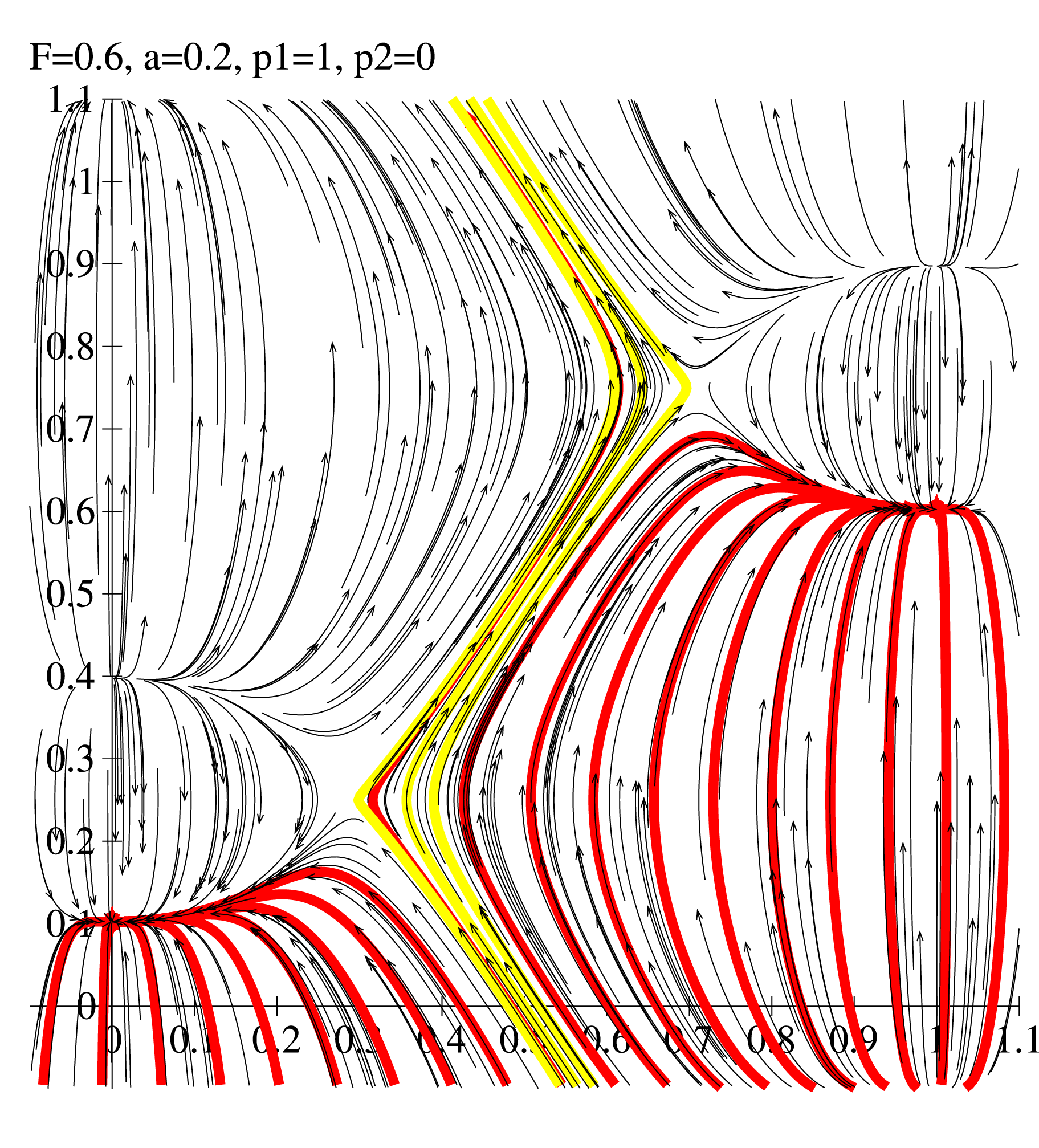}\clipfig{0.7}{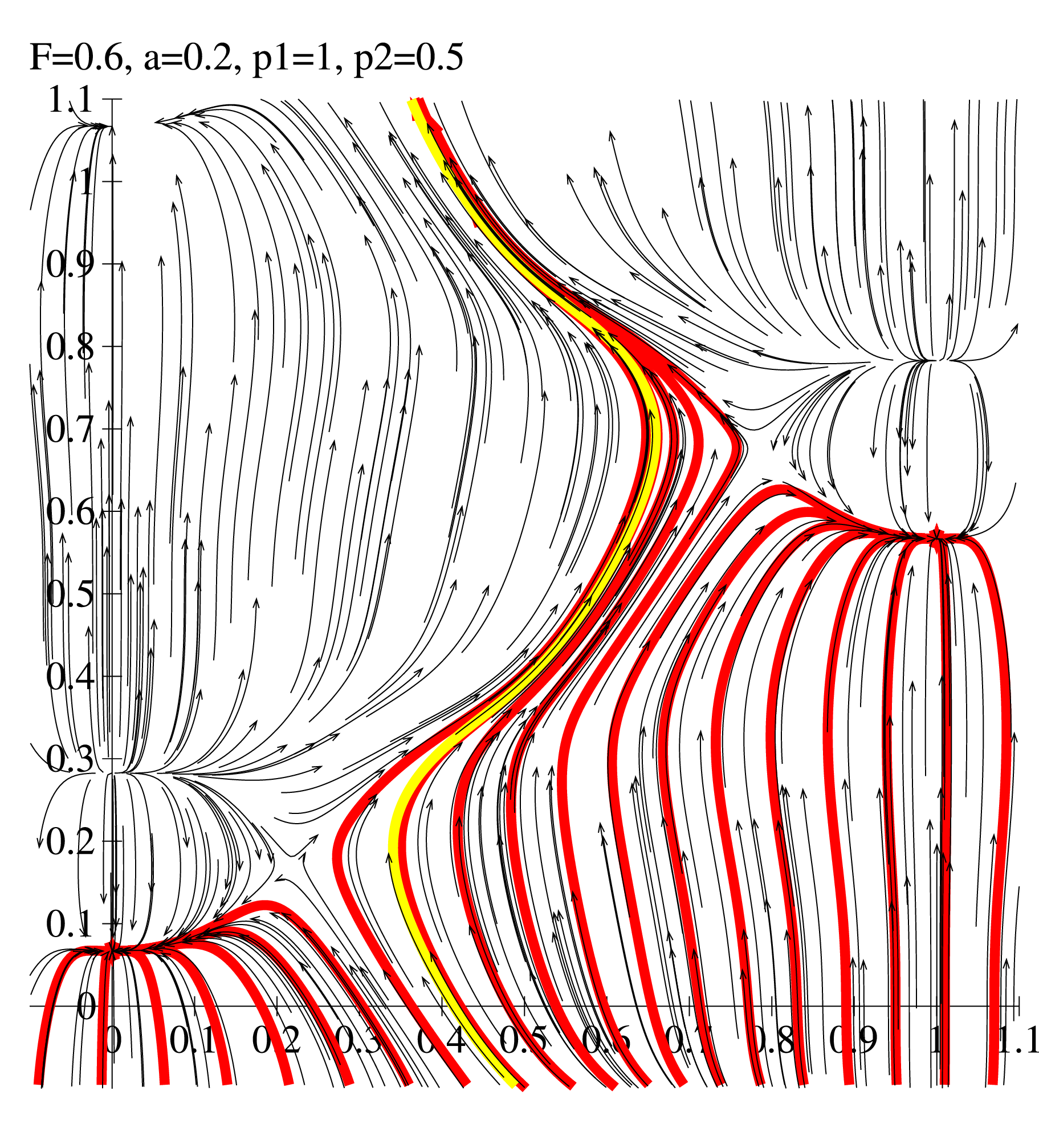}\clipfig{0.7}{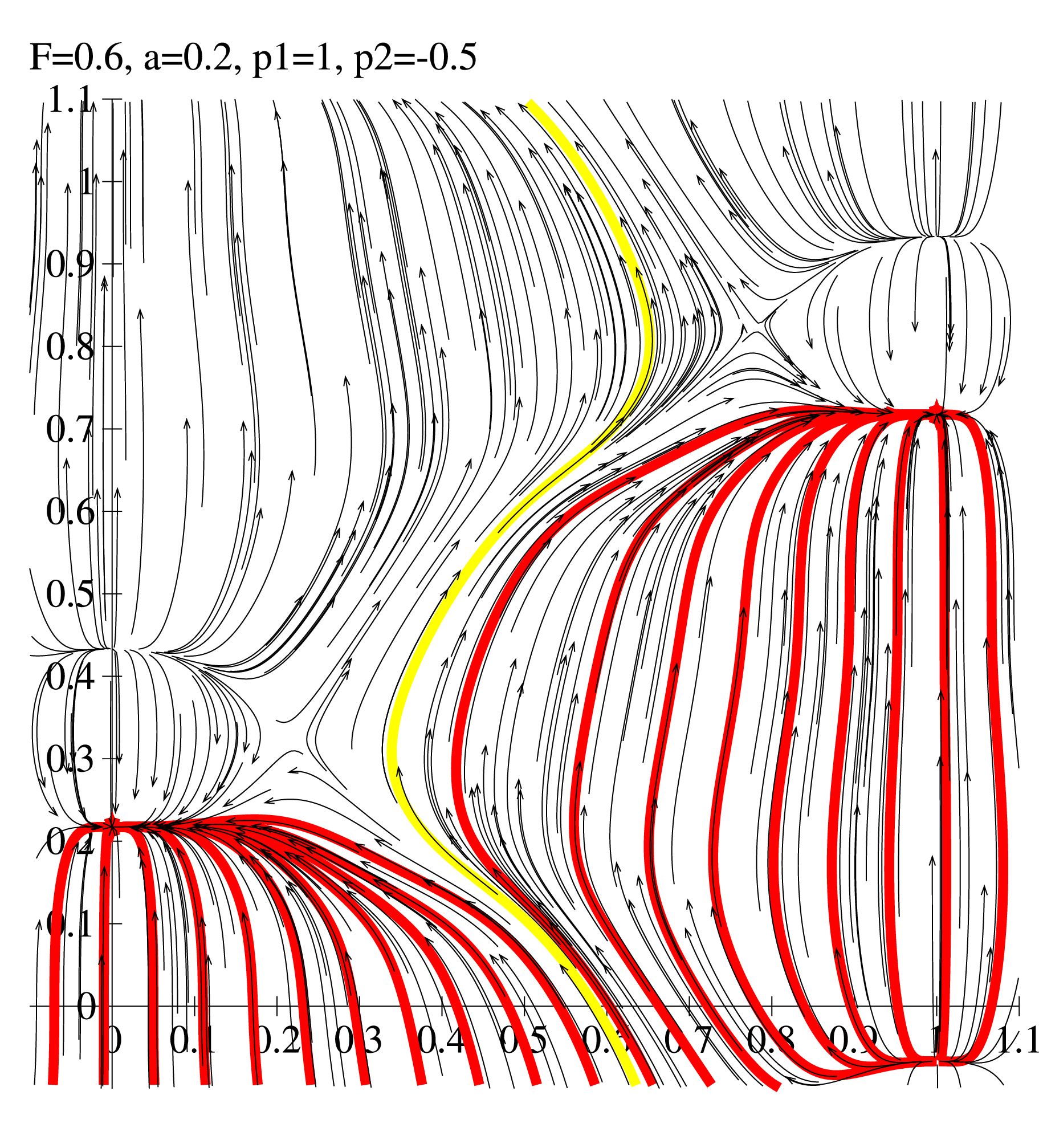}}
\caption{Viscously coupled particles ($a=0.2$) below depinning of the uncoupled particles ($F=0.6$). The
anharmonic coefficient $p_{2}$ differs from left to right: $p_{2}=0$ (left); $p_{2}=0.5$ (middle) and
$p_{2}=-0.5$ (right).  $p_{1}=1$. We plot 20 sample trajectories starting from $y=0$, and equally spaced in $x$.
One sees that for $p_{2}>0$ more trajectories converge towards the unique stable solution (yellow). In the case
of $p_{2}=0$, there is a family of periodic solutions, of which we have plotted three.} \label{f.2}
\end{figure*}
\begin{figure*}
\centerline{\clipfig{0.7}{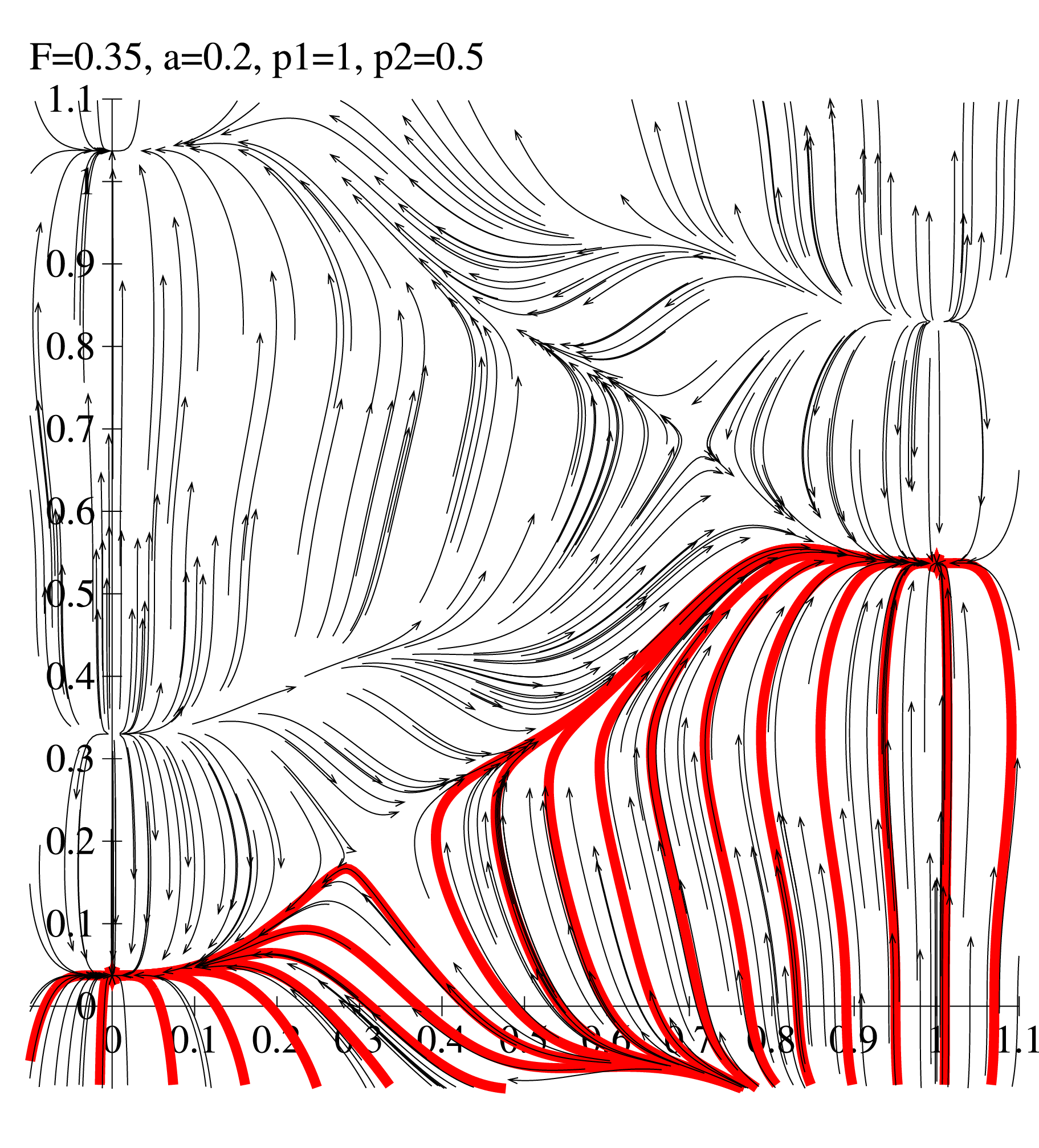}\clipfig{0.7}{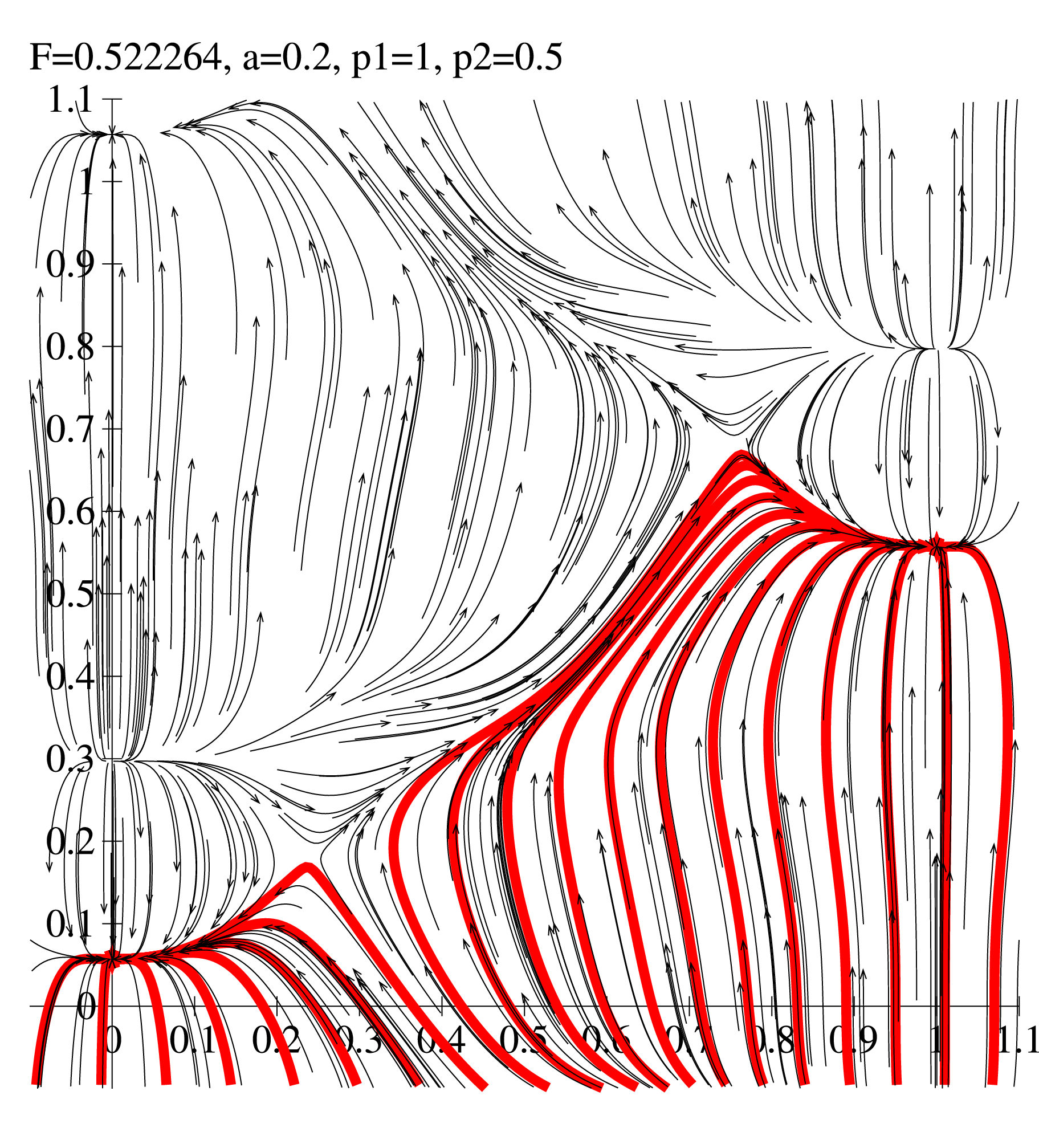}\clipfig{0.7}{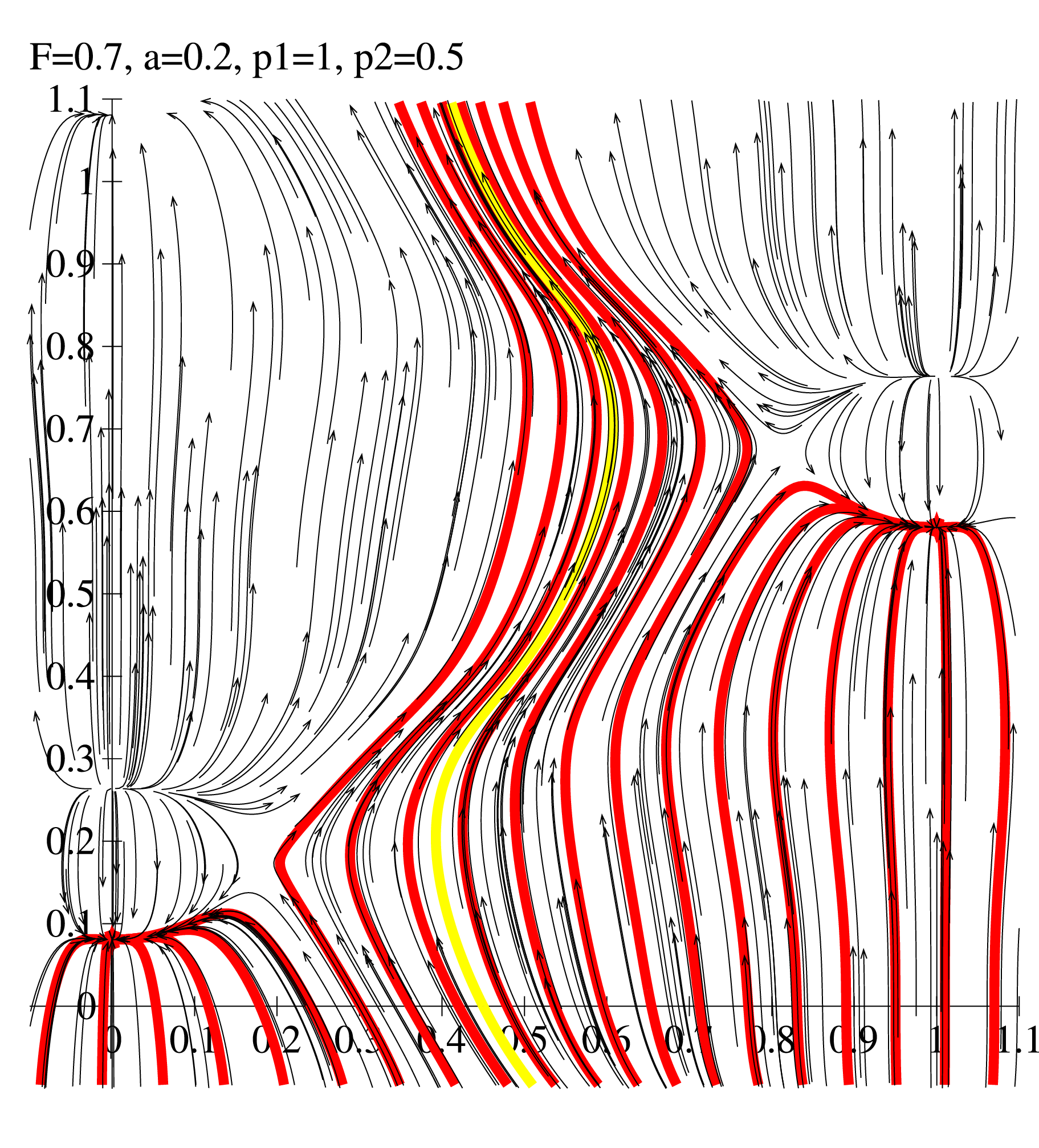}}
\caption{Viscously coupled particles ($a=0.2$) above ($F=0.7$, right), at
($F=F_{c}=0.522264 $, middle) and below  depinning ($F=0.35$, left). The anharmonic
coefficient $p_{2}=0.5$. We plot 20 sample trajectories starting from
$y=0$, and equally spaced in $x$. For $F=0.7$ (right), we plot the unique
stable solution (yellow). Even for this large $F$, one sees the
convergence to this stable solution.}
\label{f:3}
\end{figure*}

\subsection{Smooth potentials}
\label{subsec:smooth}
We now study the following model:
\begin{eqnarray}\label{k14}
\gamma \dot{u}_{1} &=& \eta (\dot{u}_{2}-\dot{u}_{1} ) +f +\hat \phi (u_{1})\\
\gamma \dot{u}_{2} &=& \eta (\dot{u}_{1}-\dot{u}_2 ) +f +\hat \phi (u_{2})
\end{eqnarray}
In this Section we adopt slightly different notations for center-of-mass and difference coordinates:
\begin{eqnarray}\label{k15}
y&=&\frac{u_{1}+u_{2}}{2}\\
x&=&u_{1}-u_{2}\ .
\end{eqnarray}
In these coordinates, the equation of motion becomes:
\begin{eqnarray}\label{k19}
 \dot y &=& F - \frac{1}{2}\left[ \phi
 (y+x/2) +\phi (y-x/2) \right] \\
 \dot x &=& a \left[ \phi (y-x/2)-  \phi (y+x/2) \right]\ ,
\end{eqnarray}
where we have defined:
\begin{eqnarray}\label{k18}
a &=& \frac{\gamma}{\gamma +2\eta}\\
\phi &=& - \frac{\hat \phi } {\gamma}\\
F &=& \frac{f}{\gamma}
\end{eqnarray}
For definiteness we consider the family of periodic-force landscapes:
\begin{equation}\label{k26}
\phi (u)= \frac{p_{1}\sin(2 \pi u)+p_{2}\sin( 4\pi
u)}{\sqrt{\frac{5}{8}-\frac 1{128 p_{2}^2 }
+p_{2}^{2}+\left(\frac{1}{4}+\frac{1}{128 p_2^{2}}\right) \sqrt{(1+32 p_2^2)}}}\ .
\end{equation}
They are
normalized such that if one takes $|p_{1}|=1$ (the standard choice made in the following) the single-particle
critical depinning force is $F_{c}^{\mathrm{sp}}=\pm 1$ (i.e.\ $\max (\phi (u))=\max (-\phi (u))= 1$) for any $p_{2}$. It
turns out that the single-harmonic case is non-generic and one needs to include at least one other harmonics, i.e.\ $p_2 \neq 0$.

We have integrated these equations numerically and plotted \footnote{We are grateful to Alan Middleton for
clarifying remarks during the analysis of these flows.} the resulting flow in Figs.\ \ref{f:1}, \ref{f.2}, and
\ref{f:3} for various values of $f$ and $p_2$. The center-of-mass coordinate $y$ is plotted along the vertical
axis, the relative displacement $x$ along the horizontal axis.

It is instructive to start with the case $\eta=0$ ($a=1$) of two uncoupled particles, given in Fig.\ \ref{f:1}. The
vertical trajectories along the $y$ axis at $x=0$ or $x=1$ correspond to the two particles either in the same
position or shifted by one period. As the force is decreased below threshold (right to left) a pair of fully-attractive and fully-repulsive fixed points appears on these axis. The total phase space for $F< F_c^{\mathrm{sp}}=1$ is
fragmented in pinned regions which flow to one of these ``pinned-phase'' fixed points, corresponding either to
the two particles pinned in the same well or pinned in two wells shifted by one period, depending on the initial
condition. Note also the other zero-force fixed point which has one attractive and one repulsive direction and
corresponds to one particle in a stable equilibrium position at the bottom of one well and the other in an
unstable equilibrium position at a hill top. This fixed point controls the separatrices of the flow. This
structure, obvious in the absence of a coupling, will persist, with some modification, for non-zero $\eta$.

Interesting physics happens when the viscous coupling $\eta$ is increased. The case $a=0.2$ is shown in Fig.\ \ref{f.2}. Exactly along the axis $x=0$ and $x=1$ the equation of motion has not changed and the same attractive
pinning fixed points are present for $F<F_c^{\mathrm{sp}}=1$. However, unbounded motion is now possible for smaller
forces $F_c < F < F_c^{\mathrm{sp}}=1$, and takes place away from the axis. The force chosen in Fig.\ \ref{f.2} is $F=0.6$.
On the left figure the case $p_2=0$ is represented. One can easily see that it is fully integrable and that each
trajectory in the central region is exactly periodic and crosses the $y=0,1$ axis at the same $x$. The region
where this flow occurs is delimited by the separatrices which meet at the  above-mentioned zero-force saddle points.
Hence one sees clearly that the phase space splits into a pinned region and a flowing periodic region. In the
case of a pure sine ($p_2=0$), this region is made of an infinity of neutral periodic trajectories (with zero
Lyapunov exponent). In the more generic case $p_2 \neq 0$, the flowing region contains a single periodic
trajectory. This trajectory is either attractive (case $p_2=0.5$, figure \ref{f.2} in the middle) or repulsive (right part of
figure \ref{f.2}, with $p_2=-0.5$). It is easy to prove from the symmetry properties of the flow  that
the Lyapunov exponent is reversed when the sign of the force landscape is reversed $\phi(u) \to - \phi(u)$ \footnote{Denoting
$u^i(t,f,p_1,p_2)$ the solution of the equation of motion - for some given but unspecified initial condition,
one sees that $u^i(t,-f,p_1,p_2)=-u^i(t,-f,p_1,p_2)$, $u^i(t,f,p_1,p_2)=1/2 + u^i(t,f,- p_1,p_2)$, and
$u^i(-t,f,-p_1,-p_2)=-u^i(-t,f,p_1,p_2)= 1/2 -u^i(-t,f,- p_1,p_2)$. This last property implies that the two
leftmost figures in Fig.\ \ref{f:2} can be deduced by symmetry and that the Lyapunov exponent on the periodic
trajectories (which are globally preserved by the symmetry) are reversed in sign. The pinned fixed points
however remain attractive and are simply exchanged by this symmetry (they are not individually preserved).}. In the repulsive
case, any particle in the region {\em apparently flowing} on the figure eventually gets pinned at some larger $y$,
after visiting a few cells; the basin of attraction of the flowing phase has measure zero. This is an example
where a non-trivial periodic stationary state exists, but is dynamically unstable. On the contrary, in the case
$p_2=0.5$ (middle of figure \ref{f:2}) a distinct flowing phase exists, and its properties are dominated by a unique
attractive periodic trajectory, and   e.g.\ the average velocity is given by the inverse period of this trajectory.

Finally Fig.\ \ref{f:3} illustrates how the periodic orbit in the middle, hence the moving phase, disappears when
the force is reduced below $F_c =0.522265$, leaving only a pinned phase for $F<F_c$.

\begin{figure*}
\centerline{\fig{0.6}{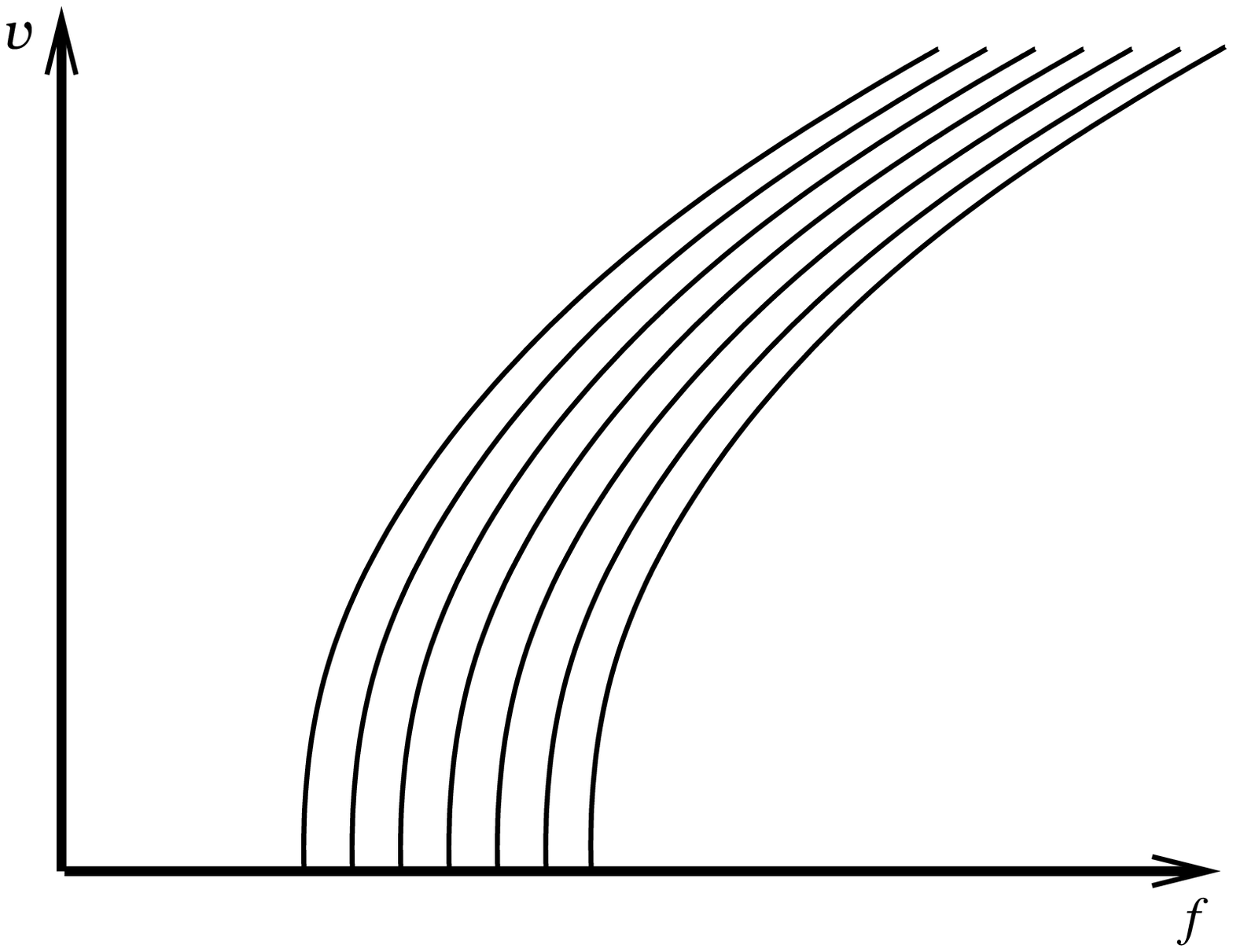}\qquad\quad\fig{0.6}{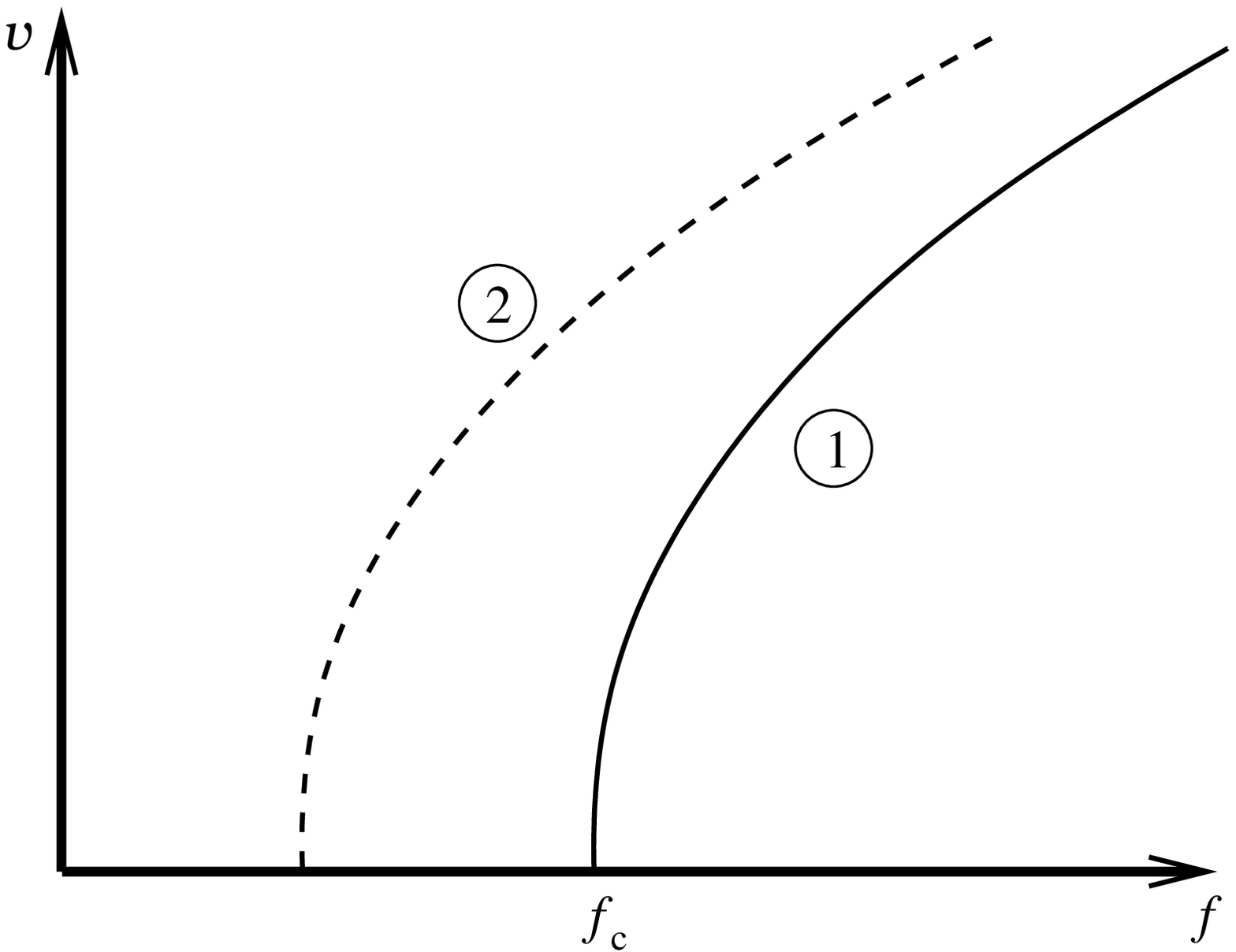}\qquad\quad\fig{0.6}{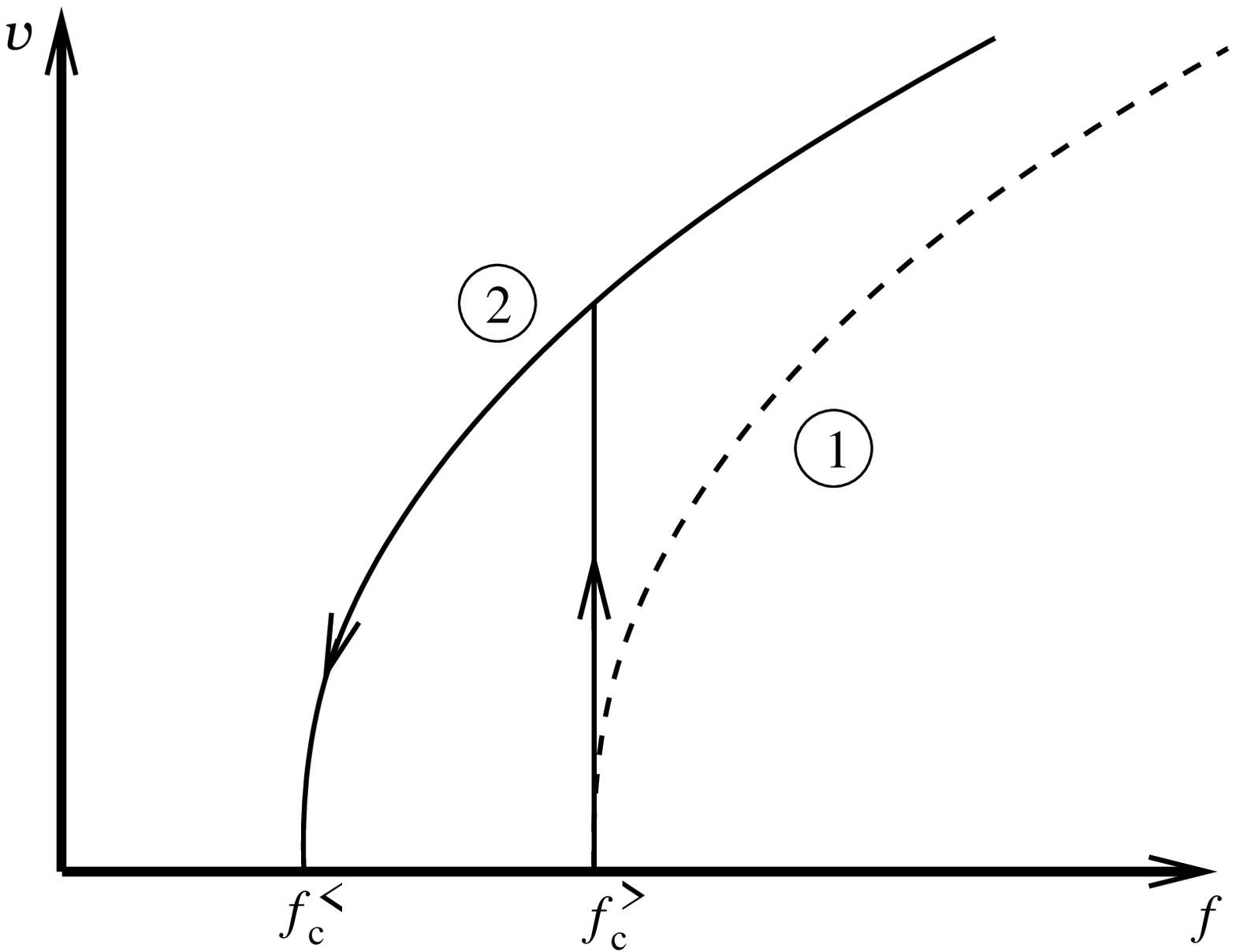}} \caption{Schematic $v(f)$ curves corresponding
to the three cases discussed in the text: (i) left: single sine force landscape (ii) middle: the non-trivial
periodic orbit $2$ is repulsive and the $v(f)$ curve is the same as for a single particle $1$ (iii): left: the
periodic orbit $2$ is attractive and the $v(f)$ curve exhibits hysteresis as discussed in the text} \label{k30}
\end{figure*}

We can now analyze the resulting $v(f)$ curve. The $v(f)$ curve for the pure-sine model is indicated
schematically on the left of Fig.\ \ref{k30} and is non-generic, as discussed above. In the case $p_2 \neq
0$ there are two branches corresponding to the two steady states, one (labelled $1$) corresponding to the
trajectory along the $x=0,1$ axis, i.e.\ the single particle $v(f)$ curve, and the second (labelled $2$)
corresponds to the periodic orbit near the middle of the figures \label{f:2}, which generally has a higher
$v(f)$ curve. If the second is repulsive ($p_2=-0.5$), then the trajectory along the $x=0,1$ axis is attractive:
the global $v(f)$ curve then coincides with the single-particle one and there is no hysteresis (middle of
Fig.\ \ref{k30}). If the second is attractive ($p_2=0.5$), the Lyapunov exponent of the periodic trajectories are
inverted \footnote{Note that while the line $x=0,1$ are always attractive in the vicinity of the pinned fixed
points for $F< 1$, it becomes -- in that case with $p_2=0.5$ -- repulsive for $F>1$ when the flow starts along this
line. This is again a consequence of the symmetry properties mentioned above which inverts the Lyapunov along a
periodic trajectory globally preserved by the symmetry.} and the global $v(f)$ curve follows the second branch.
In that case there is a hysteresis as the force is varied adiabatically; this is shown in the right figure. Upon
decreasing the force from a large value the system follows the attractive trajectory in the middle until it
disappears at $F_c$ and the velocity vanishes. But if the force is increased from a value smaller than $F_c$, it
can be seen from the left plot on Fig.\ \ref{f:3} that it first converges to a pinned fixed point along the
axis $x=0,1$. Since these fixed points remain attractive up to $F=F_c^{\mathrm{sp}}=1$, the velocity remains zero until that force and
then jumps to the stable moving state.

\begin{figure}
\Fig{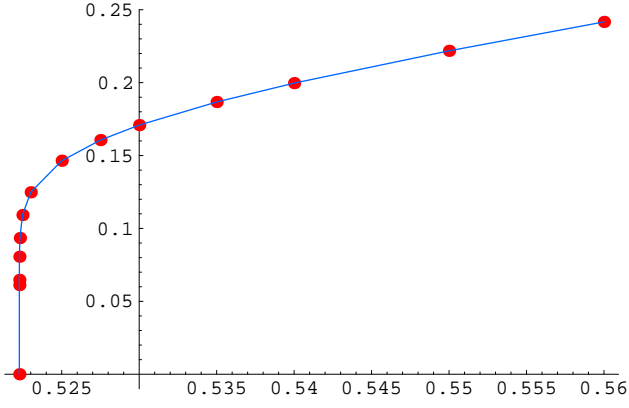}
\caption{Velocity $v$ as a function of $F$. The parameters used are $p_{1}=1$, $p_{2}=0.5$, $a=0.2$.}
\label{k30a}
\end{figure}

\begin{figure}
\Fig{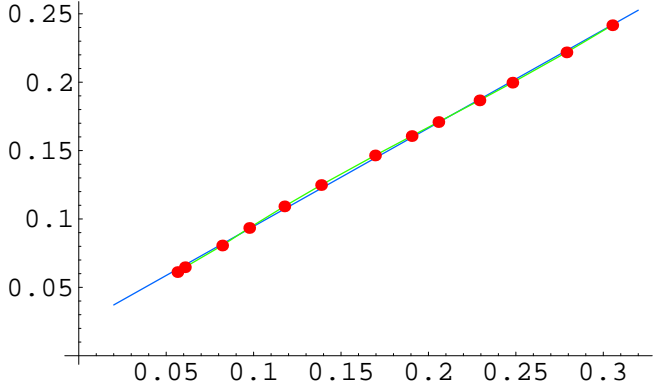}
\caption{Velocity $v$ as a function of $-1/\ln (F-F_{c})$, with
$F_{c}=0.522265$. The fit-function is $v=0.0227714 - 0.718327 /\ln
(F-F_{c})$. The linear fit is excellent. A (much worse) fit to a power-law would give an
exponent of about 0.1. The parameters used are $p_{1}=1$, $p_{2}=0.5$, $a=0.2$.}
\label{k31}
\end{figure}

The question of whether a jump exists in the descending curve can be settled by analyzing how the periodic
trajectory disappears at $F=F_c$. It can be seen from the middle plot on Fig.\ \ref{f:3}, that this occurs
abruptly, but that the period diverges at $F=F_c^+$ as the system spends more and more time near the zero-force saddle points. These hence play an important role in the transition at $F=F_c$. A simple argument indicates
that the time spent near these points increases logarithmically, as is verified by the numerical integration of the flow in Figs.~\ref{k30a}, \ref{k31}. Hence, although this system exhibits hysteresis in the case $p_2=0.5$ it does not exhibit a
velocity jump along the descending branch. Note that the critical behaviour at $F_c$ is different from the
single-particle case $v \sim (F-F_c^{\mathrm{sp}})^{1/2}$, due to the zero-force saddle-point mechanism.

%\begin{figure}
%\ClipFig{F=1a=1p1=1p2=0}
%\caption{Uncoupled particles at depinning: $F=1$, $a=1$, $p_{1}=1$,
%$p_{2}=0$. We always plot $x$ to the right and $y$ to the
%top. Separatrices are drawn in red.}
%\label{k27}
%\end{figure}
%
%\begin{figure}
%\ClipFig{F=0p85a=1p2=0}
%\caption{Uncoupled particles below depinning: $F=0.85$, $a=1$, $p_{1}=1$,
%$p_{2}=0$.}
%\label{k28}
%\end{figure}
%
%\begin{figure}
%\ClipFig{F=1p2a=1p2=0}
%\caption{Uncoupled particles above depinning: $F=1.2$, $a=1$, $p_{1}=1$,
%$p_{2}=0$.}
%\label{k29}
%\end{figure}

\subsection{Scalloped potential}

\label{subsec:scalloped}

Here we consider the two-particle toy model for a piecewise parabolic (scalloped) potential, corresponding to a
piecewise linear pinning force with jump discontinuities at the boundaries of each period. The equations of
motion for the position of the particles are
\begin{eqnarray}\label{sc1}
\gamma \dot{u}_{1} &=& \eta (\dot{u}_{2}-\dot{u}_{1} ) +f +\frac12+n-u_1\\
\gamma \dot{u}_{2} &=& \eta (\dot{u}_{1}-\dot{u}_2 ) +f +\frac12+m-u_2
\end{eqnarray}
for $n\leq u_1\leq n+1$ and  $m\leq u_2\leq m+1$, with $n$ and $m$ integers.

When $\eta=0$, the particles are decoupled, and the dynamics can be determined exactly. Each particle is pinned for $f<1/2$. For $f>1/2$ there is a unique periodic orbit
of period
\begin{equation}\label{scSP}
\frac{1}{v}=  \gamma\ln\Big(\frac{f+1/2}{f-1/2}\Big)
\end{equation}
 that diverges linearly as $f\rightarrow (1/2)^+$. No  periodic orbits exist for $f<1/2$ and the system does not exhibit hysteresis.

To consider the case $\eta\not=0$, we introduce center-of-mass and difference coordinates as in Eq.~(\ref{k15}).
%
%\begin{eqnarray}\label{sc2}
%y&=&\frac{u_{1}+u_{2}}{2}\\
%x&=&u_{1}-u_{2}
%\end{eqnarray}
%
In these new coordinates, the equations of motion are
\begin{eqnarray}\label{sc3}
\gamma  \dot{y} &=& -y+ f + \frac12 +\frac{n+m}{2}\\
\gamma \dot{x} &=& -a x + a(n-m)
\end{eqnarray}
for $\frac{n+m}{2}\leq y\leq\frac{n+m}{2}+1$ and $n-(m+1)\leq x\leq 1+n-m$, with $a$ given in Eq.~(\ref{k18}).
\begin{figure}
\fig{1}{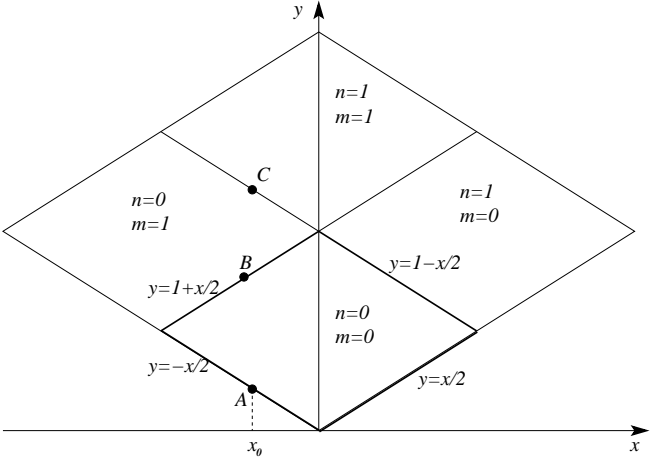}
\caption{The figure shows the boundaries of the regions where the
relative and center-of-mass velocities of two particles in a periodic scalloped potential have jumps. The horizontal and vertical coordinates are the relative and center-of-mass position of the two particles, respectively, as defined in Eqs.~(\ref{k15}). The equations for the straight lines bounding the region corresponding to $n=0$, $m=0$ are indicated in bold.}
\label{f:scallops}       % Give a unique label
\end{figure}
Our goal is to identify the stable periodic orbits for this model and calculate the corresponding period
or its inverse, the mean velocity.   The regions in coordinate space corresponding to the various periods of the pinning potential are shown in Fig.~\ref{f:scallops}. Consider a particle  that starts at point $A$  with
$[x(0),y(0)]=(x_0,-x_0/2)$ in the region of the pinning potential corresponding to $(n,m)=(0,0)$. The boundaries of this region  are defined by
$-x/2\leq y\leq 1+x/2$, for $-1\leq x\leq 0$, and $x/2\leq y\leq 1-x/2$, for $0\leq x\leq 1$.
This particle will   travel across the $(0,0)$ region to point $B$ in a time $t_1(x_0)$ and then across
the $(0,1)$ region to a point $C$ in a time $t_2(x_0)$, according to
\begin{eqnarray}\label{sc4}
A&=\displaystyle\Big(x_0,-\frac{x_0}{2}\Big)\stackrel{t_1(x_0)}{-\!\!\!-\!\!\!-\!\!\!\longrightarrow} B=\Big(x(t_1),1-\frac{|x(t_1)|}{2}\Big)\nonumber \\
&\stackrel{t_2(x_0)}{-\!\!\!-\!\!\!-\!\!\!\longrightarrow}
C=\displaystyle\Big(x(t_1+t_2),1+\frac{|x(t_1+t_2)|}{2}\Big)\qquad
\end{eqnarray}
The case of a scalloped pinning potential can be studied analytically since the equations of motion are linear within each pinning period, with jump discontinuities in the velocity at boundaries of the pinning regions shown in Fig.~\ref{f:scallops}.

\begin{figure}
\fig{1}{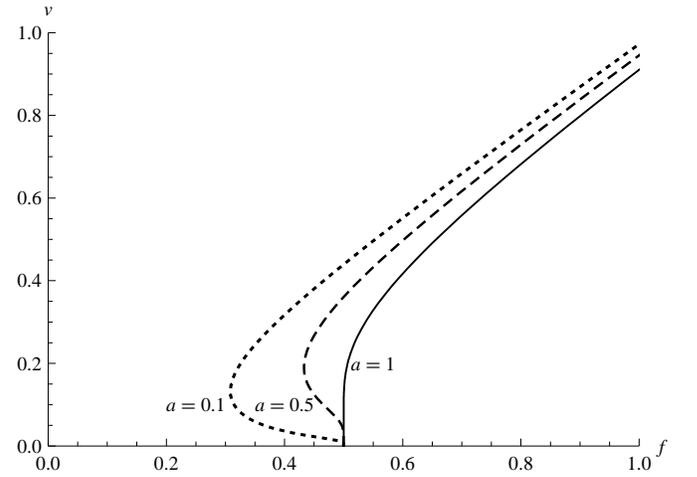}
\caption{The velocity-force characteristic obtained by inverting Eq.~(\ref{fv}) for $a=1$ (solid line), corresponding to decoupled layers,  $a=0.5$ (dashed line) and $a=0.1$ (dotted line).}
\label{f:toyscallop-1}       % Give a unique label
\end{figure}

\paragraph{Periodic orbits}
We wish to determine the values of $x_0$ that correspond to periodic orbits as defined by the fixed point
\begin{equation}\label{sc5}
x(t_1+t_2)\equiv x'(x_0)=x_0\ .
\end{equation}
The period of such orbits is $t_1+t_2$ and $v=1/(t_1+t_2)$.
It is convenient to introduce a new notation:
\begin{eqnarray}\label{sc6}
z_1(x_0)&=&e^{-t_1/\gamma}\\
z_2(x_0)&=&e^{-t_2/\gamma}
\end{eqnarray}
with
\begin{equation}\label{sc7}
v=\Big[\gamma\ln(1/z_1z_2)\Big]^{-1}
\end{equation}
The dynamics from
$(x_0,|x_0|/2)$ to $(x',1+|x'|/2)$ can be examined analytically since the equations of motion are
piecewise linear. It is determined by
\begin{eqnarray}
\label{z1eqn}z_1\Big(f+\frac{1}{2}-\frac{|x_0|}{2}\Big)-\frac{|x_0|}{2}z_1^a&=&f-\frac{1}{2}\;,\\
\label{z2eqn}fz_2-\frac{1}{2}z_2^a+\frac{|x_0|}{2}z_1^a(z_2+z_2^a)&=&f-\frac{1}{2}\;,\quad
\end{eqnarray}
with
\begin{equation}
\label{xp}x'=x_0z_1^az_2^a+\frac{x_0}{|x_0|}(1-z_2^a)\;.
\end{equation}
We now look for a periodic solution or fixed point as defined by Eq.~({\sc5}). Then Eq.~(\ref{xp}) gives
(provided $u_0\not=0$)
\begin{equation}
\label{xpfixed} |x_0|=\frac{1-z_2^a}{1-z_1^az_2^a}\;.
\end{equation}
Substituting this in Eqs.~(\ref{z1eqn}) and (\ref{z2eqn}) we
obtain
\begin{eqnarray}
&&z_1(f+1/2)-\frac{(1-z_2^a)(z_1+z_1^a)}{2(1-z_1^az_2^a)}=f-1/2\;,\\
&&z_2(f+1/2)-\frac{(1-z_1^a)(z_2+z_2^a)}{2(1-z_1^az_2^a)}=f-1/2\;.\qquad
\end{eqnarray}
These two equations are symmetric in $z_1$ and $z_2$,
indicating that the solution must satisfy $z_1=z_2=z$. There
is a fixed point $x^*$ of $x_0$ where the system undergoes a
periodic orbit of period $1/v=-2 \ln(z)$, with
\begin{eqnarray}
&&x^*=\frac{1}{1+z^a}\;,\\
\label{z1z2}&&z(f+1/2)-\frac{z+z^a}{2(1+z^a)}=f-1/2\;.
\end{eqnarray}
For any value of $a$ we  obtain $f(v)$ from
Eq.~(\ref{z1z2}), with the result
\begin{equation}
\label{fv} f=\frac{1}{2(1-z)}\Big[1+z-\frac{z+z^a}{1+z^a}\Big]\;,
\end{equation}
where
\begin{equation}
z=e^{-1/(2v)}.
\end{equation}
The  $v(f)$ curves obtained by inverting Eq.~(\ref{fv}) are
shown in Fig.~\ref{f:toyscallop-1} for a few values of $a$. These curves resemble those
obtained  in mean-field
theory and suggest the possibility of a velocity jump. However, before making any conclusions we must study the stability of these periodic orbits.
An analytic solution of Eq.~(\ref{fv}) can be
obtained for $a=1/2$.
%In this case Eq.~(\ref{z1z2}) has two roots
%for $f>f_{min}=\sqrt(3)/4$, given by
%
%\begin{eqnarray}
%z_\pm(f)=\frac{1\pm 4\sqrt{(f^2-f_{min}^2)}}{2(2f+1)}\;.
%\end{eqnarray}
%
%
For $a=1$, corresponding to $\eta=0$ (decoupled particles), $x_0$ is undetermined. For any $x_0$ one recovers the single-particle result  given in Eq.~(\ref{scSP}). For instance, for $x_0=0$ we obtain
\begin{eqnarray}\label{sc10}
z_1^*&=&\frac{f-1/2}{f+1/2}\\
z_2^*&=&1
\end{eqnarray}
which yields Eq.~(\ref{scSP}).

\begin{figure}
\fig{1}{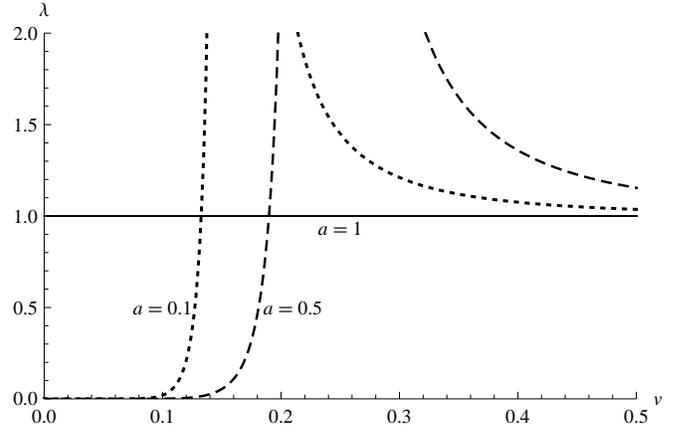}
\caption{The figure shows the Lyapunov exponent $\lambda$ as function of $v$ for
$a=0.1$ (dotted), $a=0.5$ (dashed) and $a=1$ (solid).  For $a=1$, $\lambda=1$.}
\label{lambdafig}       % Give a unique label
\end{figure}

\paragraph{Stability.}
The stability of the periodic solutions found in the preceding paragraph can be examined from the linear
response to a perturbation of the initial condition. Letting $x_0\rightarrow x_0^*+\delta x$, we define a
Lyapunov exponent $\lambda$ \footnote{By contrast to the Lyapunov exponent for a continuous-time flow defined in the
previous section, this is the Lyapunov exponent for the discrete map, hence the transition of stable to unstable occurs as
$|\lambda|$ crosses unity.}
\begin{equation}\label{scLyapunov}
x'(x_0^*+\delta x)\equiv x_0^*+\lambda\delta x
\end{equation}
where $x'(x_0)$ is given by the right-hand side of Eq.~(\ref{xp}).
We find
\begin{equation}
\label{lambdaF}
\lambda=\Big[\frac{z^{1+a}(a+z^a+2f(1+z^a))}{2fz(1+z^a)+z^{1+a}-az^a}\Big]^2\;.
\end{equation}
At the fixed point $z$ and $f$ are related by Eq.~(\ref{fv}).
Inserting this into Eq.~(\ref{lambdaF}), we obtain
\begin{equation}
\label{lambda}
\lambda=\Big[\frac{z^{1+a}(1+z^a+a(1-z))}{z(1+z^a)-az^a(1-z)}\Big]^2\;.
\end{equation}
This Lyapunov exponent is plotted in Fig.~\ref{lambdafig} for a few values of $a$. It equals 1 for $a=0$
(corresponding to $\eta\rightarrow\infty$) and for $a=1$ (corresponding to $\eta=0$). For all other values of
$a$ one finds $\lambda<1$ only for very small $z$, i.e.\ small $v$. This region corresponds to the part of the $v(f)$ curve that
has negative slope near the depinning threshold $f=1/2$. The conclusion of this analysis is that this part is
stable, while the portion with positive slope is unstable. This result is somewhat surprising in view of the
results obtained in mean-field theory. However, as was explained in the previous section, reversing the sign of
the force landscape would exchange the attractive and repulsive trajectories and result in the $v(f)$ curve more
similar to the one shown in the mean-field section. The special nature of the two-particle scalloped-force
landscape may be related to the absence of zero-force saddle points which played an important role in the case
of the smooth potential. Note that we have not looked for more complicated periodic solutions, which are
difficult to rule out.

\subsection{Discussion of toy models}

\label{sec:discussion}

We conclude from the previous two sections that a large variety of behaviors can already occur with two
viscously coupled degrees of freedom in a random-force landscape. Understanding their systematics, for instance
how one evolves from the smooth potential to the scalloped one as more harmonics are included, remains to
be done. In each case one must identify the periodic trajectories and the attractor, whose structure may become
more complex if the landscape contains more harmonics and more zero-force points. It is clear  that an even more careful systematic study is necessary  when increasing the number of degrees of freedom within
these coupled layer models. It is not clear at this stage whether chaotic attractors exist, or whether
 even multiple stable periodic attractors do coexist.

\section{Discussion}

We have studied in this paper a model of two (single-component $N=1$) elastic layers driven over a random substrate and only coupled by a viscous coupling $\gamma_{12}$, going beyond mean-field theory. We have extended the functional RG approach which allows to describe the elastic depinning in each layer in the absence of an elastic coupling to the case of a non-zero $\gamma_{12}$. We have found that the FRG fixed point which describes elastic depinning is unstable to an arbitrarily weak viscous coupling beyond a plastic scale $L_{\mathrm{pl}}$ which diverges with a universal exponent as $\gamma_{12} \to 0$. To describe the plastic physics beyond that scale we have studied the FRG to one loop in the moving state at non-zero velocity. We found that the high-velocity branch of the $v(f)$ versus $f$ curve terminates at a point $v=v_c$ where the slope is infinite. This point corresponds to a force smaller than the elastic depinning threshold, hence there is a range of values of $f$ where a pinned state coexists with a moving state. This dynamical hysteresis is very similar to the one found in mean-field theory. One could then conclude that the 1-loop FRG result nicely confirms the main features of the mean-field theory and, in addition, allows to establish precise universal results and identify the proper length scales.

This conclusion may, however, be too hurried as, surprisingly, our two-loop calculation shows some
possible problems with this picture. The calculation is based on certain assumptions (discussed in Section \ref{sec:twoloop}, i.e.\ the neglect of violations of Middleton's theorem, the neglect of higher orders in the region of small $\eta_{+}$) and more work is clearly needed to ascertain its validity. However, as a preliminary step it indicates a new universality class in the case of non-periodic objects and in the periodic case, a possible breakdown of the 1-loop picture depending on how irrelevant operators are taken into account. This possible alternative picture is the  absence of  a dynamical hysteresis and a nearly vertical $v(f)$ curve near the elastic threshold.

In an attempt to understand which effects could be missed by the mean field and 1-loop approaches, we have solved simple toy models in $d=0$. We found indeed that dynamical hysteresis may or may not be present, depending on the realization of disorder. Although in all cases considered one finds a periodic trajectory with non-zero velocity which survives below the elastic (uncoupled) threshold, this trajectory may be attractive or repulsive depending on the disorder realization. It remains to be studied in detail how these properties carry to a larger number of degrees of freedom. In any case it cannot be assumed that a single attractive periodic attractor exists and a detailed study of such attractors as the number of particles increases must be done with care before any conclusion can be drawn.

The particle models also show the importance of the zero-force points in phase space. These are couples of configurations in the two layers $(u^1_x,u^2_x)$ where all forces vanish. They are defined in the statics, hence are independent of the viscous coupling. However it is important to know their dynamical stability in presence of a viscous coupling. It is proved in Appendix \ref{app:theorem} that metastable states, i.e.\ stable states where the energy has a local minimum, remain stable, i.e.\ dynamically attractive, at non-zero viscous coupling. This means that each pinned state, i.e.\ blocking configuration for each uncoupled layer, keeps a non-zero basin of attraction when the viscous coupling is increased from zero. In the absence of a viscous coupling, when upon increasing $f$ such a pinned state becomes unstable, the next configuration is determined by the no-passing rule and Middleton's theorem, as the minimum over $u$ of all metastable configurations in the direction of the force. In the presence of a viscous coupling however, there is no guarantee that the system will not flow from there to a periodic orbit, resulting in a jump in the $v(f)$ curve. Thus even if the metastable zero-force couples remain attractive, they may not be easily {\it dynamically accessible}, i.e.\ their basin of attraction may shrink and be nearly invisible in a procedure such as force ramping. These effects, as well as the competition between the zero-force fixed points and the periodic orbits, clearly remain to be studied systematically.

It is thus a remaining challenge to understand how the FRG can describe the structure of such a complicated phase space with periodic orbits coexisting with pinned fixed points. It is probable that the FRG calculation presented here retains only averaged effects and does not adress these issues with sufficient accuracy. One possible geometry to study this in a controlled manner starting with a particle and then extending to manifolds is to use the drive by an harmonic well. Since the coupling between the layers exists only when the system is moving, one needs to go beyond the present calculation and study for instance how the avalanches in the two layers are correlated.

In a broader context, one needs to justify why effort should be devoted to clarify the behaviour of this simple
two-layer model with viscous coupling only. This model indeed neglects the competition between plastic
inter-layer and elastic intra-layer couplings, and one should question its range of validity. Since any small
realistic interaction between two identical elastic layers, such as a crystal or a CDW in the absence of
in-layer topological defects (which we have not included here) generates some small inter-layer commensurate
coupling, one should ask whether our model is stable to that. This can be analyzed by adding a force $- g_1 \sin
\left(2 \pi p (u^1_{xt}-u^2_{xt})\right)$, with $p=1$, to the equation of motion (\ref{y2}) of layer 1, and its
opposite for layer 2. Such a coupling generates an elastic coupling at small scale between the layers, and if it
is relevant in the RG sense, at large scales as well. In the latter situation the elastic coupling sould
dominate the viscous one and one expects that the system is described by elastic depinning. It was found in
mean-field models that such a coupling is always relevant in the moving phase
\cite{SaundersSchwarzMarchettiMiddleton2004}. If such a result were general, the model studied here would be
somewhat artificial, or describe only a limited range of length scales.

It is easy to compute, to 1-loop accuracy and for any $p$, the linear eigenvalue of an infinitesimal
perturbation $g_1>0$ at the unperturbed quasi-static depinning fixed point at $v=0^+$ studied here. A first (and
naive) calculation in the spirit of the statics yields:
\begin{equation}
\partial_l\, \tilde g_1 = \left[2 - 4 \pi^2 p^2 \tilde \Delta(0) \right] \tilde g_1 \ ,
\end{equation}
where $\tilde g_1$ is the dimensionless coupling. This is essentially the result obtained for the problem of a
single layer in presence of both disorder and a commensurate potential \cite{EmigNattermann1997} up to a factor
of two which accounts for the fact that the disorder exists in both layers, compared to
\cite{EmigNattermann1997}. Inserting the value of $\tilde \Delta(0)=\epsilon/36$ at the 1-loop depinning fixed
point implies that this coupling is always relevant near $d=4$ which would seem to confirm the mean-field
conclusion. It also yields a critical dimension near $d=2$, i.e.\ $4 - d_c \approx - 18/\pi^2$ below which the
coupling should become irrelevant. This conclusion is correct for the statics, but incorrect
near the depinning threshold.

At depinning at least two new effects should be taken into account. First, one sees that the coupling
$g_1$ generates in perturbation theory a correction to the critical force, which amounts to adding {\it the same}
force $- g_2 \cos \left(2 \pi p (u^1_{xt}-u^2_{xt})\right)$, with $p=1$, to the equation of motion (\ref{y2}) of
each layer. These two terms feed into each others and the correct linearized RG equation takes the form, to one loop:
\begin{equation}
 \left(\begin{array}{c}
\partial_l \tilde g_1 \\
\partial_l \tilde g_2
\end{array} \right) =
\left(\begin{array}{cc}
2 - 4 \pi^2 p^2 \tilde \Delta(0)  &  - 2 \pi p \tilde \Delta'(0^+)  \\
2 \pi p \tilde \Delta'(0^+)   & 2 - 4 \pi^2 p^2 \tilde \Delta(0)
\end{array} \right)
\left(\begin{array}{c}
\tilde g_1 \\
\tilde g_2
\end{array} \right)  \label{eigenmat}
\end{equation}
Second, and most importantly, from the two-loop solution of the standard depinning fixed point
\cite{LeDoussalWieseChauve2002}, we know that $\tilde \Delta(0)$ does not flow to a fixed point, it always
increases as $\tilde \Delta(0) \sim e^{\epsilon l} \sim L^\epsilon$ . Physically, a {\it static random force} is
generated by the quenched disorder in the limit $v=0^+$. This is due to terms in the two-loop
beta-function  which account for the irreversibility of depinning,
and is at variance with the statics \cite{hhh3}.
At
depinning however, a small coupling between layers is always irrelevant for $d<4$. This can be seen from
(\ref{eigenmat}), since $\tilde \Delta(0)$ grows while $\tilde \Delta'(0^+)$ converges to a $O(\epsilon)$ fixed
point. It justifies a posteriori the model studied in this paper. Of course at larger bare couplings it is
likely that a coupled phase will arise and it would be interesting to study that transition\cite{hhh4}.

Let us finish by recalling that one issue in the theory of plastic flow is whether one can use $P(v)$, the distribution of time-averaged individual particle velocities, as a meaningful order parameter in the thermodynamics limit. One could then distinguish two classes of plastic flow (i) flows with non-trivial $P(v)$ (e.g.\ pinned particles coexisting with flowing rivers) (ii) flows with peaked $P(v)$ (a delta function in the large-size limit) but which cannot be described by a fully elastic theory. The layered model studied here is a tractable example of class (ii) and requires, to exhibit a non-zero depinning threshold, elastic interactions inside the layers. Models for class (i) have been studied, where particles interact only through a hard core interaction  \cite{WatsonFisher1996}. It would be quite interesting to find a tractable model which encompasses 
both classes and their possible transitions.

\section{Acknowledgments}
We would like to thank Jennifer Schwarz for very useful interactions at the beginning of this work. We also acknowledge helpful discussions with Alan Middleton and Denis Bernard.
MCM was supported by the National Science Foundation through grants DMR-0305407 and DMR-0705105 and through a Rotschild-Yvette-Mayent sabbatical fellowship at the Institut Curie in Paris. She thanks the Institut Curie and ESPCI for their hospitality during the completion of some of this work. PLD and KW acknowledge support from ANR program blan05-0099-01.

%\newpage

\appendix

\section{Stability of zero-force fixed points}
\label{app:theorem}

Let us call $u^i_{x}$ a static configuration where the force is zero, i.e.
$F^i(u^i(x),x)=0$, $i=1,2$.
The equation of motion linearized around the FP is:
\begin{eqnarray}
&& \left(\begin{array}{c}
\dot u^1_{x,t} \\
\dot u^2_{x,t}
\end{array} \right) = M \left(\begin{array}{c}
u^1_{x,t} \\
u^2_{x,t}
\end{array} \right) \\
&& M =  A B =  - \frac{1}{\gamma_{11}^2-\gamma_{12}^2}
\left(\begin{array}{cc}
\gamma_{11} H_1 & - \gamma_{12} H_2 \\
- \gamma_{12} H_1 & \gamma_{11} H_2
\end{array} \right) \qquad \\
&& A^{-1} = \left(\begin{array}{cc}
\gamma_{11} &  \gamma_{12}  \\
\gamma_{12}  & \gamma_{11}
\end{array} \right) \  , \quad
B = \left(\begin{array}{cc}
- H_1 &  0  \\
0 & - H_2
\end{array} \right)
 \label{linear}
\end{eqnarray}
and we are interested in the Lyapunov exponents, or relaxation rates around the zero-force fixed point,
i.e the eigenvalues of the matrix $M$. We have
introduced the Hessian $(H_i)_{xx'} = - \nabla_x^2 \delta_{xx'} + V_i''(u^i_{x},x)\delta_{xx'}$ in each
layer, which are hermitian matrices. They have eigenvalues $\mu_{i,\alpha_i}$ and eigenvectors
$\phi_{i,\alpha_i}(x)$. In the absence of a coupling between the layers ($\gamma_{12}=0$) the eigenvalues $\mu_{i,\alpha_i}$ are proportional to
the Lyapunov exponents, i.e.\ $\lambda_{i,\alpha_i}= - \gamma_{11}^{-1} \mu_{i,\alpha_i}$. A question is how they vary
as the viscous coupling is increased. Note that one can decompose:
\begin{eqnarray}
u^i_{xt} = \sum_{\alpha_i} \phi_{i,\alpha_i}(x) u_{\alpha_i}(t)
\end{eqnarray}
\begin{widetext}
and in that basis the equation of motion reads:
\begin{eqnarray} \sum_{\alpha_1', \alpha_2'}
\left(\begin{array}{cc}
\gamma_{11} \delta_{\alpha_1,\alpha'_1} & \gamma_{12} \int_x \phi_{1,\alpha_1}^*(x) \phi_{2,\alpha_2'}(x) \\
\gamma_{12} \int_x \phi_{2,\alpha_2}^*(x) \phi_{1,\alpha_1'}(x) & \gamma_{11} \delta_{\alpha_2,\alpha'_2}
\end{array} \right)
\left(\begin{array}{c}
\dot u_{\alpha_1'}(t) \\
\dot u_{\alpha_2'}(t)
\end{array} \right) = - \left(\begin{array}{c}
\mu_{1,\alpha_1}  u_{\alpha_1}(t) \\
\mu_{2,\alpha_2}  u_{\alpha_2}(t)
\end{array} \right)
\end{eqnarray}
\end{widetext}
since the velocity coupling between layers is local in space, it becomes non-local in
the eigenstates of the two Hessians.

The matrix $M$ has several interesting properties. Although it is not Hermitian, since $A$ and $B$ do not commute,
its eigenvalues are real. Indeed consider an eigenstate $v$ such that $M \cdot v = \lambda v$. This implies
$B \cdot v = \lambda A^{-1} \cdot v$, hence:
\begin{eqnarray}
v^\dagger \cdot B \cdot v = \lambda ~~ v^\dagger \cdot A^{-1} \cdot v   \label{lambda2}
\end{eqnarray}
Since $B$ and $A$ are Hermitian (and also real symmetric) matrices, $v^\dagger \cdot B \cdot v$ and
$v^\dagger \cdot A^{-1} \cdot v$ are real, hence $\lambda$ is real.

Consider now a bare model such that $A^{-1}$ is strictly positive definite with eigenvalues $\gamma_{+} >0$, $\gamma_{-} >0$, i.e.\ $\gamma_{12}^2 < \gamma_{11}^2$. Then (\ref{lambda2}) implies that the sign of $\lambda$ is the same as the
sign of $v^\dagger \cdot B \cdot v$. Let us consider a stable (i.e.\ attractive) zero-force point with all $\mu_{i,\alpha_i} >0$, hence $B$ is strictly negative definite. In its neighborhood in phase space, in the absence of viscous coupling between layers, the system is pinned. Since $v^\dagger \cdot B \cdot v < 0$ for any non-zero $v$, the above property implies that the zero-force fixed point remains stable, i.e.\ all Lyapunov exponents remain strictly negative, as the
viscous coupling between layers $\gamma_{12}^2 < \gamma_{11}^2$ is increased, and (\ref{lambda2}) implies the bounds
\begin{eqnarray}
\frac{\mu_{\mathrm{min}}}{\gamma_{-}} < - \lambda < \frac{\mu_{\mathrm{max}}}{\gamma_{+}}
\end{eqnarray}
for a model with $\gamma_{12}<0$, and where $\mu_{\mathrm{min}}$ and $\mu_{{\mathrm{max}}}$ are the smallest and largest eigen-values of $-B$.

These eigenvalues of stability can be obtained exactly in the case where $H_1$ and $H_2$ commute. Then one can choose the same basis in both layers $\phi_{1,\alpha}(x)=\phi_{2,\alpha}(x)$. The Lyapunov exponents, i.e.\ the eigenvalues $\lambda$ in $\dot u = \lambda u$ of (\ref{linear}) can then be
organized in pairs with:%\begin{widetext}
\begin{eqnarray}
\lambda_{\alpha} &=& - \frac{\gamma_{11} (\mu_{1,\alpha}+\mu_{2,\alpha})} {2(\gamma_{11}^2-\gamma_{12}^2)}\nn\\ &&
 \pm\frac{
\sqrt{\gamma_{11}^2 (\mu_{1,\alpha}-\mu_{2,\alpha})^2 + 4 \gamma_{12}^2 \mu_{1,\alpha} \mu_{2,\alpha}}}{2(\gamma_{11}^2-\gamma_{12}^2)}\qquad
\end{eqnarray}%\end{widetext}
and one checks that as long as $\gamma_+=\gamma_{11}-\gamma_{12} >0$ a stable FP remains stable
as $\gamma_{12}^2$ is increased (this holds for the two-particle model considered above).  In general one does not expect $H_1$ and $H_2$ to commute, since
the disorders in the two layers are uncorrelated. For small interlayer coupling one can apply second-order perturbation theory:%\begin{widetext}
\begin{eqnarray}
\lambda_{1,\alpha_1} &=& - \frac{\gamma_{11}}{\gamma_{11}^2-\gamma_{12}^2} \Bigg[  \mu_{1,\alpha_1} \\
&& + \frac{\gamma_{12}^2}{\gamma_{11}^2} \sum_{\alpha_2} \frac{\mu_{1,\alpha_1} \mu_{2,\alpha_2}}{\mu_{1,\alpha_1}
-\mu_{2,\alpha_2}} |\langle 1, \alpha_1 | 2, \alpha_2 \rangle|^2 + O(\gamma_{12}^4) \Bigg]\nn
\end{eqnarray} %\end{widetext}
which always makes the smallest eigenvalue (assumed all positive) get closer to zero, but even in the
most dangerous case when this eigenvalue is near marginal, i.e.\ $\mu_{1,\alpha_1} >0$ near zero,
the second-order correction vanishes as $\mu_{1,\alpha_1} \to 0$. Hence there is no mecanism for it
to cross zero. This is not too surprising since the determinant of the matrix in (\ref{linear})
cannot change sign as $\gamma_{12}^2<\gamma_{11}^2$ is increased. Since the eigenvalues remain
real (as shown above) they cannot continuously change sign. Hence, as above, we conclude
that a stable zero-force fixed-point remains stable. Both the size (in phase space) of its basin of attraction (pinned
phase) and the Lyapunov exponent may decrease  with increasing interlayer coupling, but they do not cross zero.

Finally note that if e.g.\ $H_1$ has a marginal direction, i.e.\ $H_1 \cdot v^1=0$, then $v=(v^1,0)$ is an eigenvector of $M$ with zero eigenvalue. Hence a marginal direction remains marginal.

\section{2-loop calculations} \label{a:2loop}
In this appendix we derive the FRG
equations up to two loops using the method of ref.\
\cite{LeDoussalWieseChauve2002}. All calculations are done at zero
velocity, at the depinning transition. All static quantities like the
disorder correlator are the same as for the standard depinning
transition, and we refer to \cite{LeDoussalWieseChauve2002} for
details. Here we only calculate the corrections to friction, i.e.\
corrections to $\cgamma_{11}$ and $\cgamma_{12}$.

\begin{figure}[b]
\centerline{\Fig{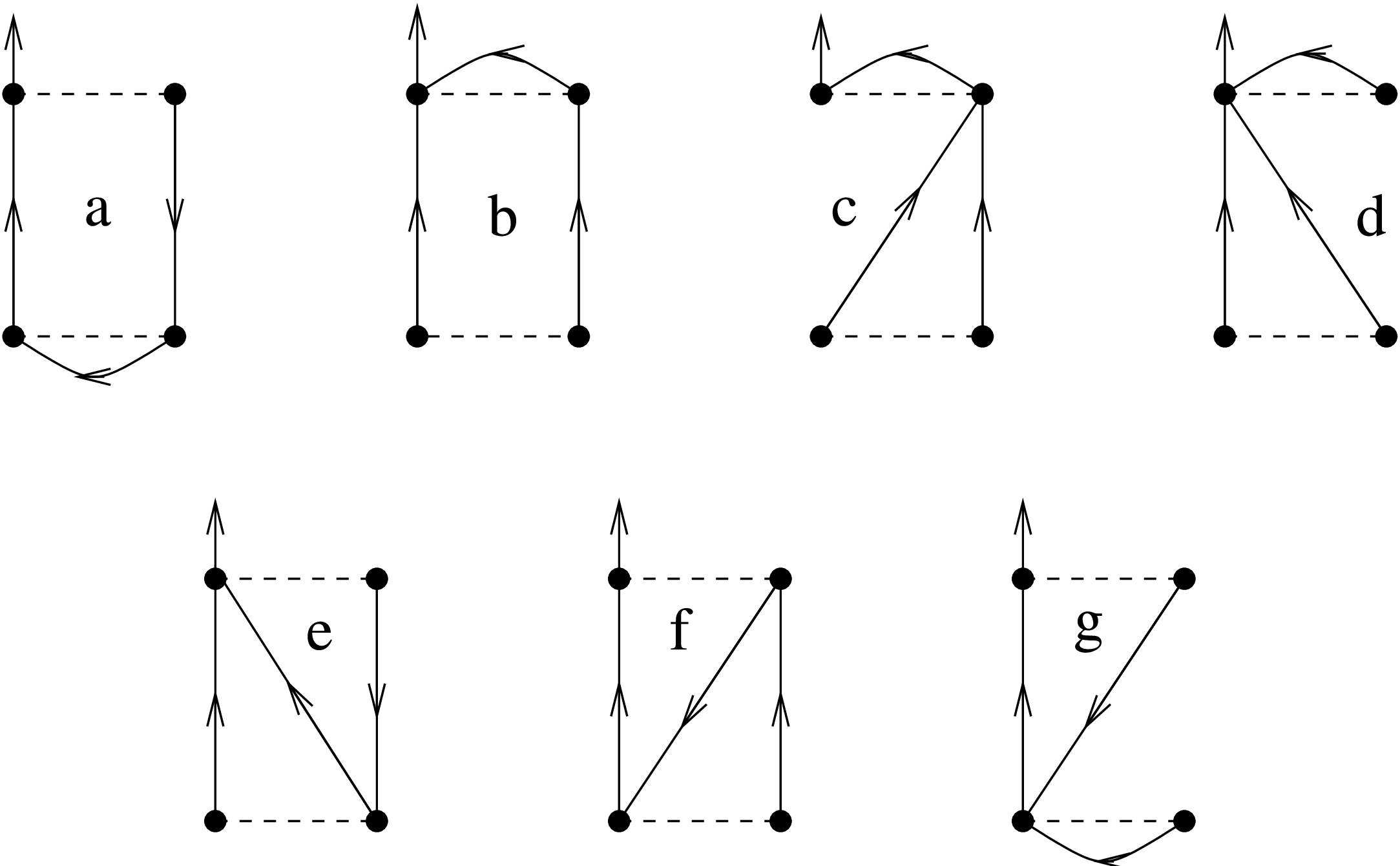}}
\caption{2-loop dynamical diagrams correcting the friction. They all
have multiplicity $8$ except (c) and (d) which have multiplicity $4$.}
\label{eta-2-loop-plas}
\end{figure}%

\subsection{1-loop order}\label{s:cor:eta}
There are no corrections to
$\cgamma_{12}$ at 1-loop order, since there exists only a single vertex,
thus one cannot get a term of the form $\int\tilde u^{2}\dot
u^{1}$. For this, one needs (at least) 2 loops.
Therefore:
\begin{equation}\label{y6}
\delta \cgamma_{11}^{\mathrm{1 loop}} = - \Delta'' (0) I_{1} \cgamma_{11}
\end{equation}
\begin{equation}\label{k32}
I_{1}:= \int_{p} \frac{1}{(p^{2}+m^{2})^{2}}
\end{equation}

\subsection{2-loop order: List of diagrams}\label{k33}
There are seven contributions,
drawn on figure \ref{eta-2-loop-plas}. Their contribution to $\cgamma$ is
symbolically
\begin{equation}
\delta \cgamma = - \frac{1}{8} \times 4 \times 2 \left[\mathrm a +\mathrm
b + \mathrm c+ \mathrm d + \mathrm e + \mathrm f + \mathrm g\right] \ .
\end{equation}
The combinatorial factor is $1/8$ from the interaction, 4 from the
time-ordering of the vertices, and an additional factor of 2 for the
symmetry of diagrams a, b, e, f and g.

The diagrams are calculated as in
\cite{LeDoussalWieseChauve2002}. When expanding the argument of
$\Delta (u_{xt}^{i}-u_{xt'}^{i})$, it is important to keep the index
of the field. Only diagrams with one disorder on one layer, and one
disorder on the other layer can give rise to a contribution to
$\cgamma_{12}$, which will be the new feature found below.

\subsection{Expressions for the diagrams}\label{y22}
The first combination is
\begin{equation}\label{y23}
\mathrm{a} + \mathrm{g} = -\Delta'' (0^{+})^{2}I_{1}^{2}
\end{equation}
as before, since the  free integration kills the inter-layer term.
In the following we give
corrections proportional to $\tilde u_{1}$. The \color{black} index $i$ runs over
both layers. Integrations over momenta and time are not written.
\begin{eqnarray}\label{y24}
&&\mathrm{b}+\mathrm{c}+\mathrm{d} = \tilde u_{1} \sum_{i} R_{1i}
(q_{1},t_1) R_{11} (q_{2},t_{2}) R_{1i}(q_{1},t_{3})\nonumber \\
&&\hphantom{\mathrm{b}+\mathrm{c}+\mathrm{d} = }\times
\left[|t_{3}-t_{1}|-|t_{3}+t_{2} -t_{1}\right] \Delta''' (0)\Delta' (0)
\dot u^{i}\nonumber \\
&&
\label{y25}\\
&&\mathrm{e} = - \frac{1}{2}\sum_{i} R_{1i}
(q_{1},t_1) R_{1i} (q_{2},t_{2}) R_{1i}(q_{3},t_{3})|t_{3}-t_{2}|\nonumber \\
&& \qquad \qquad  \times  \Delta''' (0)\Delta' (0)
\dot u^{1}
\end{eqnarray}
\begin{equation}\label{y26}
\mathrm{f} = - 2 \Delta''' (0^{+}) \Delta' (0^{+}) I_{A} - 2 \Delta''
(0^{+})^{2} I_{A}
\end{equation}
Integrating  over times yields the diagrams
presented in the next two subsections.
They  involve the following non-trivial momentum integrals:
\begin{eqnarray}\label{c1}
I_{\cgamma }&:=&  \int\limits_{q_1,q_2}
\frac{1}{( q_1^2+m^{2}) ( q_2^2+m^{2})^{2} (q_2^2 + q_3^2+2m^{2})}\nn \\
&=& \left(\frac{1}{2\E^2}+\frac{1-2\ln 2}{4\E} \right) (\E I_1)^{2}
+\mbox{finite}
\end{eqnarray}
\begin{eqnarray}\label{c2}
I_A&:=&\int \frac{\rmd^d q_1}{(2\pi)^d} \frac{\rmd^d q_2}{(2\pi)^d}
\frac{1}{q_1^2+m^2}\frac{1}{q_2^2+m^2} \frac{1}{(
(q_1{+}q_2)^2+m^2)^2}\nn\\
&=& \left(\frac1{2\E^2} + \frac1{4\E} +O(\E^2) \right) (\E I_1)^2
\ .
\end{eqnarray}
They are calculated in \cite{LeDoussalWieseChauve2002}.
\begin{equation}\label{c3}
\int_{q_{1}q_{2}}\frac{1}{( q_{1}^{2}+m^{2})( q_{2}^{2}+m^{2})^{2}( q_{1}^{2}+
q_{2}^{2}+2 m^{2})} = \frac{1}{2}I_{1}^{2}\ ,
\end{equation}
as can be seen by symmetrizing in $q_1$ and $q_2$.

\subsection{Corrections to $\cgamma_{12}$}\label{y13}
\begin{equation}\label{y14}
\delta \cgamma_{12}^{\mathrm  a +\mathrm  g} = 0
\end{equation}
%%%
\begin{widetext}
\begin{eqnarray}\label{2.11}
\delta \cgamma_{12}^{ \mathrm b +\mathrm c + \mathrm d} &=&\int_{q_{1}q_{2}}
\frac{\cgamma_{12}^2 \left(\left(q_1^2+q_2^2\right)
\cgamma_{11}^2+\left(q_2^2-q_1^2\right) \cgamma _{12}^2\right) \Delta'(0^+)
\Delta'''(0^+)}{2 q_1^2 q_2^2 \left(q_1^2+q_2^2\right) \cgamma _{11}
\left(\left(q_1^2+q_2^2\right)^2
\cgamma_{11}^2-\left(q_1^2-q_2^2\right)^2 \cgamma _{12}^2\right)}
\end{eqnarray}
\end{widetext}
%%%
In principle, (\ref{2.11}) should be written with massive
propagators. We have put $m=0$ for notational compactness.
It is easy to see that for generic values of $\cgamma_{11}$ and
$\cgamma_{12}$ (\ref{2.11}) has no subdivergence for either $q_{1}\to
\infty$ or $q_{2}\to \infty $, only if they become large together.
Properly regularized, it therefore has only a single pole in
$\epsilon$, and this pole is universal, i.e.\ independent of the
regularization scheme. For such an integral, which moreover is
homogenous in $q_{1}$ and $q_{2}$, the pole  can be expressed  as
\begin{equation}\label{magic}
\int_{q_{1},q_{2}} f (q_{1}^{2}, q_{2}^{2},m^{2}) = \frac{(\epsilon
I_{1})^{2}}{\epsilon} \int _{0}^{\infty} \rmd (q_{2}^{2}) f
(1,q_{2}^{2},0)\, q_2^{2} +O (\epsilon^{0})
\end{equation}
This is proven using conformal mappings of the different sectors, and
was established in \cite{WieseDavid1995,DavidWiese1996,WieseDavid1997,WieseHabil}. Accepting that the integral
is indeed universal, a quick  way of deriving (\ref{magic})
is as follows
\begin{equation}\label{k34}
\int_{q_{1},q_{2}} f (q_{1}^{2}, q_{2}^{2},m^{2})
\approx \int_{q_1<\Lambda}\int_{q_{2} } f (q_{1}^{2}, q_{2}^{2},0) \approx   F
\frac{\Lambda^{\epsilon}}{\epsilon } \ ,
\end{equation}
where the ``$\approx$'' indicates up to terms of $O(\epsilon)$.
To obtain the residue $F$, we derive w.r.t.\
$\Lambda$, and then set $\Lambda=1$:
\begin{equation}\label{k35}
F \approx  S_{D} \int_{q_{2} } f (q_{1}^{2}=1, q_{2}^{2},0)
\end{equation}
Since the integral is finite, we can take the limit of $\epsilon \to0$
or $d\to4$. This gives the result (\ref{magic}) up to an overall
normalization, which is also easily checked.

It is now straightforward to integrate (\ref{2.11}) using (\ref{magic}):
%{ \tiny
%\begin{equation}\label{delta-eta-12-bcd}
%\delta \cgamma_{12}^{\mathrm{a}+\mathrm{b}+\mathrm{c}}=\frac{ \cgamma _{12}
%\Delta'(0^+) \Delta
%   ^{(3)}(0)}{8 \epsilon }\log \left(\frac{\left(\cgamma _{11}+\cgamma
%   _{12}\right)^2}{\left(\cgamma _{11}-\cgamma _{12}\right)^2}\right)
%\end{equation}}
\begin{equation}\label{delta-eta-12-bcdp}
\delta \cgamma_{12}^{\mathrm{a}+\mathrm{b}+\mathrm{c}}=\frac{ \cgamma _{12}
\Delta'(0^+) \Delta
   ^{(3)}(0)}{4 \epsilon }\ln \left|\frac{\cgamma _{11}+\cgamma
   _{12}}{\cgamma _{11}-\cgamma _{12}}\right|
\end{equation}
Note that physically one has to restrict to $\cgamma_{+}>0$ and $\cgamma_{-}>0$, thus all results are to be
taken in this domain only. The absolute-value therefore represents nothing but a notational commodity.

The other two diagrams are trivial:
\begin{eqnarray}\label{y20}
\delta \cgamma_{12}^{\mathrm{e}} &=& 0\\
\delta \cgamma_{12}^{\mathrm{f}} &=& 0 \ .
\end{eqnarray}

\subsection{Corrections to $\protect\cgamma_{11}$}\label{y7} Grouping
diagrams, which partially cancel, we find for the corrections to
$\cgamma_{11}$: %{\red \tiny(signs are such that on the r.h.s.\ there appear the
%diagrams with an additional minus signs)}
\begin{equation}\label{y8}
\delta \cgamma_{11}^{\mathrm  a +\mathrm  g} =  \cgamma_{11} \Delta''(0^+)^2  I_1^2
\end{equation}
\begin{widetext}
\begin{eqnarray}\label{k36}
\delta \cgamma_{11}^{\mathrm b +\mathrm c + \mathrm d}
&=&\int_{q_{1}q_{2}} \frac{\left(2 \left(q_1^2+q_2^2\right)^2 \cgamma
_{11}^4+\left(-2 q_1^4+3 q_2^2 q_1^2+q_2^4\right) \cgamma _{12}^2 \cgamma
_{11}^2+q_2^2 \left(q_1^2-q_2^2\right) \cgamma _{12}^4\right) \Delta'(0^+)
\Delta'''(0^+)}{2 q_1^2 q_2^4 \left(q_1^2+q_2^2\right) \cgamma _{11}
\left(\left(q_1^2+q_2^2\right)^2
\cgamma_{11}^2-\left(q_1^2-q_2^2\right)^2 \cgamma _{12}^2\right)}\nonumber\\
 &&\stackrel{\cgamma_{12}=0}{-\!\!\!-\!\!\!\longrightarrow} \int_{q_{1}q_{2}}\frac{\cgamma
_{11} \Delta'(0^+) \Delta'''(0^+)}{q_1^2 q_2^4
\left(q_1^2+q_2^2\right)}
\end{eqnarray}
%\end{widetext}
$\delta \cgamma_{11}^{\mathrm b +\mathrm c + \mathrm d}$ is a little bit
more complicated to calculate analytically, since it has a
subdivergence, which   has to be subtracted if we want to use the
magic relation (\ref{magic}). We observe that subtracting the term at
$\cgamma_{12}=0$, the diagram has no longer a subdivergence. In order to
proceed, one then uses (\ref{magic}), and performs a  partial fraction decomposition (the
variable is $q_{2}^{2}$) of the
remaining term, leading
to integrals known by Mathematica. The final result is
\begin{eqnarray}\label{y9}
\delta \cgamma_{11}^{\mathrm b +\mathrm c + \mathrm d} &=&  \frac{\cgamma
_{11} \Delta'(0^+) \Delta'''(0^+)}2 \, I_{1}^{2}
%\nonumber \\&&
+ \frac{ \cgamma _{12} \Delta'(0^+) \Delta'''(0^{+})}{4 \epsilon }\log \left|\frac{\cgamma _{11}+\cgamma
_{12}}{ \cgamma _{11}-\cgamma _{12}}\right|\qquad~~
\end{eqnarray}
%{\tiny This can be proven as follows: First, a very strong check is to Taylor-expand in $\cgamma_{12}$ to
%order 60, and compare the integrated series with $\delta \cgamma_{12}^{\mathrm{b}+\mathrm{c}+\mathrm{d}} $,
%which conincide. Second, the integrand at $q_{1}^{2}=1$, multiplied by $q_{2}^{2}$, see (\ref{magic}) above, %canbe partial fraction decomposed, and integrated. The result is the same. }
%\begin{widetext}
The next diagram is
\begin{eqnarray}\label{y9a}
\delta \cgamma_{11}^{\mathrm{e}} &=&\int_{q_{1}q_{2}} \frac{\cgamma _{11}
\left(\left(q_2^2+q_3^2\right)^2 \left(q_2^4+q_3^4\right) \cgamma
_{11}^2-\left(q_2^8-2 q_3^2 q_2^6-2 q_3^4 q_2^4-2 q_3^6
q_2^2+q_3^8\right) \cgamma _{12}^2\right) \Delta'(0^+) \Delta'''(0^+)}{2
q_1^2 q_2^4 q_3^4 \left(q_2^2+q_3^2\right)
\left(\left(q_2^2+q_3^2\right)^2 \cgamma
_{11}^2-\left(q_2^2-q_3^2\right)^2 \cgamma _{12}^2\right)}\ .
\end{eqnarray} \end{widetext}
We have used the abbreviations  $\vec q_{3}:=\vec q_{1}+\vec
q_{2}$,  and $q_{3}:=|\vec q_{3}|$.
Again, this diagram has a subdivergence (double pole), which we want
to subtract. Let us again try the term at $\cgamma_{12}\to 0$:
\begin{eqnarray}\label{k36p}
\delta \cgamma_{11}^{\mathrm{e}}&\stackrel{\cgamma_{12}\to 0}{-\!\!\!-\!\!\!\longrightarrow} & \int_{q_{1}q_{2}}
\frac{\left(q_2^4+q_3^4\right) \cgamma _{11} \Delta'(0^+) \Delta
'''(0^{+})}{2 q_1^2 q_2^4 q_3^4 \left(q_2^2+q_3^2\right)} \nonumber \\
&=& \int_{q_{1}q_{2}}
\frac{ \cgamma _{11} \Delta'(0^+) \Delta
'''(0^{+})}{ q_1^2 q_2^4  \left(q_2^2+q_3^2\right)}
\end{eqnarray}
$\delta \cgamma_{11}^{\mathrm{e}}$ can be rewritten as:
\begin{eqnarray}\label{y10}
\delta \cgamma_{11}^{\mathrm{e}} &=& \Delta' (0) \Delta''' (0)
\int_{q_{1}q_{2}} \Bigg[\frac{\left(q_2^4+q_3^4\right) \cgamma _{11}}{2 q_1^2
q_2^4 q_3^4 \left(q_2^2+q_3^2\right)} \nonumber \\
&& + \frac{2 \cgamma _{11} \cgamma
_{12}^2}{q_1^2 \left(q_2^2+q_3^2\right)^3 \cgamma _{11}^2-q_1^2
\left(q_2^2-q_3^2\right)^2 \left(q_2^2+q_3^2\right) \cgamma _{12}^2} \Bigg]\nonumber \\
&=& \Delta' (0) \Delta''' (0) \int_{q_{1}q_{2}}  \Bigg[\frac{ \cgamma _{11}}{
q_1^2 q_2^4 \left(q_2^2+q_3^2\right)}  \nonumber \\
&& + \frac{2 \cgamma _{11}
\cgamma_{12}^2}{q_1^2 \left(q_2^2+q_3^2\right)
\left(\left(q_2^2+q_3^2\right)^2
\cgamma_{11}^2-\left(q_2^2-q_3^2\right)^2 \cgamma _{12}^2\right)} \Bigg]\nonumber \\
\end{eqnarray}
The last integral is
\begin{eqnarray}\label{y11}
 &&\int_{q_{1}q_{2}} \frac{1}{q_1^2 \left(q_2^2+q_3^2\right)
\left(\left(q_2^2+q_3^2\right)^2
\cgamma_{11}^2-\left(q_2^2-q_3^2\right)^2 \cgamma _{12}^2\right)}
\nonumber \\
&&\qquad =
\frac{\log \left|\frac{\cgamma _{11}+\cgamma _{12}}{\cgamma _{11}-\cgamma
_{12}}\right|}{4 \epsilon \cgamma _{11} \cgamma _{12}}+\frac{\log
\left|1-\frac{\cgamma _{12}^2}{\cgamma _{11}^2}\right|}{4 \epsilon \cgamma
_{12}^2}
\end{eqnarray}
%{  \tiny({\bf the absolute-value in the last log is tentative!})}
\begin{widetext}\noindent
One way to prove this is as follows. Introduce Schwinger-parameters to
write the l.h.s.\ of (\ref{y11}) as
\begin{equation}\label{y28}
\int_{q_{1},q_{2}}\int_{s_{1}>0, s_{2}>0, s_{3}>0,s_{4}>0} \rme^{-s_1
\left(q_2+q_3\right)^2-\left(q_2^2+q_3^2\right) s_2-s_3 \left(\cgamma
_{11} q_2^2+\cgamma _{12} q_2^2+q_3^2 \cgamma _{11}-q_3^2 \cgamma
_{12}\right)-s_4 \left(\cgamma _{11} q_2^2-\cgamma _{12} q_2^2+q_3^2 \cgamma
_{11}+q_3^2 \cgamma _{12}\right)} \rme^{-s_{2}}
\end{equation}
where we have introduced a mass for  $s_{2}$ only
(using again universality of the leading pole in $\epsilon$).
Then integrate over the $q_{i}$'s:
\begin{equation}\label{y29}\int_{s_{1}>0, s_{2}>0, s_{3}>0,s_{4}>0}  \rme^{-s_{2}}
\left(-\left(s_3-s_4\right)^2 \cgamma _{12}^2+s_2 \left(2
s_1+s_2\right)+\left(s_3+s_4\right) \cgamma _{11} \left(2
\left(s_1+s_2\right)+\left(s_3+s_4\right) \cgamma
_{11}\right)\right)^{-d/2}
\end{equation}
\end{widetext}
Rescale all $s_{i}$ with $i\neq 2$ by $s_{2}$ and integrate over
$s_{2}$. Then go to new variables $s_{3}\to (s+t)/2$, $s_{4}\to
(s-t)/2$. Our integral becomes
\begin{equation}\label{2.27}
\int_{s>0}\int_{-s}^{s}\rmd t \int_{s_{1}>0} \frac{\Gamma (\epsilon
)}{2 \left(\left(s \cgamma _{11}+1\right) \left(2 s_1+s \cgamma
_{11}+1\right)-t^2 \cgamma _{12}^2\right)^2}
\end{equation}
The result can be simplified to (\ref{y11}). A tricky point are logs
halfway. Expanding (\ref{2.27}) in $\cgamma_{12}$, we circumvent the
problem and can check the first terms of the Taylor series.
The complete result for $\delta \cgamma_{11}^{\mathrm{e}}$ is
(up to finite terms)
\begin{eqnarray}\label{y30}
\delta \cgamma_{11}^{\mathrm{e}} &=&  \Delta' (0) \Delta''' (0) \bigg[ \cgamma _{11}
I_{\cgamma }
\nonumber \\
&&
 +\frac{\cgamma_{12} }{2\epsilon}
\log \left|\frac{\cgamma _{11}+\cgamma _{12}}{\cgamma _{11}-\cgamma
_{12}}\right|
 +\frac{\cgamma_{11} }{2\epsilon}
\log
\left|1-\frac{\cgamma _{12}^2}{\cgamma _{11}^2}\right| \bigg]\qquad \quad
\end{eqnarray}
%{  \tiny We performed a non-trivial check: We integrated the (Taylor-expanded) integral (\ref{y11}), where
%$q_{1}^{2}$ had been replaced by $q_{2}^{2}$, using (\ref{magic}). This checked the code and normalisations.}

\noindent The final diagram is
\begin{widetext}
%%%%%%%%%%%  diagram f %%%%%%%%%%%%
\begin{eqnarray}\label{y12}
\delta \cgamma^{\mathrm{f}}_{11}&=& \int_{q_{1}q_{2} }\frac{\cgamma _{11}
\left(\left(q_1^2+q_2^2\right) \Delta'(0^+) \Delta'''(0^+)
q_3^2+q_1^2 \left(q_2^2+q_3^2\right) \Delta ''(0)^2\right)}{q_1^4
q_2^4 q_3^4} = 2 \cgamma _{11} \int_{q_{1}q_{2} }  \frac{ \Delta'(0^+)
\Delta'''(0^+)+\Delta'' (0)^{2}}{q_{1}^{2}q_{2}^{4}q_{3}^{2}}\nonumber \\
&=& 2 \cgamma_{11} \left[{ \Delta'(0^+)
\Delta'''(0^+)+\Delta'' (0)^{2}} \right] I_{A} \ .
\end{eqnarray}
This gives the flow equations given in the main text.
%\end{widetext}
%\twocolumngrid
\end{widetext}

%\bibliography{citation,cristina,mcm,driven}
%\bibliography{citation,mcm}
%\bibliography{citation}
%\bibliography{citation,mcm,driven}

%\bibliography{citation,mcm,cristina}

\begin{thebibliography}{100}
\bibitem{hhh4}
{Additional
effects from KPZ-like terms generated from the partial breakdown of the statistical tilt symmetry (in the
relative displacement) will be studied elsewhere.}
\end{thebibliography}

\bigskip

\tableofcontents

\end{document}